\documentclass[%
reprint,
superscriptaddress,
nofootinbib,
 amsmath,amssymb,
prx,
floatfix,
]{revtex4-2}

\usepackage{mathtools}

\usepackage{lipsum,bbm,dsfont} 

\usepackage{graphicx}
\usepackage{dcolumn}
\usepackage{bm}
\usepackage[hidelinks,colorlinks=true,linkcolor=blue,citecolor=blue]{hyperref}
\usepackage{xcolor}
\usepackage{dsfont}
\usepackage{graphicx}
\usepackage{adjustbox}

\usepackage{caption} 
\usepackage{array}
\newcolumntype{P}[1]{>{\centering\arraybackslash}p{#1}}

\usepackage{braket}

\newcommand{\dket}[1]{\lvert #1 \rangle\!\rangle}

\begin{document}

\preprint{APS/123-QED}









\title{Superpositions of Quantum Gaussian Processes}

\author{Lorenzo Braccini}
\affiliation{Department of Physics and Astronomy, University College London, Gower Street, WC1E 6BT London, United Kingdom}

\author{Sougato Bose}
\affiliation{Department of Physics and Astronomy, University College London, Gower Street, WC1E 6BT London, United Kingdom}

\author{Alessio Serafini}
\affiliation{Department of Physics and Astronomy, University College London, Gower Street, WC1E 6BT London, United Kingdom}

\date{\today}
\begin{abstract}

We generalise the Gaussian formalism of Continuous Variable (CV) systems to describe their entanglement with Discrete Variable (DV) systems, leading to superpositions of CV Gaussian states. A new class of CV-DV entangled states, named Gaussian-Branched Cat States (GBCSs), yields an analytical formalism to describe quantum hybrid systems. GBCSs~are fully characterised by their superposed phase-space parameters: sets of generalised complex first moments and covariance matrices, along with the DV reduced density matrix (phases and contrasts). These states arise in all the instances where Gaussian dynamics, operations, and measurements are performed conditionally on a DV state. The time evolution of the GBCS phase-space parameters allows one --  via a new set of equations in closed form -- to analytically treat a large set of unitary and open dynamics, generated by Gaussian Hamiltonians labelled by DV eigenvalues. Conditional operations, such as displacements and rotations, and Gaussian measurements (homodyne/heterodyne) jointly with DV projectors, can be both described as maps on GBCS's parameters. A phase-space perturbation theory is given to extend the analysis to non-orthogonal DV super-operators, e.g. DV decay. We showcase our general formalism with two paradigmatic examples of experimental modelling: (i) a dispersively coupled qubit to a driven parametric amplifier; (ii) a levitated nanoparticle undergoing Stern-Gerlach matter-wave interferometry in a diffusive environment. Both examples highlight the generation of novel Wigner negativities through qubit measurements. 

\hspace{0.5cm}

\end{abstract}

\maketitle

\section{Introduction \label{sec:intro}}

The interplay between Discrete-Variable (DV) quantum systems, such as qubits, and Continuous-Variable~(CV) ones, such as quantum modes, is pivotal to most of the current research in quantum technology and experiments, ranging from the detection of quantum correlations to state-of-the-art quantum computers~\cite{hensen_loophole_2015,hybrid_Liu_2026}. Hybrid CV-DV systems can be used to engineer a vast number of quantum protocols, from non-classical state preparation in the CV to computation in the DV.

In CV systems, quantum Gaussian processes \textit{per se} have found application in a wide number of protocols~\cite{serafini_quantum_2017,quantum_braunstein_2003,weedbrook_gaussian_2012}, such as entanglement generation~\cite{peres_horodecki_2000,giedke_entanglement_2003,garttner_entanglement_2023}, quantum key distribution~\cite{rudolph_teleportation_2001,grosshans_cryptography_2002} and quantum control routines~\cite{wiseman_optimal_2005,wiseman_quantum_2010}. Describing quantized bosonic field operators, CVs have found experimental implementations in quantum optics~\cite{walls_quantum_1994,bowen_experimental_2003,bowen_teleport_2003}, circuit electrodynamics~\cite{blais_circuit_2021,flurin_generating_2012,fedorov_unconditional_2021}, atomic systems~\cite{leibfried_quantum_2003}, collective spins for atomic ensembles~\cite{hammerer_quantum_2010}, and levitated nanoparticles~\cite{ballestero_levitodynamics_2021,delic_cooling_2020,tebbenjohanns_motional_2020,magri_real_2021}. The Gaussian structure of states, interactions, noises, operations, and measurements allows for an exact description at the level of the phase space (avoiding truncations), dispensing one with the cumbersome task of solving quantum Gaussian processes in the infinite-dimensional Hilbert space~\cite{serafini_quantum_2017}. In fact, Gaussian processes can be formulated in terms of evolutions and maps of a vector of expectation values (first moments) and a matrix of uncertainties (covariance matrix). Uniquely determining a Gaussian state, these state parameters have a clear operational description as they are directly related to quantities that are often measured in the experimental practice. 

\begin{figure}[!b]
    \centering
    \includegraphics[width=1\linewidth]{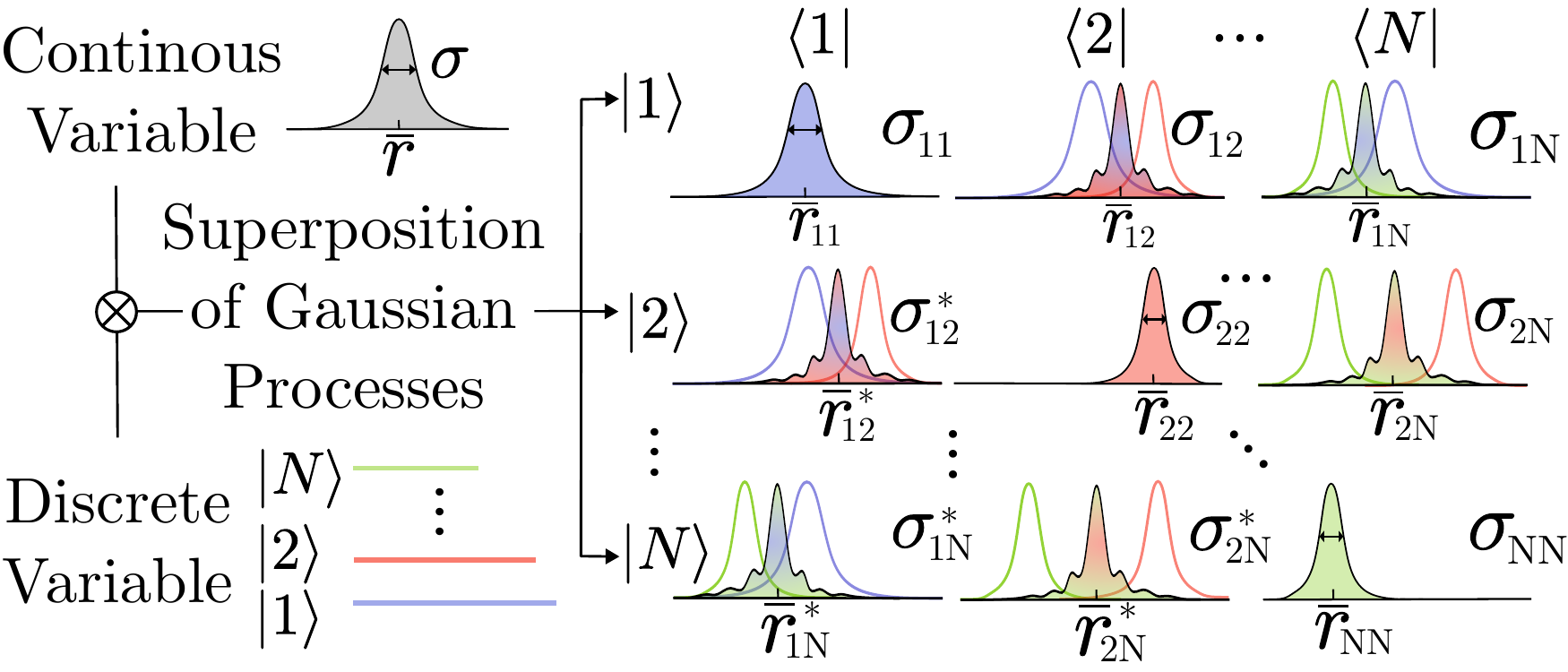}
    \caption{Representation of a superposition of Gaussian processes generating a Gaussian-Branched Cat State.}
    \label{fig:enter-label}
\end{figure}

\newpage However, in CV systems, non-Gaussian states are crucial to unlock most quantum advantages, since Gaussian states can in principle be mimicked by classical probability distributions: for instance, non-Gaussian elements are needed for universal quantum computation~\cite{lloyd_quantum_1999,Jeong_Effective_2002,Ralph_Quantum_2003,albarelli_resource_2018,Eriksson2024} and error correction~\cite{gottesman_encoding_2001,campagne-ibarcq_quantum_2020,dahan_creation_2023}. Non-Gaussian states represent excellent quantum sensors, which could test the macroscopic limit of quantum mechanics and non-classical gravitational effects~\cite{romero-isart_large_2011,roda_macroscopic_2024,bose_spin_2017,marletto_gravitationally-induced_2017}. A classic way to generate non-Gaussianity and delocalisation is to create a distinguishable superposition of CV Gaussian states, often named Schr\"odinger cat states~\cite{Shore_jaynes_1993}. They naturally arise in CV-DV hybrid systems from fundamental light-matter interactions and, typically, these hybrid states are of the form
\begin{equation}
    \ket{\psi}= \frac{1}{\sqrt{2}}\left( \ket{g_\uparrow} \ket{\uparrow} + \ket{g_\downarrow}\ket{\downarrow} \right) ,
\end{equation}
where $\ket{\uparrow}$ and $\ket{\downarrow}$ are some DV states and $\ket{g_i}$ labels some CV Gaussian state.

CV-DV cat states have arisen in classic foundational experiments~\cite{Monroe_Schrodinger_1996,brune_observing_1996} and are still central to recent protocols~\cite{cavailles_demosntration_2018,schrodinger_bild_2023,Realization_Xiaozhou_2025}, as they possess quantum advantages for both technological applications and foundational tests~\cite{Sanders_2012,Munro_weekforce_2002,Stob_violation_2007,Leghtas_Hardware_2013,Mirrahimi_Dynamically_2014,Bergmann_quantum_2016,wbhn-v1sw,Reglade_quantum_2024}. It is remarkable how often quantum systems have shown the capability of generating such states, for instance, in the fields of non-linear quantum optics~\cite{gerry_quantum_1997,serafini_minimum_2004} and optomechanics~\cite{bose_preparation_1997,bose_scheme_1999,qvarfort_time-evolution_2020,marchese2026cascadedoptomechanicalsensingsmall}, atomic~\cite{machluf_coherent_2013,Amit_T_2019,margalit_realization_2020,keil_stern_2021} and matter-wave~\cite{bose_qubit_2006,Romero_Isart_quantum_2011,scala_matter_2013} interferometry, and quantum field theory~\cite{Funai_gaussian_2025}. Their applications represent the building blocks of modern quantum computers in numerous architectures~\cite{hybrid_Liu_2026,xiang_hybrid_2013}, including superconducting~\cite{strong_wallraff_2004,coherent_chiorescu_2004,blais_quantum_2007,coupling_majer_2007}, solid-state~\cite{strong_samkharadze_2018,strong_samkharadze_2018,coherent_mi_2018,coherent_landig_2018,strong_kubo_2010,strong_chen_2018,realization_liu_2023}, and ion-trap systems~\cite{cirac_quantum_1995,sorensen_quantum_1999,leibfried_quantum_2003,haljan_spin_2005,Fluhmann2019,gan_hybrid_2020}.  The frequent appearance of these states is due to the fact that, often, the leading-order CV-DV non-linearities -- hence, the ones common in practical set-ups~\cite{florian_ion_2001,webb_resilient_2018,kaku2025suddendecoherenceresonantparticle,tobar2025quantummetasurfacesprobesvacuum,Ayyash_driven} -- are the ones that are quadratic in the CV observables $\hat{X}$ and $\hat{P}$ (or $\hat{a}$ and $\hat{a}^\dag$) and functions of a single DV operator. We shall term this set of  Hamiltonians as operator-valued Gaussian interactions. 

In some instances, both CV and DV subsystems play a central role in the hybrid system, while, in other instances, the presence of either subsystem is hidden behind effective Hamiltonians as direct qubit-qubit~\cite{Kentaro_crosscross_2021} or non-Gaussian CV interactions, such as Kerr non-linearities~\cite{semiao_kerr_2009,Nigg_black_2012,Leghtas_deterministic_2013}. Yet, fundamentally and operationally, such interactions are almost always mediated by CV subsystems, such as positions of atoms or states of resonators. Solutions for experimental predictions of these hybrid systems are often achieved numerically by truncating the CV Hilbert space (in the $n$ basis or other appropriate spectral decompositions) or limited to specific analytical examples~\cite{hybrid_Liu_2026}.

Despite the obvious importance of these states, to the best of our knowledge, no generalisation of the Gaussian CV formalism exists to analytically treat entangled CV-DV states under open-system dynamical conditions, measurements and operations, which generate superpositions of Gaussian states.
We therefore introduce ``Gaussian-Branched Cat States'' (GBCS): $n$-mode and $N$-dimensional DV entangled states, where the terms of the superposition are tensor products of Gaussian states and a set of orthogonal qubit states. GBCS admit an operational description where a phase space of $n$ CV degrees of freedom is associated with each element of an $N\times N$ DV density matrix forming, so to speak, a ``superposition of phase spaces''. In this representation, any GBCS is uniquely characterised by sets of $2n$-dimensional complex vectors and matrices (containing the first moments and covariance matrix of each branch), and by the reduced density matrix of the DV. In practice, GBCS arise in all protocols where Gaussian operations, dynamics, and measurements, are performed conditionally on the state of a DV subsystem, alongside any local quantum Gaussian CV processes and arbitrary DV operations, which we shall term \textit{``Superpositions of Quantum Gaussian Processes''}. 

\newpage
Being based on Gaussian building blocks, the chosen representation of GBCS allows one to directly apply a large set of 
Gaussian phase-space results, firstly developed in quantum optics~\cite{serafini_quantum_2017} and summarized in Sec.~\ref{sec:intro_phase}. In Sec.~\ref{sec:sup_phase_space}, the most general GBCS is defined via its phase-space parametrisation. At the dynamical level, GBCS solve without truncation the most general operator-valued Gaussian interaction between $n$ modes and an $N$-level system via the time evolution of its superposed phase-space parameters, formulated in Sec.~\ref{sec:dynamics} for both unitary and open dynamics. The two central case studies of this work are the dispersive coupling ($\hat{a}^{\dag} \hat{a} \otimes \hat{\sigma}_z$) and the Stern-Gerlach coupling ($\hat{X} \otimes \hat{\sigma}_z$), describing qubits in off-resonance time-dependent electromagnetic fields and gradients, respectively~\cite{xiang_hybrid_2013}.  Sec.~\ref{sec:operations} deals with conditional operations, such as rotations and displacements, which -- with the inclusion of local qubit and/or Gaussian operations -- have been shown to be universal for computation of either the CV and for the DV subsystems~\cite{Hybrid_loock_2008,krastanov_universal_2015,Heeres_Implementing_2017,Eickbusch_fast_2022,grosso_controlled_2025}. This state parametrisation naturally gives a phase-space picturization of such non-Gaussian processes and provides a geometrical understanding of Wigner fringes,  illustrated in Sec.~\ref{sec:wigner}.  
Gaussian and qubit (ideal and noisy) joint measurements can be described as maps of the phase-space parameters (Sec.~\ref{sec:measure}). In Sec.~\ref{sec:perturbation}, we place GBCS in the context of a perturbation expansion, thus allowing one to describe non-diagonal DV super-operators such as DV decaying noise. By analytically describing the leading-order non-linearity and noises, along with the measurement and operations accessible in experiments, the proposed formalism is applicable to solving structured dynamics and make predictions for experiments in realistic conditions. In Sec.~\ref{sec:physical_examples}, our methodology is applied to the modelling of physical examples: (i) the dispersive coupling of a qubit with a parametric amplifier; (ii) a Stern-Gerlach matter-wave interferometer for a levitated nanoparticle in a diffusive environment. This allows us to give comparisons to other known methods for the experimental modelling of hybrid systems. Conclusions are drawn in Sec.~\ref{sec:conclusion}. 

\section{Introduction to Gaussian States \\ \& Phase Space Methods \label{sec:intro_phase}}

In this section, with the purpose of introducing notation and laying out the Gaussian techniques that will be used, a review of Gaussian phase space methods for CV systems is presented (for more details and derivation, see Ref. \cite{serafini_quantum_2017}). As we shall see, this representation of Gaussian quantum states allows for the analytical treatment of a large set of dynamics, measurements, and operations, which will be key to the remainder of this work. 

The expectation value of a quantum operator $\hat{O}$ is denoted with $\braket{\hat{O}} = \text{Tr} [ \hat{O} \hat{\varrho} ]$, where $\hat{\varrho} $ is the  density matrix of the system under exam (operators will wear a hat, $\hat{\cdot}$, while phase space quantities will not). Consider a set of $n$ continuous degrees of freedom, with canonical operators $\hat{X}_j$ and $\hat{P}_j$. The Hilbert space associated with the $n$-dimensional CV system (modes) is $\mathcal{H} = L^2 (\mathds{R}^n)$. The dimensionless canonical operators $\hat{x}_j=\hat{X}_{j}\sqrt{\frac{m_j\omega_j}{\hbar}}$ and $\hat{p}_j=\hat{P}_{j}\sqrt{\frac{1}{m_j\omega_j\hbar}}$, where $m_j$ and $\omega_j$ are the mass and frequency associated with the degree of freedom, have commutation relations $[\hat{x}_j, \hat{p}_k] = i\delta_{jk}$.  Thus, it is possible to define the vector of operators
\begin{equation}
    \hat{r} := \begin{pmatrix}
        \hat{x}_1 &
        \hat{p}_1 &
        \hat{x}_2 & 
        ... &
        \hat{x}_n &
        \hat{p}_n 
    \end{pmatrix}^{\rm T} \; , 
\end{equation}
such that the commutator relations read $[ \hat{r}, \hat{r}^{\rm T} ] = i \Omega_n $, where  $\Omega_n = \bigoplus_{i = 1}^n \Omega_1$ is the $n$-dimensional symplectic form, with
\begin{equation}
    \Omega_1 = \begin{pmatrix}
        0 & 1 \\
        -1 & 0
    \end{pmatrix} \;.
\end{equation}
It may be useful to recall the definition of the symplectic group as the set of real matrices $S \in Sp_{2, \mathds{R}}$ such that $S \Omega S^{\rm T} = \Omega$. Furthermore, the symplectic inner product is antisymmetric, i.e., $r_1^{\rm T} \Omega r_2 = - r_2^{\rm T} \Omega r_1 $ for any pair of vectors, and $\Omega^{-1} = \Omega^{\rm T} = - \Omega$. The ladder operators are defined as $\hat{a}_i = (\hat{x}_i + i \hat{p}_i)/ \sqrt{2}$, and one can move from one representation to the other by the use of the transformation $U_n = \bigoplus_{i = 1}^n U_1$ such that the vector  $\hat{a} = U_n \hat{r} = (\hat{a}_1, \hat{a}_1^\dag, ... , \hat{a}_n, \hat{a}_n^\dag)^{\rm T}$, with
\begin{equation}
\label{eq:change_basis}
    U_1 = \frac{1}{\sqrt{2}}
   \begin{pmatrix}
        1 & i \\
        1 & -i
    \end{pmatrix}   \implies  \begin{pmatrix}
        \hat{a} \\
        \hat{a}^\dag
    \end{pmatrix} = U_1 \begin{pmatrix}
        \hat{x} \\
        \hat{p}
    \end{pmatrix}\;.
\end{equation}
In this compact notation, the first and second moments of a quantum state $\hat{\varrho}$ are defined as the expectation values $\bar{r} := \text{Tr} [ \hat{r} \hat{\varrho}]$ and $\sigma := \text{Tr} [ \{ (\hat{r}-\bar{r}), (\hat{r}-\bar{r})^{\rm T} \} \hat{\varrho} ] $, respectively, where $\{ \cdot, \cdot \}$ represents anti-commutation. The former is a $2n$-dimensional real vector and the latter is a $2n \times 2n$ matrix, also known as the covariance matrix, which is real, symmetric, and positive semi-definite. For instance, in the case of $n=1$ mode, such parameters are
\begin{equation}
\label{eq:def_first_second}
    \bar{r}  = \begin{pmatrix}
        \braket{x}\\
        \braket{p}
    \end{pmatrix} , \hspace{1cm}
    \sigma = \begin{pmatrix}
        2\Delta_{x^2} & \Delta_{\{x,p\}} \\
        \Delta_{\{x,p\}} & 2\Delta_{p^2}
    \end{pmatrix} \;,
\end{equation}
where $\Delta_A = \braket{\hat{A}^2} - \braket{\hat{A}}^2$. In this formalism, the Heisenberg uncertainty principle reads $\sigma + i \Omega \geq 0$.
 
The first and second moments uniquely define a quantum Gaussian state of the $n$ modes through its phase-space description. The phase-space method relies on the set of displacement operators
\begin{equation}
    \hat{D}_{\xi} = {\rm e}^{i \xi^{\rm T} \Omega \hat{r}} \; , 
\end{equation}
where $\xi \in \mathds{R}^{2n}$, which are the operators that displace the vacuum to a coherent state. The displacement operators $\hat{D}_\xi$ are orthonormal with respect to the Hilbert--Schmidt inner product, such that any bounded operator $\hat{O}$ on the space can be written according to the Fourier–Weyl relation
\begin{equation}
    \hat{O} = \frac{1}{(2\pi)^n} \int_{\mathds{R}^{2n}} \hspace*{-0.2cm}\text{d} \xi\,  \text{Tr} \left[ \hat{D}_{-\xi} \hat{O} \right] \hat{D}_{\xi} \;. 
\end{equation}
If the operator $\hat{O}$ is a density matrix $\hat{\varrho}$, the relationship above defines the characteristic function $\chi(\xi) = \text{Tr} \left[ \hat{D}_{-\xi} \hat{\varrho} \right]$, which represents the quantum state via
\begin{equation}
    \hat{\varrho} =  \frac{1}{(2 \pi)^n} \int_{\mathds{R}^{2n}} \text{d} \xi  \; \chi (\xi) \hat{D}_{\xi} \;,
\end{equation}
where $r$ is the vector of the $2n$ phase-space variables. The characteristic function is generally complex and, by definition, $\text{Tr} [\hat{\varrho}] = \chi (0) = 1$. The characteristic function of a general Gaussian state ($\chi_g$) is fully determined by its first moments and covariance matrix, as per
\begin{equation}
\label{eq:chara_general_gaussian}
    \chi_g(\xi) = \exp \left( -\frac{1}{4} \xi^{\rm T} \Omega^{\rm T} \sigma   \Omega \xi + i \xi^{\rm T} \Omega^{\rm T} \bar{r} \right) \;.
\end{equation}
The Wigner function can be computed as the Fourier transform of the characteristic one,\footnote{i.e., $\mathcal{W}(\xi)= \frac{1}{2^n \pi^{2 n}} \int_{\mathbb{R}^{2 n}} d \xi' {\rm e}^{i \xi'^{\rm T} \Omega \xi} \chi (\xi')$} leading to 
\begin{equation}
\label{eq:wigner_general_gaussian}
    \mathcal{W}_g (\xi) = \frac{2^n}{\pi^n \sqrt{\text{det} \sigma }}{\rm e}^{-\left(\xi - \bar{r} \right)^{\rm T}  \sigma ^{-1} \left(\xi - \bar{r} \right)} \;,
\end{equation}
which is always a real and positive Gaussian if the state is Gaussian (a possible definition of a Gaussian state). 

Crucially, phase space methods allow us to map the problem of evaluating the Markovian time evolution of the quantum state $\hat{\varrho}$ of $n$ infinite-dimensional Hilbert spaces, i.e. $L^2({\mathbbm R^n})$, to a Partial Differential Equation (PDE) for $\chi (\xi)$, equivalent to Schrödinger's for a closed system or to a diffusive master equation under white noise. Seeking conciseness of the expressions, let $\tilde{\xi} = \Omega \xi$. Specifically, from the definition of the characteristic representation of operators, one finds the mappings
\begin{align}
    \partial_{\tilde{\xi}^p} \chi (\tilde{\xi}) &\longleftrightarrow  \frac{i}{2} (\hat{r}^p  \hat{\varrho} + \hat{\varrho} \hat{r}^p ) \;,
    \label{eq:phase_space_corresp1} \\
    \tilde{\xi}^p \chi (\tilde{\xi}) &\longleftrightarrow  \Omega^{qp} (\hat{r}^q \hat{\varrho} -  \hat{\varrho} \hat{r}^q) \;, \label{eq:phase_space_corresp2}
\end{align}
where the superscripts ($p$ and $q$) label the elements of the vectors and matrices, and the Einstein notation is understood (and will be throughout). These (well known) identities and their repetitive use (see Appendix~\ref{app:map_phase_space}) can be used to map unitary and Lindbladian quantum evolutions to PDEs governing the evolution of $\chi$ (and of the Wigner function, which is its Fourier transform). When restricting to the Gaussian dynamics -- that is, to linear or quadratic Hamiltonians and linear couplings to the environment -- of Gaussian states, the PDE has solutions of the form of Eq.~(\ref{eq:chara_general_gaussian}), with a set of ODEs that govern the time evolution of the phase-space parameters ($\sigma$ and $\bar{r}$). Such ODEs are often analytically integrable, providing an exact solution to Gaussian quantum dynamics. 

For instance, the most general Gaussian Hamiltonian of $n$ modes is
\begin{align}
    \frac{\hat{H}_g}{\hbar \omega} &=   \frac{1}{2} \hat{r}^{\rm T} H_m \hat{r} - r_m^{\rm T} \hat{r}  \;,
\end{align}
where $H_m$ is a $2n \times 2n$ symmetric unitless matrix, capturing the kinetic and local quadratic potential, as well as entangling interactions between the modes, and $r_m$ is a $2n$-dimensional vector representing `forces' (both momentum and position kicks). 
Consider the time in frequency units, that is, $t \to \tau = \omega t$. Then, the unitary evolution of a Gaussian quantum state $\hat{\varrho}_g$ is given by the von Neumann equation 
\begin{equation}
    \frac{\partial \hat{\varrho}_g}{\partial \tau} = \frac{i}{\hbar \omega} \left[ \hat{\varrho}_g, \hat{H}_g\right] \;. 
\end{equation}
By moving to the phase space representation (Eqs.~\ref{eq:phase_space_corresp1}, \ref{eq:phase_space_corresp2}), the PDE for the time evolution of the characteristic function has solution of the form in Eq.~(\ref{eq:chara_general_gaussian}), with
\begin{equation}
\label{eq:ODE_gaussian}
    \dot{\sigma} = \Omega H_m \sigma - \sigma H_m \Omega \; ,\hspace{0.5cm} \dot{\bar{r}} =  \Omega H_m \bar{r} - \Omega r_m  
    \;,
\end{equation}
where $\dot{A}$ represents the time derivative of quantity $A$. One can solve both equations by defining the symplectic transformation at time $\tau$ generated by $H_m$ as 
\begin{equation}
\label{eq:sympletic_OG}
 S (\tau) = {\rm e}^{\tau \Omega H_m}\;. 
\end{equation}
Then, the solutions to Eq.~(\ref{eq:ODE_gaussian}) for an initial Gaussian state with phase-space parameters $\sigma_0$ and $\bar{r_0}$ are
\begin{align}
\label{eq:cov_gaussian}
\sigma (\tau) &= S(\tau) \sigma_0 S^{\rm T} (\tau) ,\\
     \bar{r}(\tau)  \nonumber
     &= \bar{r}_0(\tau) - ( \tilde{r}_m (\tau) - \tilde{r}_m) \;,
\end{align}
where we defined $\tilde{r}_m =  H_m^{-1} r_m$ and $r_i(\tau) = S (\tau) r_i$ (with $i=0,m$), which solves every quadratic and linear Hamiltonian through a single symplectic transformation\footnote{If $H$ is not invertible, as would be the case for the free Hamiltonian $\hat{p}^2$, this solution would only hold on the rank of $H$ (multiplied by $\Omega$), in which case $H^{-1}$ would have to be taken as a pseudoinverse (i.e., as the inverse on the rank and zero elsewhere). The general solution in the null space would require polynomial terms in time and cannot be written in terms of a single exponential (this aspect is of course common to all dynamics in the quadratic regime). Notice also that exact solutions are Gaussian and available for such cases too and that the framework of GBCS that we are about to introduce would still apply to all of them.}. A similar procedure can be applied to Gaussian quantum dynamics that are open (Markovian), by mapping the problem of solving Lindblad-type differential equations to a set of ODEs governing $\bar{r}$ and $\sigma$ (the latter is governed by Lyapunov equations describing decay and diffusive processes). In this instance, the solutions can be expressed always in integral form. Continuous Gaussian monitoring and linear \textit{feedback} of Gaussian states result instead in non-linear Riccati equations for the covariance matrix. Instances of these will arise in unitary dynamics in this work.

Any deterministic Gaussian CP map -- such as classical mixing, losses, attenuators and amplifiers -- can be expressed via the phase space representation as the map 
\begin{equation}
    \bar{r} \to X \bar{r} + d\hspace{0.5cm} \sigma \to X \sigma X^{\rm T} + Y 
\end{equation}
where $X$ and $Y$ are the two matrices respecting $Y + i \Omega \geq i X \Omega X^{\rm T}$ (to preserve the Heisenberg uncertainty principle) and defining specific maps and $d$ is a displacement.

The phase space representation also allows for the description of measurements often performed in experiments (more details are given in Sec.~\ref{sec:measure}). A general class of Gaussian measurements is represented by the set of general-dyne detections, with Positive Operator-Valued Measure (POVM):
\begin{equation}
\label{eq:generadyne}
{\mathds{1}} = \frac{1}{(2\pi)^n} \int_{\mathbbm{R}^{2n}}dr_{o} \hat{D}_{-r_o} \varrho_o \hat{D}_{r_o} ,
\end{equation}
where $n$ is the number of measured modes, $r_{o}$ is the $2n$-dimensional vector of (continuous) measurement outcomes. $\varrho_o$ is a Gaussian state with null first moments and covariance matrix $\sigma_o$: by varying $\sigma_o$ among the physical covariance matrices (i.e., such that $\sigma_o+i\Omega\ge 0$), this class allows one to capture several measurement procedures, such as homodyne and heterodyne measurements, in both ideal and noisy conditions. From the phase-space representation of a Gaussian state, the probability distribution of the outcome $r_o$ of a general-dyne detection is 
\begin{equation}
    \label{eq:prob_generadyne_gaussian}
    P(r_o) =  \frac{{\rm e}^{ - (\bar{r} - r_o)^{\rm T} ( \sigma + \sigma_o )^{-1}  (\bar{r} - r_o) }}{\pi^n  \sqrt{ \text{det} [ \sigma + \sigma_o ] }} \;.
\end{equation}


\section{ Gaussian-Branched Cat States \\ \& Superpositions of Phase Spaces \label{sec:sup_phase_space}}

The Gaussian phase-space method can be extended to CV-DV hybrid systems by introducing Gaussian-Branched Cat States (GBCSs). In this section, the main ideas and methodology of this work are presented:  a quantum superposition of phase spaces is used to give the general form of GBCSs. This is laid out for the general case of an $N$-dimensional DV subsystem entangled with $n$-dimensional CV ones, followed by the explicit case of the joint system of a single qubit ($N = 2$) with $n$ modes, to provide a compact example. 

\begin{figure*}[t]
\begin{equation}
\label{eq:density_one_qubit}
    \hat{\varrho} = \frac{1}{(2 \pi)^n} \int_{\mathds{R}^{2n}} \text{d} \tilde{\xi} \;  \hat{D}_{\tilde{\xi}} \otimes \; \begin{pmatrix}
         p_+ \exp \left[-\frac{1}{4} \tilde{\xi}^{\rm T} \sigma^{\text{on}}_+  \tilde{\xi}  +  i  \tilde{\xi}^{\rm T} \bar{r}^{\text{on}}_+   \right] & q  \exp \left[ -\frac{1}{4} \tilde{\xi}^{\rm T} \sigma^{\text{off}}  \tilde{\xi}  +  i  \tilde{\xi}^{\rm T} \bar{r}^{\text{off}}  -   \mathcal{C}  + i  \phi  \right]   \\[8pt]
       q^* \exp \left[ -\frac{1}{4} \tilde{\xi}^{\rm T} \sigma^{\text{off}\;*}  \tilde{\xi}  +  i  \tilde{\xi}^{\rm T} \bar{r}^{\text{off}\;*}  -   \mathcal{C}  - i  \phi  \right]    & p_- \exp \left(-\frac{1}{4} \tilde{\xi}^{\rm T} \sigma^{\text{on}}_-  \tilde{\xi}  +  i  \tilde{\xi}^{\rm T} \bar{r}^{\text{on}}_-   \right) 
       \end{pmatrix}
\end{equation}\caption*{The most general single-qubit by $n$ modes GBCS.\label{fig:density_one_qubit}}
\end{figure*}

Mathematically, the state of a joint system of $n$ CV degrees of freedom and an $N$-dimensional DV is described by the density matrix operator $\hat{\varrho}$ acting on the joint Hilbert space $\mathcal{H}_m \otimes\mathcal{H}_q \sim L^2(\mathds{R}^n) \otimes \mathds{C}^{N} $, where -- here and in the following -- the superscripts $m$ and $q$ refer to quantities related to the CV (modes) and DV (qudit) subsystems, respectively. Without loss of generality, given one DV operator $\hat{J}_z$, its eigenstates $\hat{J}_z \ket{j} = j \ket{j}$ -- which span the DV Hilbert space, with $j \in [ 0, N- 1]$~-- can be used to expand the density matrix of the joint system
\begin{equation}\label{rhojk}
    \hat{\varrho} = \sum_{j, k  = 0}^{N-1}  \hat{\varrho}_{jk} \otimes \ket{j} \bra{k} \;,
\end{equation}
where $ \hat{\varrho}_{jk} = \braket{j|\hat{\varrho}|k}  $ are operators acting on the CV Hilbert space, labelled by the eigenvalue of~$\ket{j}$. The operators $\hat{\varrho}_{jj}$ in diagonal entries are unnormalized density matrices of $n$ modes; however, off-diagonal operators are not Hermitian and, hence, not physical, though the overall hermiticity implies $(\hat{\varrho}_{jk})^\dag = \hat{\varrho}_{kj} $. We associate a phase space with each $ \hat{\varrho}_{jk} $ via the composite Fourier–Weyl relation 
\begin{equation}
\label{eq:density_phase_space}
    \hat{\varrho} = \frac{1}{(2 \pi)^n} \sum_{j, k  = 0}^{N-1} \int_{\mathds{R}^{2n}} \text{d} \xi \; \chi_{jk} (\xi)  \hat{D}_{\Omega^{\rm T} \xi} \otimes \ket{j} \bra{k} \;,
\end{equation}
where we define the \textit{Branched Characteristic Functions}: \begin{equation}
\label{eq:chara_jk_def}
    \chi_{jk} (\xi) = \text{Tr}_m [ \hat{D}_{\Omega \xi}  \hat{\varrho}_{jk} ] \;.
\end{equation}
Formally, this is possible as the set of operators $\{\hat{D}_{r} \otimes \ket{j} \bra{k} \} $ with $j,k  \in [ 0, N-1] $ and $\xi \in \mathds{R}^{2n}$ is a complete orthonormal basis of the space of operators acting on the CV-DV Hilbert space, i.e. the joint space of trace-class operators $\mathfrak{S} (\mathcal{H}_m \otimes\mathcal{H}_q)$, with respect to the Hilbert--Schmidt product.
 A \textit{Gaussian-Branched Cat State} is a CV-DV entangled state with Gaussian branched characteristic functions:
\begin{equation}
\label{eq:chara_jk}
    \chi_{jk} (\tilde{\xi}) =  \exp \left( -\frac{1}{4} \tilde{\xi}^{\rm T} \sigma_{jk} \tilde{\xi}  +  i  \tilde{\xi}^{\rm T} \bar{r}_{jk} + s_{jk}   \right) \;,
\end{equation}
where $\sigma_{jk}$ are $2n\times 2n$ symmetric matrices, $\bar{r}_{jk}$ are $2n$-dimensional vectors and  $s_{jk} $ are scalars, all labelled by all combinations of the discrete eigenvalues ($j$ and $k$). We will collectively name them phase-space parameters, which fully characterise a GBCS and, rather loosely, we will refer to all $\bar{r}_{jk}$ and $\sigma_{jk}$ as first and second moments, though it should be borne in mind that the off-diagonal parameters do not correspond to physical first and second moments. As we shall see, they are typically complex quantities and, given that $\hat{\varrho}$ is Hermitian, it follows that $\sigma_{jk}^* = \sigma_{kj}$, $\bar{r}_{jk}^* = \bar{r}_{kj}$, and~$s_{jk}^* = s_{kj}$. 

Let us then describe the general properties of GBCS. The matrix $\hat{\chi} (\tilde{\xi} ) = \sum_{jk}\chi_{jk} (\tilde{\xi} ) \ket{j} \bra{k}$ is a Hermitian operator acting on the DV Hilbert space as a function of the phase space variable $\tilde{\xi}$. Of particular interest is the case of $\tilde{\xi} = 0$, which, according to the definition of $\chi_{jk}$ (Eq.~\ref{eq:chara_jk_def}), is the Qudit Reduced Density Matrix (QRDM) $\hat{\varrho}^{q} := \text{Tr}_m [\hat{\varrho}] = \hat{\chi} (0) $, with components
\begin{equation}
     \varrho^{q}_{jk} = \exp(s_{jk}) = \exp \left(- \mathcal{C}_{jk} + i  \phi_{jk} \right),
\end{equation}
where we defined $\mathcal{C}_{jk}$ and $ \phi_{jk} $ as the purely real and imaginary parts of $s_{jk}$, respectively, and we will name them as \textit{contrasts}\footnote{More precisely, $\mathcal{C}_{jk}$ are the exponents of the decay in contrast, which represent sources of decoherence in the QRDM. However, they include apparent decoherence induced by the entanglement with a superposition of non-overlapping Gaussian CV states. When considered dynamically, they can show coherence revival when the CV wavefunction is recombined, see Sec.~\ref{sec:physical_examples}.} and \textit{phases} of the QRDM. Notice that, as will be clear from the final examples, the quantities $\mathcal{C}_{jk}$ provide an expedient and systematic way to identify processes where the qubit's coherence is preserved, spoilt or revived during a superposed Gaussian process. Note that  $\mathcal{C}_{jk} = \mathcal{C}_{kj}$ and $\phi_{jk} = -\phi_{kj}$.

Furthermore, one can find the characteristic function of the modes reduced density matrix, $\hat{\varrho}^m := \text{Tr}_q [\hat{\varrho}] $, such that $\chi^m (\xi) = \sum_{j} \chi_{jj} (\xi)$. On the diagonal terms, the phase-space quantities $\sigma_{jj}$ and $r_{jj}$ are real and they respectively represent physical covariance matrices and vectors of first moments, while $\varrho^q_{jj} = {\rm e}^{r_{jj}^{(0)}} = p_{jj}$ are real probabilities. Then, it is possible to interpret $\chi^m$ as the statistical mixture of $n$-mode Gaussian states with first and second moments $r_{jj}$ and $\sigma_{jj}$, weighted according to the probability distribution $\{ p_{jj} \}$ encoded in the diagonal elements of the QRDM. 

It is insightful to explicitly describe the case of $n$ modes with a qubit.  In this case, the DV Hilbert space $\mathds{C}^{2}$ can be represented by a basis $\ket{\pm 1}$, for example, of the Z-Pauli operator $\hat{\sigma}_z \ket{\pm 1} = \pm \ket{\pm 1}$. The most general $n$-mode single-qubit GBCS is given in Eq.~(\ref{eq:density_one_qubit}), where the QRDM quantities are, generically, $p_\pm \in \mathds{R}^+$ with $p_+ + p_- = 1$, $q \in \mathds{C}$, and $|q {\rm e}^{-\mathcal{C}}| \leq \sqrt{p_{+} p_{-}} $. Furthermore, we define the quantities $\sigma^{\text{on}}_\pm = \sigma_{\pm1 , \pm 1}$ and $\sigma^{\text{off}} = \sigma_{+ 1 , - 1} =  \sigma_{- 1 , + 1}^*$ as the on- and off-diagonal covariance matrices, and proceed similarly for the first moments $r^{\text{on/off}}$, thus, applying the symmetry imposed by the hermiticity of the joint state.

Note that, in the general case, our formalism is apt to describe several physical systems. As canonically DV variable, this is single large DV physical subsystem (where $N = 2 J +1 $, where $J$ could be the total angular momentum), as well as, an ensemble of $N_q$ qubits, where $N = 2^{N_q}$. For a large spin $J$, for instance, the $\hat{J}_z$ eigenvalues $m \in [-J, J]$ correspond to $j = m + J \in [0, 2J]$ in our notation. Similarly, for $N_q$ qubits, the strings $\vec{s} = (s_1,s_2, \dots, s_{N_q}) \in \{\pm 1\}^{N_q}$ correspond to $j = \sum_{i=1}^{N_q} (1+s_i)\, 2^{\,i-2}$. Ultimately, for $N\rightarrow\infty$, our framework could also apply to non-linear, purely CV composite systems, where one subsystem (for instance, a mechanical resonator coupled to a cavity mode~\cite{bose_preparation_1997,bose_scheme_1999,qvarfort_time-evolution_2020,marchese2026cascadedoptomechanicalsensingsmall}) is described in the Fock basis.

\section{Dynamics \label{sec:dynamics}}

Any `operator-valued' Gaussian dynamics, which is generated by interactions that are Gaussian Hamiltonians ``labelled'' by the DV state, can be analytically solved by an evolving GBCS. In fact, this treatment and state parametrisation allow one to map the dynamics of the CV-DV system -- originally described as a Lindblad-type equation for $\hat{\varrho}$ (or $\hat{r}$, in the Heisenberg picture) -- to the time evolution of the phase-space parameters of a GBCS. Although this class of Hamiltonians may appear restrictive, this type of interactions arises frequently in experimental practices (as discussed in Sec.~\ref{sec:intro}). 

Let us introduce the most general Gaussian Operator-Valued Hamiltonian. Specifically, the ``operator-valued'' trait refers to the operational dependence of the Hamiltonian only on a single Hermitian DV operator, such that 
$\hat{H}=H (\hat{r}, \hat{J}_z)$, implying that $\hat{H}$ is diagonal in the DV eigenbasis of $\hat{J}_z=\sum_{j} j \ket{j}\bra{j}$. In this sense, these are the interaction Hamiltonians that commute with the local Hamiltonian of the DV, i.e. the DV energy is conserved during these dynamics and no transitions occur. Further, we will make the assumption that the Hamiltonian is a second-order polynomial in $\hat{r}$, giving rise to Gaussian dynamics within each branch of a DV superposition. Thus, the most general operator-valued Gaussian Hamiltonian~is 
\begin{equation}
\label{eq:Hamiltonian_max}
    \hat{H}_g (\hat{r}, \hat{J}_z) = \hbar \omega \sum_{j  = 0}^{N-1}\left(  \frac{1}{2} \hat{r}^{\rm T} H_{j} \hat{r} - r_{j} ^{\rm T} \hat{r} +  \frac{1}{2} H_j^0 \right)\otimes \ket{j} \bra{j}     \;,
\end{equation}
where $H_j$ are $N$ real, invertible\footnote{While, it is worth noticing that, if $H_j$ are not invertible, the following discussion holds at the level of ODEs.} symmetric matrices of dimension $2n \times 2n$, $r_{j}$ are $N$ real $2n$-dimensional vectors, and $H_j^0$ are scalars, all classically labelled by the eigenvalues $j$.

In the specific case of a single qubit, by choosing the qubit operational basis as ${\mathbbm 1}$ and $\sigma_z$, we adopt a different parametrisation such that the general qubit operator-valued Gaussian Hamiltonian $H_g (\hat{r}, \hat{\sigma}_z)$ can be written as
\begin{align}
\label{eq:Hamiltonian_max_qubit}
    \hspace{-0.3cm}\frac{\hat{H}_g (\hat{r}, \hat{\sigma}_z)}{\hbar \omega} &=  \left(  \frac{1}{2} \hat{r}^{\rm T} H_m \hat{r} - r_m^{\rm T} \hat{r}  \right) \otimes \mathds{1}\nonumber \\
    &\hspace{1cm} + \left( \frac{1}{2} \hat{r}^{\rm T} H_q \hat{r} - r_q^{\rm T} \hat{r} + \frac{1}{2} H_q^0 \right) \otimes \hat{\sigma}_z
\end{align}
where $H_m$ and $H_q$ are two $2 n \times 2n $ symmetric matrices describing the modes and operator-valued quadratic potentials, $r_m$ and $r_q$ are $2n$-dimensional vectors, representing forces which can be ``unconditional'' or operator-valued by the qubit, and $H_q^0$ is the zero energy splitting of the qubit. This recovers the general form of Eq.~(\ref{eq:Hamiltonian_max}), with $j = \pm 1$ via $H_m  = (H_{+ 1} + H_{-1})/ 2$ and $ H_q  = (H_{+ 1} - H_{-1})/ 2$; and similarly for $r_m$ and $r_q $. 

Let us now lay out how we shall solve the quantum dynamics via GBCS, and thus understand the limitation imposed by  the required assumptions of the Hamiltonian. The diagonality assumption allows us to derive the time evolution of the branched characteristic functions (Eq.~\ref{eq:chara_jk_def}) by mapping the Lindblad-type equations to sets of $N^2$ uncoupled PDEs of $\chi_{jk}$ (which would be coupled if the Hamiltonian were not DV-diagonal). The Gaussianity assumption implies that the PDEs are quadratic in the phase-space variables and their partial derivatives. We finally show that the solutions to the uncoupled Gaussian PDEs are the branched characteristic functions of a time-dependent GBCS (Eq.~\ref{eq:chara_jk}). Specifically, we find that a GBCS is a solution of a particular quantum dynamics if and only if the time evolution of its phase-space parameters, $\sigma_{jk}(\tau)$, $\bar{r}_{jk}(\tau)$,  and $s_{jk} (\tau)  $,  are solutions of a set of coupled ODEs of the form
\begin{align}
   \dot{\sigma}_{jk} &= f(\sigma_{jk}), \;\;
   \dot{\bar{r}}_{jk} = g(\sigma_{jk}, \bar{r}_{jk}), \;\;
   \dot{s}_{jk} = h(\sigma_{jk}, \bar{r}_{jk}),  \nonumber
\end{align}
 where $f$, $g$ and $h$ are functions depending on the considered type of dynamics. The explicit forms of the sets of ODEs and the integral form of their solutions (when generically derivable) represent one of the central results of this work, presented in Sec.~\ref{sec:unitary} for unitary dynamics and in Sec.~\ref{sec:open} for open dynamics. In Appendices \ref{app:map_phase_space}, \ref{app:unitary_general}, and~\ref{app:noise}, the mapping from Hilbert-space dynamics to phase-space parameters is employed to derive our dynamical equations. If either the assumption of diagonality for the qubits or Gaussianity for the CV system is dropped, the dynamics does not have a general analytical solution, as the algebra does not close~\cite{wei_james_1963,legget_dynamics_1987,circuit_niemczyk_2010,superconducting_yoshihara_2017,ralph_dynamical_2018,roda_macroscopic_2024}. In Sec.~\ref{sec:perturbation}, we overcome this limitation by discussing perturbation theory with GBCS.

 Observe that, given the structure of the couplings between these sets of ODEs, one can already infer how physical effects in these systems propagate throughout the composite state (for instance, diffusion in the covariance matrices will affect the QRDM, while dephasing of the QRDM would not affect the CV). Furthermore, since $\dot{s}_{jk}$ is independent from ${s}_{jk}$, its expression can be interpreted as the rates of decoherence and phase shift of the DV, as per $\Gamma_{jk} = - \text{Re} [\dot{s}_{jk} ]$ and $\dot{\phi}_{jk} = \text{Im} [\dot{s}_{jk} ]$.

 \begin{table*}[!]
    \centering
    \renewcommand{\arraystretch}{1.7}
    \begin{tabular}{ |p{4.0cm}||P{6cm}|P{6cm}|}
        \hline
        Phase-Space Parameters  & \multicolumn{2}{c|}{ODEs for Unitary Dynamics with Operator-valued Gaussian Interactions} \\ 
        \hline
        Second Moments  & \multicolumn{2}{c|}{\ensuremath{ \dot{\sigma}_{jk} =   \frac{1}{2} \left[  \Omega  ( H_j + H_k )  \sigma_{jk} - \sigma_{jk}  ( H_j + H_k ) \Omega     
  -  i   \sigma_{jk} (H_j - H_k) \sigma_{jk}  + i  \Omega (H_j - H_k) \Omega \right]}}\\
       First Moments & \multicolumn{2}{c|}{
       \ensuremath{
       \dot{\bar{r}}_{jk} =\frac{1}{2}\left[ \left(  \Omega ( H_j + H_k )  - i   \sigma_{jk} (H_j - H_k) \right)  \bar{r}_{jk} 
     - \Omega (r_j + r_k) - i  \sigma_{jk} (r_j - r_k) \right] }} \\
     QRDM Exponent & \multicolumn{2}{c|}{\ensuremath{ \dot{s}_{jk} =   - \frac{i}{2}  \bar{r}_{jk}^{\rm T} (H_j - H_k)  \bar{r}_{jk}  + i (r_j - r_k)^{\rm T}  \bar{r}_{jk} 
   -  \frac{i}{4} \text{Tr} [(H_j - H_k)\sigma_{jk}] -   \frac{i}{2} (H_j^0 -H_k^0}) } \\
    \hline
    \end{tabular}
    \caption{ODEs governing the time evolution of the phase-space parameters of a GBCS (Eq. \ref{eq:chara_jk}) undergoing an operator-valued Gaussian unitary dynamics with Hamiltonian matrices $H_j$ and force vectors $r_j$ (Eq. \ref{eq:Hamiltonian_max}).}
    \label{tab:unitary}
\end{table*}

The set of initial conditions is needed to find the particular solution to describe the evolution of a specific initial quantum state. Let us first consider an initial product state of the form $\hat{\varrho}(0) = \hat{\varrho}_m^{(g)} (0) \otimes \hat{\varrho}^q (0) $, where $\hat{\varrho}_m^{(g)} (0) $ is a general Gaussian state of the $n$ modes and $\hat{\varrho}^q (0)$ is a general qubit density matrix. The initial first and second moments ($r_0$ and $\sigma_0$) of $\hat{\varrho}_m^{(g)} (0) $, with the elements of the initial QRDM ($\varrho^q_{jk} (0)$), form a complete set of initial conditions for a particular solution of the time-evolution ODEs governing the time evolution of the phase-space parameters. This choice allows one to describe a large set of experiments, as often the initial state of the modes is indeed Gaussian -- for such are coherent, thermal and squeezed states. In the following, we shall give concrete examples of exact dynamics for product initial states.  Yet, if a GBCS is the initial state of a quantum dynamics, its phase-space parameters are still valid as initial conditions ($\sigma_{jk}(0)$, $\bar{r}_{jk}(0)$ and  $s_{jk}(0)$). This allows repetitive use of the dynamics, with the inclusion of intermediate operations and measurements (see Secs.~\ref{sec:operations} and~\ref{sec:measure}).


\subsection{Unitary Dynamics\label{sec:unitary}}

The unitary evolution of the system is given by the von Neumann equation \begin{equation}
\label{eq:vonNeumann_master}
    \frac{\partial \hat{\varrho}}{\partial \tau} =  \frac{i}{\hbar \omega} \left[ \hat{\varrho} , \hat{H}_g (\hat{r}, \hat{J}_z)  \right] \, .
    \end{equation}
The evolution of the branched $\hat{\varrho}_{jk}$ under this globally unitary dynamics is neither trace-preserving nor Hermitian (see Appendix~\ref{app:unitary_general}), leading to a vast, rich structure~\cite{Real_Bender_1998,ashida_nonHermitian_2020,third_mcdonald_2023}, analogous to that of generalised master equations~\cite{RevModPhys.70.101,RevModPhys.81.1665,PRXQuantum.5.020201}. Among the class of operator-valued Gaussian Hamiltonians, a main distinction can be made between quadratic and linear qubit-modes interactions, which will be analysed separately in the two sections that follow. 


\subsubsection{Quadratic Interactions \label{subsec:quadratic}}

The former is the general case of Eq.~(\ref{eq:Hamiltonian_max}) and, by performing the map from Hilbert to superposed phase space (Appendix \ref{app:unitary_general}), we prove that the GBCS with characteristic functions of Eq.~(\ref{eq:chara_jk}) is a solution of the von Neumann equation with Hamiltonian of Eq.~(\ref{eq:Hamiltonian_max}) if and only if $\sigma_{jk} (\tau)$, $\bar{r}_{jk} (\tau)$, and $  s_{jk}(\tau) $ are solutions of the ODEs in Table \ref{tab:unitary}. The quadratic coupling induces a complex Riccati term (i.e. $i \sigma (H_j - H_k) \sigma$) in the evolution of the covariance matrix, which does not allow for a general closed integral solution. In Sec.~\ref{subsec:riccati}, we provide a systematic way to compute specific analytical solutions of the Riccati equation -- and thus of the dynamics -- via a linear system (at the cost of doubling the variables). 

It is worth noticing that the ODEs of Table~\ref{tab:unitary} hold for both time-independent and time-dependent Hamiltonians. In fact, this treatment can be trivially extended to time-dependent Hamiltonians by considering $H_j (\tau)$, $r_j (\tau)$, and $H_j^{0} (\tau)$, which are matrices and vectors of time-dependent couplings, while keeping the same form of ODEs~\cite{braccini_exponential_2024}. 

Furthermore, one can study steady-state solutions of (stable) hybrid systems: by setting $\dot{\sigma}_{jk}= 0$ and $\dot{\bar{r}}_{jk}= 0$  and solving for the resulting algebraic equation, the covariance matrices and first moments of the steady state can be computed and used to find the form of $\dot{s}_{jk}$. 



\begin{table*}[!]
    \centering
    \renewcommand{\arraystretch}{1.7}
    \begin{tabular}{ |p{4.0cm}||P{6cm}|P{6cm}|}
    \hline
        Phase-Space Parameters & \multicolumn{2}{c|}{Solution of Unitary Dynamics with Operator-Valued Linear Interactions} \\ 
        \hline
        \hline
        Covariance Matrices   & \multicolumn{2}{c|}{\ensuremath{
   \sigma (\tau) := \sigma_{jk}(\tau)  = S_m (\tau) \sigma_0 S_m^{\rm T} (\tau)}}\\
       First Moments & \multicolumn{2}{c|}{
       \ensuremath{\bar{r}_{jk} (\tau) = r_0(\tau) - \frac{1}{2}  \left( \tilde{r}_j(\tau) - \tilde{r}_j + \tilde{r}_k(\tau) - \tilde{r}_k  \right) -  \frac{i}{2} \sigma (\tau)   \Omega  \left[ (\tilde{r}_j (\tau) - \tilde{r}_j)  - (\tilde{r}_k (\tau) - \tilde{r}_k ) \right]   }} \\
     QRDM Contrasts & \multicolumn{2}{c|}{
       \ensuremath{\mathcal{C}_{jk} (\tau) = \frac{1}{4} \left[ (\tilde{r}_j (\tau) - \tilde{r}_j)  - (\tilde{r}_k (\tau) - \tilde{r}_k ) \right]^{\rm T} \Omega^{\rm T} \sigma (\tau)   \Omega  \left[ (\tilde{r}_j (\tau) - \tilde{r}_j)  - (\tilde{r}_k (\tau) - \tilde{r}_k ) \right]  }}\\
    QRDM Phases & \multicolumn{2}{c|}{\ensuremath{ \phi_{jk} (\tau) = - (\tilde{r}_j - \tilde{r}_k)^{\rm T} \Omega \left[  ( \tilde{r}_0 (\tau) - \tilde{r}_0) -  \frac{1}{2} \left( \tilde{r}_j (\tau) - \tilde{r}_j + \tilde{r}_k (\tau) -  \tilde{r}_k  \right) \right]  }} \\
 & \multicolumn{2}{c|}{\ensuremath{ \hspace{2cm}  + \frac{\tau}{2} \left[ (\tilde{r}_j - \tilde{r}_k)^{\rm T} H_m (\tilde{r}_j + \tilde{r}_k) - (H_j^{0}-H_k^{0})\right] }} \\
    \hline
    \end{tabular}
    \caption{Time evolution of the phase space parameters of a GBCS (Eq.~\ref{eq:chara_jk}) undergoing unitary evolution with operator-valued forces (Eq.~\ref{eq:Hamiltonian_force_only}), where $\tilde{r}_a = H^{-1}_m r_a$, $S_m (\tau) = \exp (  \Omega H_m \tau)$, and $\tilde{r}_a (\tau) = S_m (\tau) \tilde{r}_a$, with $a \in 0,j,k$. The initial state is an $n$-mode Gaussian state with first and second moments $r_0$ and $\sigma_0$ and a general QRDM $\varrho^{q}_{jk}(0)$.}
    \label{tab:force}
\end{table*}

For clarity, let us explicitly write the ODEs (and when possible their solution) for the on- and off-diagonal terms in the single-qubit case, thus providing one with the general solution of the dynamics in terms of the state given by Eq.~(\ref{eq:density_one_qubit}). On the diagonal elements, i.e. $\ket{\pm 1}\bra{\pm 1}$,  the ODEs take the form 
\begin{align}
\label{eq:general_unitary_ODE_on}
    \dot{\sigma}^{\text{on}}_{\pm} &=  \Omega  (H_m \pm H_q)  \sigma^{\text{on}}_{\pm} -  \sigma^{\text{on}}_{\pm} ( H_m \pm H_q) \Omega   \nonumber \;,  \\ 
    \dot{\bar{r}}^{\text{on}}_{\pm} &=  \Omega  (H_m \pm H_q)   \bar{r}^{\text{on}}_{\pm} - \Omega (r_m \pm r_q) \;, 
\end{align} 
and the diagonal elements of the QRDM are unchanged. Eq.~(\ref{eq:general_unitary_ODE_on}) represents two Gaussian dynamics of the same initial Gaussian state under two Gaussian Hamiltonians labelled with $\pm$ (as per Eq.~\ref{eq:ODE_gaussian}). Analogously to Sec.~\ref{sec:intro_phase}, the general solution to these equations (see Appendix~\ref{app:only_force}) can be compactly written by defining the symplectic transformations
\begin{equation}
\label{eq:symplectic_pm}
    S_{\pm}(\tau) = \exp( \tau \Omega (H_m \pm H_q)  ) \;,
\end{equation}
as well as the four vectors $r_i^{\pm} = (H_m \pm H_q)^{-1} r_i$ and their time evolution $ r_i^{\pm} (\tau) =S_{\pm}(\tau)  r_i^{\pm}$, with $i = \{m, q\}$. Then, a Gaussian state with initial moments $r_0$ and $\sigma_0$ and a qubit with density matrix element $\varrho_{jk}^q(0)$ evolve to 
\begin{align}
\label{eq:time_diagonal_general}
    \sigma^{\text{on}}_{\pm} (\tau) &= S_\pm (\tau) \sigma_0 S^{\rm T}_\pm (\tau), \\
     \bar{r}^{\text{on}}_{\pm 1}(\tau) &=   S_\pm (\tau) r_0 - \left[ (r_m^{\pm}(\tau) - r_m^{\pm}) \pm (r_q^{\pm}(\tau) - r_q^{\pm}) \right], \nonumber 
\end{align}
while the QRDM remains constant. As expected, the on-diagonal terms of a GBCS evolution are described by a statistical mixture of Gaussian processes which preserve the Gaussianity and normalization of the states associated with the diagonal elements of the density matrix. 

For the off-diagonal terms, the ODEs read
\begin{align}
\label{eq:general_unitary_ODE_off}
     \dot{\sigma}^{\text{off}} &=   \Omega  H_m  \sigma^{\text{off}} -  \sigma^{\text{off}}    H_m   \Omega   - i  \Omega H_q  \Omega -   i   \sigma^{\text{off}} H_q \sigma^{\text{off}}  \;, \nonumber \\
     \dot{\bar{r}}^{\text{off}} &=  \left( \Omega  H_m  -  i   \sigma^{\text{off}} H_q \right) \bar{r}^{\text{off}}   -  \Omega r_m + i \sigma^{\text{off}} r_q \;,   \\
     \dot{\mathcal{C}} &= \,\text{Im}\!\left[ (\bar{r}^{\text{off}})^{\rm T} H_q \bar{r}^{\text{off}} + \frac{1}{2}\text{Tr}\!\left(H_q \sigma^{\text{off}}\right)  - 2 r_q^{\rm T} \bar{r}^{\text{off}} \right] \;, \nonumber \\
     \dot{\phi} &= -\,\text{Re}\!\left[ (\bar{r}^{\text{off}})^{\rm T} H_q \bar{r}^{\text{off}} + \frac{1}{2}\text{Tr}\!\left(H_q \sigma^{\text{off}}\right)  - 2 r_q^{\rm T} \bar{r}^{\text{off}} \right] - H_q^0 \;. \nonumber
\end{align}

Let us note that the off-diagonal phase-space parameters indeed evolve to complex values: the covariance matrix via the non-linear Riccati term ($\sigma^{\text{off}} H_q \sigma^{\text{off}}$) and the first moments via the coupling of the covariance matrix (as well as the QRDM). These features are also proper to the general case for $j\neq k$, as per Table~\ref{tab:unitary}.  

\subsubsection{Linear Interactions \label{subsec:force}}

Let us now consider the most general Hamiltonian involving a purely linear coupling between CVs and DVs:
\begin{equation}
\label{eq:Hamiltonian_force_only}
    \frac{\hat{H}_l (\hat{r}, \hat{J}_z)}{\hbar \omega} = \frac{1}{2} \hat{r}^{\rm T} H_{m} \hat{r} - \sum_{j = 0}^{N -1}\left(  r_{j} ^{\rm T} \hat{r} -  \frac{1}{2} H_j^0 \right)\otimes \ket{j} \bra{j}     \;,
\end{equation}
whose associated ODEs have analytical solutions that can always be written in integral form by a single transformation governing the quadratic evolution of all superposed phase spaces. From Table \ref{tab:unitary}, one finds the simplified ODEs,
\begin{align}
    \dot{\sigma}_{jk} &=  \Omega  H_m  \sigma_{jk} -  \sigma_{jk}  H_m \Omega \;, \nonumber  \\
    \dot{\bar{r}}_{jk} &=     \Omega  H_m  \bar{r}_{jk}  -   \frac{1}{2} \Omega (r_j + r_k) - \frac{i}{2}  \sigma_{jk} (r_j - r_k) \;, \nonumber  \\
    \dot{s}_{jk} &=   i  (r_j - r_k)^{\rm T}   \bar{r}_{jk}   -  \frac{i}{2} (H_j^0-H_k^0) \;.
\end{align}
As shown in Appendix~\ref{app:only_force}, by defining and computing the single symplectic transformation $S_m (\tau) = \exp(  \Omega H_m \tau)$, the solution of these ODEs can be expressed in integral form as in Table \ref{tab:force}, for an initial Gaussian state with first and second moments $r_0$ and $\sigma_0$ and an initial QRDM $\varrho_{jk}^q (0)$. 
Observe that all covariance matrices of the density matrix are trivially the same, i.e. $\sigma_{jk}(\tau)  = \sigma (\tau) = S_m (\tau) \sigma_0 S_m^{\rm T} (\tau) \; \forall j,k$: given that the coupling is only linear, their evolution is given by a Gaussian dynamics of the form (\ref{eq:cov_gaussian}), as expected. 

Let us define the vectors $\tilde{r}_i =  H_m ^{-1} r_i$ and their time evolution $ \tilde{r}_i (\tau) =S_m(\tau)  \tilde{r}_i$,  with $i = \{j,k,m, q, 0\}$. For the single-qubit case of Eq.~(\ref{eq:density_one_qubit}), the first moments evolve according to
\begin{align}
\label{eq:r_on_force_sol}
     \bar{r}^{\text{on}}_{\pm}(\tau) &=   r_0 (\tau) -  (\tilde{r}_m(\tau) - \tilde{r}_m) \mp (\tilde{r}_q(\tau) - \tilde{r}_q), \\ 
     \bar{r}^{\text{off}} (\tau) &= r_0(\tau) -  \left( \tilde{r}_m(\tau) - \tilde{r}_m \right) - i \sigma (\tau)   \Omega  \left( \tilde{r}_q (\tau) - \tilde{r}_q \right)  , \nonumber
\end{align}
where the former represents the two Gaussian evolutions with different forces, while the latter is a complex quantity, which includes the covariance matrix itself. The time evolution of the QRDM is then given by 
\begin{equation}
\label{eq:QRDM_1}
    \hspace{-0.2cm}\varrho_q (\tau) = \begin{pmatrix}
        p_+ & q {\rm e}^{ - \mathcal{C} (\tau) + i  \phi (\tau) } \\
       q^* {\rm e}^{ - \mathcal{C} (\tau) - i  \phi(\tau)}   & p_- \\ 
    \end{pmatrix} \;,
\end{equation}
where the contrasts and phases at time $\tau$ read
\begin{align} 
\label{eq:phi_c_force_sol}
     \mathcal{C} (\tau) &=   \left(\tilde{r}_q (\tau) - \tilde{r}_q \right)^{\rm T} \Omega^{\rm T}  \sigma (\tau)   \Omega  \left(\tilde{r}_q (\tau) - \tilde{r}_q \right) \nonumber \;,\\
      \phi (\tau) &=  \tau (2 \tilde{r}_q H_m \tilde{r}_m - H_q^0) \\
      & \hspace{0.5cm}- 2 r_q^{\rm T} \Omega \left[ \tilde{r}_0 (\tau) - \tilde{r}_0 - ( \tilde{r}_m (\tau) - \tilde{r}_m ) \right]   \nonumber \;,
\end{align}
respectively. Since $\sigma(\tau)>0$ one may infer that $ \mathcal{C} (\tau) =0$ if, and only if, $\tilde{r}_q (\tau) = \tilde{r}_q$, implying a full revival of qubit coherence when $S_m(\tau) = \mathds{1}$~\cite{braccini_exponential_2024}. Furthermore, note that the coherent phase is independent of the covariance matrix and only senses the forces. 
This solves in integral form the most general unitary dynamics of a Gaussian state evolving in Gaussian potentials and linearly coupled to a qubit.

\begin{table*}[!]
    \centering
    \renewcommand{\arraystretch}{2}
    \begin{tabular}{ |p{3.8cm}|P{7cm}|P{7cm}|}
        \hline
        Phase-Space Parameters  & \multicolumn{2}{c|}{Solution of Open Dynamics with Operator-valued Gaussian Interactions} \\ 
        \hline
        \hline
   \multicolumn{3}{|c|}{(a) Gaussian Interactions ODEs}  \\
    \hline
        Covariance Matrices   & \multicolumn{2}{c|}{\ensuremath{ \dot{\sigma}_{jk} =   \frac{1}{2} \left[  \Omega  ( H_j + H_k +2 E)  \sigma_{jk} - \sigma_{jk}  ( H_j + H_k +2 {\rm E}^{\rm T}) \Omega  -  i   \sigma_{jk} (H_j - H_k) \sigma_{jk}  - i\Omega (H_j - H_k) \Omega \right]  + D}}\\
       First Moments & \multicolumn{2}{c|}{
       \ensuremath{
       \dot{\bar{r}}_{jk} =\frac{1}{2}\left[ \left(  \Omega ( H_j + H_k +2 E)  - i   \sigma_{jk} (H_j - H_k) \right)  \bar{r}_{jk} + 
      \Omega (2 d - r_j - r_k) + i  \sigma_{jk} (r_j - r_k) \right] }} \\
     QRDM Exponent & \multicolumn{2}{c|}{\ensuremath{ \dot{s}_{jk} =   - \frac{i}{2}  \bar{r}_{jk}^{\rm T} (H_j - H_k)  \bar{r}_{jk}  + i (r_j - r_k)^{\rm T}  \bar{r}_{jk} 
   -  \frac{i}{4} \text{Tr} [(H_j - H_k)\sigma_{jk}]   -   \frac{i}{2} (H_j^0 -H_k^0)  - \frac{1}{2} (\Gamma_z^{jj} + \Gamma_z^{kk} ) + \Gamma_z^{jk}} } \\
   \hline 
   \multicolumn{3}{|c|}{(b) Linear Interactions Integral Forms}  \\
    \hline
    \hline
        Covariance Matrix   & \multicolumn{2}{c|}{\ensuremath{ \sigma(\tau) = S_A(\tau) \sigma_0 S_A^{\rm T}(\tau) +  \int_{0}^{\tau} \text{d}t S_A(\tau - t) D S^{\rm T}_A(\tau - t) }}\\
       Vectors or First Moments & \multicolumn{2}{c|}{
       \ensuremath{\; \bar{r}_{jk} (\tau) =  S_A (\tau) r_0 +   \left(S_A(\tau) - \mathds{1} \right)  \left[ d -  \frac{1}{2}    (r_j + r_k)\right] - 
    \frac{i}{2} S_A (\tau) \sigma_0 \left( S_A^{\rm T} (\tau) - \mathds{1}  \right) \Omega (A^{\rm T})^{-1}  (r_j - r_k) }} \\
    & \multicolumn{2}{c|}{
       \ensuremath{\hspace{2cm} - \frac{i}{2}  \int_0^{\tau} \text{d}t'  S_A(\tau- t') D \left( S^{\rm T}_A(\tau - t') - \mathds{1} \right) \Omega (A^{\rm T})^{-1} (r_j - r_k)}} \\
     QRDM Contrasts  & \multicolumn{2}{c|}{ \hspace{-0.4cm} \ensuremath{ \mathcal{C}_{jk} (\tau)  =  \frac{1}{4} (r_j - r_k)^{\rm T}  A^{-1} \Omega^{\rm T} \left( S_A(\tau) - \mathds{1} \right) \sigma_0 \left(S^{\rm T}_A(\tau) - \mathds{1} \right) \Omega  (A^{-1} )^{\rm T} (r_j - r_k) + \tau \left[\frac{1}{2} (\Gamma_z^{jj} + \Gamma_z^{kk} ) - \Gamma_z^{jk} \right] }} \\
 & \multicolumn{2}{c|}{\ensuremath{ \hspace{1.4cm}  + \frac{1}{4}(r_j - r_k)^{\rm T}  A^{-1}    \Omega^{\rm T} \left[  \int_0^\tau \text{d} t'    \left(S_A(\tau - t') - \mathds{1} \right) D \left(S_A^{\rm T}(\tau - t') - \mathds{1} \right)   \right]  \Omega  (A^{-1} )^{\rm T} (r_j - r_k) }} \\ 
     QRDM Phases  & \multicolumn{2}{c|}{ \hspace{-3.6cm} \ensuremath{ \;  \phi_{jk} (\tau) =  - (r_j - r_k)^{\rm T} A^{-1} \Omega \left( S_A (\tau)  - \mathds{1} \right) \left[ r_0  + A^{-1} \left( d  - \frac{1}{2} (r_j + r_k) \right) 
    \right]}} \\
 & \multicolumn{2}{c|}{\ensuremath{ \hspace{1.1cm}  + \tau \left[ (r_j - r_k)^{\rm T} A^{-1} \left(d - \frac{1}{2} (r_j + r_k) \right) -  \frac{1}{2} (H_j^0 -H_k^0) \right] }} \\
 \hline
    \end{tabular}
    \caption{(a) ODEs governing the time evolution of the phase-space parameters of a GBCS (Eq. \ref{eq:chara_jk}) undergoing an operator-valued Gaussian open dynamics with Hamiltonian matrices $H_j$, force vectors $r_j$, driving term $d$, decay matrix $E$, diffusion matrix $D$, and dephasing rates $\Gamma_z^{jk}$ (Eq. \ref{eq:linbladian}).  (b) time evolution of the phase-space parameters of a GBCS (Eq.~\ref{eq:chara_jk}) undergoing open Gaussian evolution with operator-valued forces: where $A = H_m + E$ and $S_A (\tau) = \exp (  \Omega A \tau)$. The initial state is an $n$-mode Gaussian state with first and second moments $r_0$ and $\sigma_0$ and a general QRDM $\varrho^{q}_{jk}(0)$. }
    \label{tab:open}
\end{table*}
 
\subsubsection{No Interactions: Gaussian Dynamics of GBCS}

In experimental practice and designing protocols, after the creation of a GBCS via, for instance, a conditional operation (discussed in Sec.~\ref{sec:operations}), it is often convenient to evolve the superposed state via a Gaussian dynamics. This subset of GBCS dynamics, which can be trivially included in the formalism, is the one with no interaction between the CV and DV subsystems and Hamiltonian
\begin{equation}
\label{eq:Hamiltonian_no_int}
    \frac{\hat{H}_g (\hat{r}) \otimes \hat{H}_q(\hat{J}_z)}{\hbar \omega} = \frac{1}{2} \hat{r}^{\rm T} H_{m} \hat{r} - r_{m} ^{\rm T} \hat{r}  + \frac{1}{2} \sum_{j = 0}^{N - 1}   H_j^0  \otimes \ket{j} \bra{j} \nonumber    \;.
\end{equation}
In this case, each branch of the superposition evolves according to a single Gaussian dynamics described by Eq.~(\ref{eq:cov_gaussian}) under different initial conditions $\sigma_{jk}^{0}$ and $\bar{r}_{jk}^{0}$, depending on the branch. The QRDM builds the phases $\phi_{jk} = i \tau/2 (H_j^{0} - H_k^{0})$.


\vspace{-0.4cm}
\subsection{Open Quantum Dynamics \label{sec:open}}

Our formalism requires little adjustment to include effects of open quantum dynamics, presented in this section. The most general Markovian and Gaussian operator-valued open dynamics with diagonal decoherence in the qubit coupling basis is described by the Lindblad equation  
\begin{align}\label{eq:linbladian}
    \hspace{-0.1cm} \frac{\partial \hat{\varrho}}{\partial \tau} &=  \frac{i}{\hbar \omega} [ \hat{\varrho}, \hat{H}_g (\hat{r}, \hat{J}_z)] +  i [ \hat{\varrho}, d^{\rm T} \hat{r}]   + \mathcal{D}[\hat{r}] (\hat{\varrho})  + \mathcal{D}[\hat{J}_z] (\hat{\varrho}) ,
\end{align}
where 
\begin{align}
    \mathcal{D}[\hat{r}] (\hat{\varrho})  &= \sum_{a, b = 1}^{2 n}  B^{a b} \left(  \hat{r}^a \hat{\varrho} \hat{r}^b  - \frac{1}{2} \{ \hat{\varrho}, \hat{r}^a  \hat{r}^b \} \right), \label{lindo}  \\
    \mathcal{D}[\hat{J}_z] (\hat{\varrho}) &= \sum_{j, k  = 0}^{N-1} \Gamma_z^{jk} \left(\hat{P}_{j} \hat{\varrho} \hat{P}_{k}   - \frac{1}{2} \{ \hat{\varrho}, \hat{P}_{j} \hat{P}_{k} \}  \right),\label{linda} 
\end{align}
where $B$ is a $2n \times 2n$ dimensionless matrix of coefficients representing the strength of the noise in the $n$ modes~\cite{serafini_quantum_2017}, $d$ is the driving, $\Gamma_z^{jk}$ are the unitless DV decay rates and $\hat{P}_{j} = \ket{j}\bra{j}$~\cite{PhysRevLett.114.010501}. The additional Lindbladian terms of Eq.~(\ref{eq:linbladian}) encode all the noises that are diagonal in the DV subsystem and which have jump operators linear in the CV canonical operators ($\hat{r}$). They often represent experimental leading-order noises in CV-DV systems, such as cavity decay, CV diffusion, and qubit dephasing. 

To ensure that $\hat{\varrho}$ is a Hermitian operator under such a dynamics, it follows that also $B$ has to be Hermitian, and can be, in general, rewritten as 
\begin{equation}
    B = \frac{1}{2} \Omega^{\rm T} D \Omega - i  E \; ,
\end{equation}
where $D$ and $E$ are symmetric and antisymmetric $2n \times 2n$ matrices representing, respectively, the diffusion and drift of the dynamics. We explicitly note that if the Lindbladian terms are known in the ladder operator basis (with noise matrix $B_a$ and driving $d_a$), one can derive the noise matrix in the canonical operators basis as $B = U^{\dag} B_a U$ and $d = U^{\dag} d_a$, where $U$ is given in Eq.~(\ref{eq:change_basis}).  At the fundamental and microscopic level, these terms arise via the linear interaction between the system's modes and a Markovian bath of a large number of modes with first and second moments given by $r_{\text{in}}$ and $\sigma_{\text{in}}$. Given the coupling Hamiltonian between the system and the bath $\hat{r}_s H_c \hat{r}_{\text{in}}$, from the Hamiltonian matrix $H_C$, it is possible to derive $D = \Omega H_C \sigma_{\text{in}} H_C^{\rm T} \Omega^{\rm T}$, $E =H_C \Omega H_C^{\rm T}/2$, and $d = H_C r_{\text{in}}$, which imply $D \geq 0$, since $\sigma_{\text{in}} \geq 0$~\cite{serafini_quantum_2017}. 

In Appendix~\ref{app:noise}, the additional noisy terms are added to the ODEs for the unitary dynamics (Table \ref{tab:unitary}), to find that the phase-space parameters of a GBCS undergoing a general open Markovian operator-valued Gaussian dynamics evolve according to Table \ref{tab:open}. First of all, we note that, from the form of the coupling between the sets of ODEs, the DV noise does not affect the dynamics of the CV system. Yet, the noise in the modes changes the time evolution of the QRDM, as $\dot{s}_{jk}$ has an explicit dependence on $\sigma_{jk}$ and $\bar{r}_{jk}$. Specifically, the differences with respect to the unitary case due to CV noise are: (a) the inclusion of an asymmetric part in the Hamiltonian (typical of decays, embodied by $E$); (b) a linear increase in the covariance matrices (due to the diffusive term including $D$); and (c) the linear term in the first moments (the driving term $d$). 

The DV noise leads to a decay with rate $\gamma_{jk} := \frac{1}{2} (\Gamma_z^{jj} + \Gamma_z^{kk} ) - \Gamma_z^{jk} $. It may be insightful to state the explicit form of DV noise for physical systems:
(i)~for a single qubit dephasing, described by the jump operator $\sqrt{\Gamma_z/2} \hat{\sigma}_z$, has $\Gamma_z^{jk} = \Gamma_z/2 (2j-1) (2k- 1)$ and $\gamma_{jk}  = \Gamma_z (j - k)^2$, where we recall that the unitless $\Gamma_z  = 1/(\omega T_2)$, i.e., the inverse of the dephasing time in frequency units; (ii)~for a large spin $J$ dephasing, with the jump operator $\sqrt{2 \Gamma_z} \hat{J}_z$, which implies $\Gamma_z^{jk} = 2 \Gamma_z (j-J) (k- J)$ and $\gamma_{jk}  = \Gamma_z (j - k)^2$; and (iii)~for $N_q$ qubits, with $N_q$ jump operators $\sqrt{\Gamma_{z}^{i}/2} \hat{\sigma}^{i}_z$ one finds the update to the QRDM via 
\begin{equation}
    {\rm e}^{-\gamma_{jk}\tau}
=
\left[
\bigotimes_{i=1}^{N_q}
\begin{pmatrix}
1 & {\rm e}^{-\Gamma_z^{(i)}\tau} \\[4pt]
{\rm e}^{-\Gamma_z^{(i)}\tau} & 1
\end{pmatrix}
\right]_{jk} \; .
\end{equation}

Remarkably, when restricting to operator-valued linear interactions (Eq.~\ref{eq:Hamiltonian_force_only}), the solution to the open dynamics can be expressed in integral form as in the case of unitary dynamics. In fact, being the coupling only linear, one can define a single, non-symmetric matrix $A = H_m + E$, which governs the quadratic evolutions of all phase spaces, yielding under exponentiation a (non-symplectic) transformation $S_A(\tau) = \exp( \Omega A \tau)$~\cite{serafini_quantum_2017}. Thus, as in the case of unitary evolutions, the covariance matrices $\sigma_{jk}$ evolve all according to the Lyapunov equation
\begin{equation}
\label{eq:Lyapunov_sol}
    \dot{\sigma} =  \Omega A \sigma - \sigma A^{\rm T} \Omega + D 
\end{equation}
The general integral form of its solution, alongside $\bar{r}_{jk}(\tau)$, $\mathcal{C} (\tau)$, and $\phi (\tau)$ are given in Table~\ref{tab:open}, such that the branched characteristic functions of Eq.~(\ref{eq:chara_jk}) are solutions of the open quantum dynamics under any linear operator-valued interaction. Furthermore, for $E=0$, another simplification can be found in  Table~\ref{tab:open_app} (and derived in Appendix~\ref{tab:open_app}), where we note that the noise does not affect the phases (as they do not depend on $D$).

\subsection{Linear Solutions to Riccati Equations \label{subsec:riccati}}

The Riccati term present in quadratic coupling is the only term that, in this work, does not allow for a general solution in integral form for all the phase-space parameters. Thankfully, Riccati equations are quite common, especially in control theory of classical~\cite{Kalman_new_1960,AbouKandil2003} and quantum systems~\cite{wiseman_optimal_2005,wiseman_quantum_2010,serafini_quantum_2017} (as well as in time-dependent and dissipative quantum systems~\cite{CRUZ201544,CRUZ2016609,Schuch2018}),
so that a few available approaches to their solution exist~\cite{Bini2012,blanco_interplay_2019}. Here, we discuss systematic paths to find solutions of the Riccati equation appearing in Tables~\ref{tab:unitary}~and~\ref{tab:open}, by solving an equivalent linear system for twice the number of variables (see Appendix~\ref{app:doubling} for proof and details). 

The map from a Riccati equation to a linear system of ODEs is achieved by considering the ansatz for the covariance matrices of the form $\sigma_{jk} = \mathds{Y}_{jk} \mathds{X}_{jk}^{-1}$, where $\mathds{X}_{jk} $ and $ \mathds{Y}_{jk}$ are two time-dependent $2n$-dimensional matrices. One then finds that $\sigma_{jk}$ is a solution of the Riccati equation if 
\begin{align}
    \frac{\text{d}}{\text{d} \tau}\begin{pmatrix}
        \mathds{X}_{jk} \\
        \mathds{Y}_{jk}
    \end{pmatrix} =  \mathds{A}_{jk}
    \begin{pmatrix}
        \mathds{X}_{jk} \\
        \mathds{Y}_{jk}
    \end{pmatrix} \nonumber
\end{align}
where 
\begin{equation}
    \mathds{A}_{jk}  = \frac{1}{2} \begin{pmatrix}
           ( H_j + H_k) \Omega  &  i ( H_j - H_k) \\
          i \Omega ( H_j - H_k)  \Omega^{\rm T }   & \Omega  ( H_j + H_k   )
          \end{pmatrix}
\end{equation}
under the initial conditions $\mathds{Y}_{jk}(0) = \sigma_0$ and $\mathds{X}_{jk}(0) = \mathds{1}$, to enforce that  $\sigma (0) = \sigma_0$ and that the solution is the physical one (or similarly $\mathds{Y}_{jk} = \sigma^0_{jk}$ for an initial GBCS). Being a linear system, the solution can be expressed via the transformation $\mathds{S}_{jk} (\tau) = \exp \left( \mathds{A}_{jk}  \tau \right) $ and is given by $(\mathds{X}_{jk}, \mathds{Y}_{jk})^{\rm T} (\tau) = \mathds{S}_{jk} (\tau) ( \mathds{1}, \sigma_0)^{\rm T}$. Hence, the analytical solution exists and can be found if it is possible to compute $\mathds{S}_{jk}$ and $\mathds{X}_{jk}$ is invertible.

In the specific case of a single qubit (with only $\sigma^{\text{off}} = \mathds{Y} \mathds{X}^{-1}$ having the Riccati terms and Hamiltonian of Eq.~\ref{eq:Hamiltonian_max_qubit}), the linear system reads 
\begin{equation}
    \frac{\text{d}}{\text{d} \tau}\begin{pmatrix}
        \mathds{X} \\
        \mathds{Y}
    \end{pmatrix} =
    \begin{pmatrix}
           H_m \Omega &     i  H_q \\
          i \Omega H_q \Omega^{\rm T}  & \Omega H_m 
    \end{pmatrix}\begin{pmatrix}
        \mathds{X} \\
        \mathds{Y}
    \end{pmatrix}  \;.
\end{equation}
This can in turn be used to solve the other phase-space parameters coupled to the time evolution of the covariance matrix (see Appendix~\ref{app:doubling}). This procedure can be used to compute specific analytical solutions (as we shall see in the examples of Sec.~\ref{sec:operations} and Sec.~\ref{sec:physical_examples}) and semi-analytical in terms of a restricted number of linear systems. 

\section{Operations \& Deterministic Maps \label{sec:operations}}

In practice, it is often useful to describe operations that may not be dynamical and are instead applied on a state at a given time (while they may arise from fundamental dynamics)~\cite{hybrid_Liu_2026}. Alongside the class of local operations on the DV (such as a rotation) and Gaussian operations on the CV (such as a beam-splitter), GBCS can also be used to describe the Gaussian operations conditional on the qubit states: this is achieved as maps of the phase-space parameters characterising a GBCS. Two key operations of the latter class are conditional displacements and rotations, and they will be used as pedagogical examples in Sec.~\ref{sec:wigner} and whose dynamical and experimental implementations are discussed in Sec.~\ref{sec:physical_examples}. The interplay between these two conditional operations, allowing one to prepare GBCS, along with Gaussian CV processes and DV operations is key to achieving sophisticated quantum states and tasks~\cite{hybrid_Liu_2026}. In Appendix~\ref{app:operation}, we provide the reader with proofs concerning the following sections.

\subsection{Gaussian Operations \label{subsec:gaussian_operations}}

A vast toolbox of experimentally useful operations are those that preserve the Gaussian structure of states, which have been extensively studied~\cite{serafini_quantum_2017}, and are immediately applicable on GBCS, given their phase space representation. As introduced in Sec.~\ref{sec:intro_phase}, a general Gaussian completely positive and trace-preserving map $\Phi_g$ is uniquely determined by two matrices $X$ and $Y$, such that $Y+ i \Omega \geq i X \Omega X^{\rm T}$. As in the case of Gaussian states, by applying the action of the dual Gaussian map on the displacement operators, used as the CV basis, one finds that a GBCS evolves to the new parameters $\sigma_{jk} \to X  \sigma_{jk} X^{ {\rm T}} + Y$ and $
    \bar{r}_{jk}  \to X \bar{r}_{jk}$, while the QRDM is unchanged. 
Examples of these maps are: (a) classical mixing with $X= \mathds{1}$ and $Y >0 $, (b) attenuators with $X = \cos \theta \mathds{1}$ and $Y = (\sin \theta)^2 (2 N_p + 1) \mathds{1}$, and (c) amplifiers with $X = \cosh r \mathds{1}$ and $Y = (\sinh r)^2 (2 N_p + 1) \mathds{1}$, where $N_p\ge0$ parametrises the noise induced by the operations. The specific instance of Gaussian unitary operations is captured by $X = S \in Sp_{2, \mathds{R}}$ and $Y = 0$, including, for instance, CV rotations, beam splitters, local squeezing, and two-mode squeezing~\cite{serafini_quantum_2017} (see Appendix~\ref{app:operation} for their explicit forms). 

\subsection{Qubit Operations}

Often, in order to perform protocols, operations are independently implemented on the DV, for instance, rotations of qubits. A completely positive DV operation is defined by a set of Kraus operators $K^i = K^i_{jk} \mathds{1} \otimes \ket{j}\bra{k}$, labelled by $ i \in [1, r]$ with $r \leq N^2$ (where we also include the trivial identical action on the CV subsystem). The specific instance of a quantum channel, which is also trace-preserving, requires $\sum_{i = 1}^{r}  K^{(i) \;\dag}_{jk} K^{(i)}_{kl} = \mathds{1}$. The update of GBCS can be trivially computed by its definition and the action of the channel $\Phi_{q} [\hat{\varrho}] = \sum_{i = 1}^{r}  K^{i \; \dag} \hat{\varrho} K^i$, such that the characteristic functions are updated according to 
\begin{equation}
    \chi_{jk} (\tilde{\xi})\to \sum_{i = 1}^{r}  K^{(i) \;\dag}_{jm}  \chi_{m n}(\tilde{\xi}) K^{(i)}_{n k} 
\end{equation}
where we used the Einstein notation on rotated indexes. As is apparent, this mixes the original branched characteristic functions.

For instance, the rotation around the $x$-axis of a single qubit has only one Kraus operator 
\begin{equation}
    R_x(\theta) = {\rm e}^{-i\theta\sigma_x/2} = \begin{pmatrix} \cos(\theta/2) & -i\sin(\theta/2) \\ -i\sin(\theta/2) & \cos(\theta/2) \end{pmatrix} ,
\end{equation}
updating  the characteristic functions to 
   $ \chi_{jk} (\tilde{\xi})\to R^{jm}_x(\theta) \chi_{m n}(\tilde{\xi}) R^{nk}_x(\theta)  $. Hence, each branched characteristic function becomes a sum of Gaussians. By virtue of the fact that the PDEs are linear, the dynamics of the resulting state can still be analysed in terms of the ODEs and their solutions, feeding in each~ODE $N^2$ Gaussian initial conditions (i.e. $\sigma_{jk}(0)$, $\bar{r}_{jk}(0)$ and $s_{jk}(0)$).

\subsection{Conditional Operations}

The set of quadratic operations conditional on the state of a DV subsystem -- including, for instance, conditional rotations, beam splitters, and displacements -- requires utmost care, especially due to the required handling of unnormalized operators~\cite{hybrid_Liu_2026}. A general conditional unitary quadratic operation may be written as
\begin{align}
    \hspace{-0.2cm}\Phi^{\mathcal{C}}[\hat{\varrho}] &= \hspace*{-0.2cm} \sum_{j, k  = 0}^{N-1} \int_{\mathds{R}^{2n}} \frac{\text{d} \xi}{(2 \pi)^n} \chi_{jk} (\xi) \hat{D}_{d_j} \hat{S}_{\theta_j} \hat{D}_{\xi} \hat{S}_{\theta_k}^\dag \hat{D}_{d_k}^\dag  \otimes \ket{j} \bra{k} , \nonumber
\end{align}
where $\hat{S}_{\theta_j} = \exp(i  \hat{r}^{\rm T} C_j  \hat{r}/2)$ and $\hat{D}_{d_j} = \exp(i  d_j^{\rm T} \Omega \hat{r}^{\rm T} )$ are two sets of $N$ exponential operators quadratic in the canonical operators, parametrised by some matrix $C_j$ and $d_j$. For linear operations, i.e., conditional displacement with $C_j = 0$, one can proceed by computing the update rule of the characteristic function $\chi_{jk}$ from its trace definition and compositions of displacement operators, allowing us to find the general updated rule (Sec.~\ref{subsubsec:constional_displacement}). However, for quadratic operations, there is not a general composition rule for $\hat{S}_{\theta_j}$ operators, and (also in the specific instances where compositions are possible) the trace is not computable as these operators are not trace norm. Thankfully, by knowing the generators of the transformations (i.e. the matrix $C_j$), one can use the ODEs of Table~\ref{tab:unitary}  (under $H_j \to C_j$, $r_j \to 0$ and $\tau = 1$) to obtain the update rule in differential form and compute specific instances for a given state. Extensions to non-unitary operations can be achieved via Table~\ref{tab:open}. 

\begin{figure*}[!t]
    \centering
    \includegraphics[width=1\linewidth]{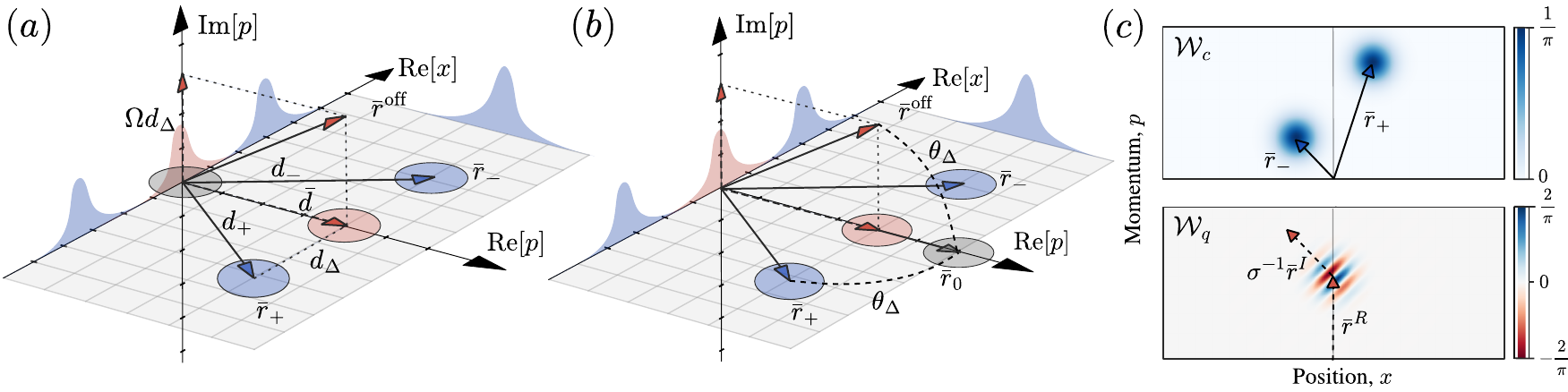}
    \caption{Representation of (a) a conditional displacement with vectors $d_\pm = \left( \bar{x} \pm \delta \bar{x} , \bar{p} \right)$ of the ground state creating a superposition in position and (b) a conditional rotation of angle $\theta_\Delta$ creating the same GBCS from a coherent state. (c) The Classical and Quantum component Wigner function given of the GBCS with arbitrary (asymmetric) displacement.} 
    \label{fig:conditional_dis}
\end{figure*}

\subsubsection{Conditional Rotations}

\begin{figure*}[!t]
    \centering
    \includegraphics[width=1\linewidth]{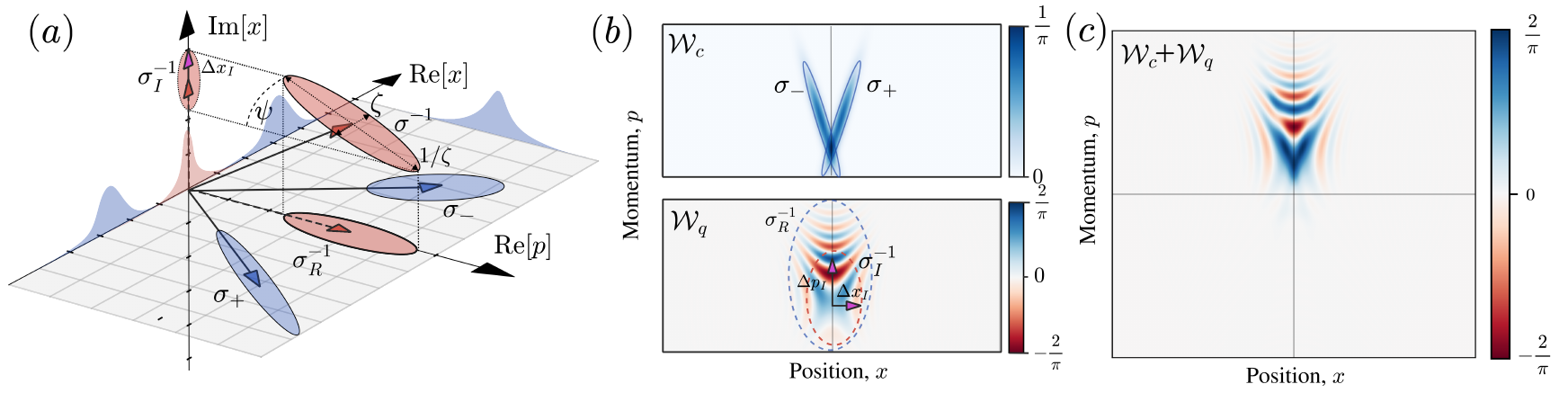}
    \caption{Conditional rotation of the squeezed state creating a GBCS: (a) representation of different squeezing for each branch and complex off-diagonal covariance matrix, (b) classical and quantum contribution of the Wigner, and (c) the total Wigner function.} 
    \label{fig:conditional_rot}
\end{figure*}

The previous point is better illustrated by a specific example: the conditional rotation of a single CV mode by angles $\theta_{\pm}$, written as $\hat{S}_{\theta_\pm}  = \exp ( i \theta_\pm/2 (\hat{x}^2 + \hat{p}^2))$. This is the case of $H_{\pm} = \theta_{\pm} \mathds{1} $. On the diagonal elements, the two Gaussian unitary transformations give the update rules: $\sigma^{\text{on}}_\pm = \to R_{\theta_\pm}\sigma^{\text{on}}_\pm  R_{\theta_\pm}^{\rm T}$ and  $\bar{r}^{\text{on}}_\pm  \to  R_{\theta_\pm} \bar{r}^{\text{on}}_\pm $ with $R_{\theta_\pm} = \exp(\theta_\pm \Omega)$, both phase space rotations (Sec.~\ref{subsec:gaussian_operations}). The off-diagonal ODEs read
\begin{align}
\label{eq:ODE_conditional_rot}
    \dot{\sigma}^{\text{off}} &= \bar{\theta} \left[\Omega \sigma - \sigma \Omega \right] + i \theta_{\Delta} \left( \mathds{1} - \sigma^{\text{off}} \sigma^{\text{off}} \right) ,\nonumber \\ 
    \dot{\bar{r}}^{\text{off}} &= \left[ \bar{\theta} \Omega -  i \theta_{\Delta} \sigma \right]\bar{r}^{\text{off}} , \\ 
    \dot{s}^{\text{off}} & = - i \theta_{\Delta} \left[ {r}^{\text{off} \; \rm T}{r}^{\text{off}} +  \text{Tr}[\sigma^{\text{off}} ]/2\right] , \nonumber
\end{align}
where $\dot{s}^{\text{ off}}$ is the off-diagonal exponent of the QRDM and we defined $\theta_{\Delta} = (\theta_+ - \theta_-)/2$ and $\bar{ \theta} = (\theta_+ + \theta_-)/2$. For an initial coherent state, with $\sigma_0 = \mathds{1}$ and $r_0 = (x_0 , p_0)^{\rm T}$, the evolution of the covariance matrix is trivial, i.e. $\sigma^{\text{off}} (\tau) = \mathds{1} \forall \tau$, and one recovers the known results  $r^{\text{off}} (\tau) = {\rm e}^{-i \theta_\Delta } r_0$ and $\varrho_q^{\text{off}} = \exp \left(  - |r_0|^2 \left( 1 - {\rm e}^{-i 2 \theta_\Delta} \right)/2 -  i \theta_\Delta \right)$. Looking for a non-trivial evolution of the covariance matrix, one can compute the conditional rotation of a squeezed state -- with $\sigma_0 = \text{diag} (1/s, s)$ -- by solving the Riccati equation via the doubling method (Sec.~\ref{subsec:riccati}), finding the update rule
\begin{align}
\label{eq:cov_rotation}
     \sigma_0 &\to    R_{\bar{\theta}} \begin{pmatrix}
 {\rm e}^{i \psi} |\zeta | & 0 \\
 0 &  {\rm e}^{-i \psi}/|\zeta|
    \end{pmatrix} R_{\bar{\theta}}^{\rm T}  
\end{align}
where $ R_{\bar{\theta}}$ is the rotation of the mean angle and, by using the polar decomposition of the complex covariance matrix, one finds the squeezing strength $\zeta^2 = 
 \frac{\cos ^2(\theta_{\Delta} )+s^2 \sin ^2(\theta_{\Delta} )}{\sin ^2(\theta_{\Delta} )+s^2 \cos ^2(\theta_{\Delta} )} $ and the complex phase $\psi = \arctan \left( \frac{1}{2} (s -\frac{1}{s}) \sin(2 \theta_\Delta) \right) $.  In Appendix~\ref{app:operation}, we derive the update for the first moment and elements of the QRDM, which can be visualised as in  Fig.~\ref{fig:conditional_rot} over complex evolution.

\subsubsection{Conditional Displacements\label{subsubsec:constional_displacement}}

To create GBCS, displacements conditional on the qubit states are often performed. A general conditional displacement is given by $D_{d_1,...,d_N} = \sum_{i}^{N}  D_{d_j}  \ket{j} \bra{j}$, where the $d_j$'s are $N$ vectors parametrising the displacements. 
The conditional structure of this operation allows one to write its action on a GBCS as 
\begin{align}
\label{eq:displacement_channel}
    \Phi^{D}[\hat{\varrho}] &= \frac{1}{(2 \pi)^n} \sum_{j, k  = 0}^{N-1}  \int_{\mathds{R}^{2n}} \text{d} \xi \; \chi_{jk} (\xi)  \hat{D}_{d_j} \hat{D}_{\xi} \hat{D}_{d_k}^\dag  \ket{j} \bra{k} .
\end{align}
By applying the composition rule of displacements ($ \hat{D}_{r_1} \hat{D}_{r_2} =  {\rm e}^{\frac{i}{2} r_2^{\rm T} \Omega r_1} \hat{D}_{r_1+r_2}$) and redefining the variable of integration, one can recast Eq.~(\ref{eq:displacement_channel}) into Eq.~(\ref{eq:chara_jk_def}), thus finding the map of the GBCS parameters under general conditional displacements:  
\begin{align}
\label{eq:cond_dis_update}
    \bar{r}_{jk} &\to \bar{r}_{jk} - \frac{1}{2} (d_j +d_k) - \frac{i}{2} \sigma_{jk} \Omega  (d_j - d_k), \nonumber\\
    \mathcal{C}_{jk} &\to \mathcal{C}_{jk} + \frac{1}{4} \left( d_j - d_k\right)^{\rm T} \Omega^{\rm T} \sigma_{jk}  \Omega \left( d_j - d_k\right),\nonumber \\
    &\hspace{1.5cm} +  (d_j - d_k)^{\rm T} \Omega \; \text{Im} [\bar{r}_{jk}]  \nonumber \\
    \phi_{jk} &\to \phi_{jk} - \frac{1}{2}  d_j^{\rm T}  \Omega d_k  +  (d_j - d_k)^{\rm T} \Omega \text{Re} [ \bar{r}_{jk}] ,
\end{align}
where we separated the real and complex parts of the first moments $\bar{r}_{jk} = \text{Re} [ \bar{r}_{jk}]  + i \text{Im} [\bar{r}_{jk}]$. As expected, conditional displacements do not affect the covariance matrix ($\sigma_{jk} \to \sigma_{jk}$). 

For the case of a single qubit,  $\hat{D}_{d_1, d_{-1}} = \hat{D}_{d_+} \ket{1}\bra{1} + 
    \hat{D}_{d_{-}} \ket{-1}\bra{-1} $ for the two displacement vectors $d_\pm$. One finds that $\bar{r}^{\text{on}}_{\pm} \to \bar{r}^{\text{on}}_{\pm} - d_\pm$ and 
\begin{align}
\label{eq:cond_dis_update_qubit}
    \bar{r}^{\text{off}} &\to \bar{r}^{\text{off}} - \bar{d} - i  \sigma^{\text{off}} \Omega  d_\Delta, \nonumber\\
    \mathcal{C} &\to \mathcal{C} +  d_{\Delta}^{ \rm T} \Omega^{\rm T} \sigma^{\text{off}}  \Omega d_\Delta  \nonumber + 2 d_\Delta^{\rm T} \Omega  \text{Im} [\bar{r}^{\text{off}}] \nonumber \\
    \phi &\to \phi - \frac{1}{2}  d_+^{\rm T}  \Omega d_{-} + 2  d_\Delta^{\rm T} \Omega\text{Re} [\bar{r}^{\text{off}}] ,
\end{align}
where $\bar{d} = (d_+ + d_-)/2$ and  $d_\Delta = (d_+ - d_-)/2$ are the average and difference in the displacements. The visualisation of these complex update rules is in Fig.~\ref{fig:conditional_dis}.

\section{Geometry of Wigner Functions \& Interpretation for Complex Moments \label{sec:wigner}}

The interplay between the Wigner function of GBCS and the phase-space parametrisation allows us to provide the reader with a geometrical description of the Wigner negativity common in cat states and, hence,  with an interpretation for the real and imaginary components of the first and second moments. By computing the Fourier transform of Eq.~(\ref{eq:chara_jk}), the branched Wigner functions of a general GBCS are given by
\begin{equation}
\label{eq:wigner_jk}
    \mathcal{W}_{jk} (\xi) = \frac{2^n \varrho_{jk}^q}{\pi^n \sqrt{\text{det} \sigma_{jk} }}\exp \left( -\left(\xi - \bar{r}_{jk} \right)^{\rm T}  \sigma_{jk} ^{-1} \left(\xi - \bar{r}_{jk} \right) \right).
\end{equation} 
On the diagonal terms, they are real and positive. On the off-diagonal terms, $\mathcal{W}_{jk}$ are complex which, when summed over $j$ and $k$, interfere with each other, giving rise to the fringes typical of GBCS.

To illustrate this, let us consider the case of $n = 1$ and $N=2$ (i.e. the state of Eq.~\ref{eq:density_one_qubit}). For the phase-space picturization of GBCS, two quantities are relevant: the  Wigner functions of the classical components, $\mathcal{W}_c := \mathcal{W}_+  + \mathcal{W}_{-}$, and the `quantum interference', given by $\mathcal{W}_q :=   \mathcal{W}^{\rm off} + \mathcal{W}^{\rm off\; *} = 2 \text{Re} [\mathcal{W}^{\rm off}]$. The former is the statistical mixture of the Wigner functions of two Gaussian states with moments $\sigma_{\pm}$ and $ \bar{r}_{\pm}$, weighted according to the on-diagonal probabilities ($p_{\pm}$). This contribution to the Wigner function can be used to build physical intuition, representing the two CV Gaussian processes that the system undergoes: the first moments are the \textit{semi-classical} interferometric paths taken with uncertainties captured by the corresponding covariance matrix. However, such a representation, which is restricted to the real plane, does not give a full description of the quantum state (for instance, omitting the relative phase between the branches). 

The quantum interference component -- which encodes the truly quantum features of GBCS -- finds its geometrical interpretation with the off-diagonal first and second moments evolving in the complex plane. The form of the quantum component can always be expressed as
\begin{align}
\mathcal{W}_q (\xi) &= \mathcal{N} \mathcal{G} (\xi) \cos \left[ \Phi_W(\xi) \right], 
\end{align}
where $\mathcal{G}(\xi)$ is a smearing Gaussian distribution, $\Phi_W(\xi) $ is the Wigner phase of interference (function of $\xi$),  $\mathcal{N}$ is a normalization constant, and $\Phi_0$ is an additional constant phase. Let us define the real and complex components of the first moments and the inverse of the covariance matrix: namely, $\bar{r}^{\text{off}} = \bar{r}_R + i \bar{r}_{I}$ and $\sigma^{\text{off} -1} = \sigma_R^{-1} + i \sigma_{I}^{-1} := \text{Re} [\sigma^{-1}]+ i \text{Im} [\sigma^{-1}]$\footnote{Where we note that, for instance, $\sigma_R^{-1}$ is \textit{not} the inverse of the real part of the covariance matrix, but is the real part of the inverse of the covariance matrix. They are equivalent only if $\sigma^{\text{off}}$ is diagonal.}. Then, one finds the general 
\begin{align}
\mathcal{G} (\xi) &= \exp \left( {-(\xi -\bar r_R)^{\mathsf T} \sigma_R^{-1} (\xi -\bar r_R)- 2 \tilde \xi^{\mathsf T} \sigma_{I}^{-1}  \bar r_I} \right) \nonumber \\
\mathcal{N} &= \frac{4|q|}{\pi\sqrt{|\det(\sigma)|}}\,\exp \left( - \mathcal{C} + \bar r_I^{\mathsf T}\sigma_R^{-1}\bar r_I + \bar r_R^{\mathsf T} \sigma_{I}^{-1}  \bar r_I \right) \nonumber \\
\Phi_W (\xi) &= (\xi-\bar r_R)^{\mathsf T} \sigma_{I}^{-1} (\xi-\bar r_R)-2(\xi -\bar r_R)^{\mathsf T} \sigma_R^{-1}  \bar r_I + \Phi_0 \nonumber \\
\Phi_0 &=  \bar{r}_I^{\mathsf T} \sigma_{I}^{-1} \bar{r}_I - \frac{1}{2} \arg ( \det(\sigma)) + \phi
\end{align}

First, let us consider the case when the superposition is only present in the first moments and $ \sigma^{\text{off}} \in \mathds{R}$. Then, one can understand that the real components of the off-diagonal first moments ($\bar r^{R}$) are the centre of the Gaussian envelope that smears the fringes. The complex components $\bar r^{I}$ of the first moments are what cause the interference fringes, which, in this case, are linear in $\xi$. The fringes are fully characterised by the Wigner wave-vector $k_W = 2 \sigma^{-1}\bar{r}_{I}$ with the zero at $\bar{r}_R$. Furthermore, the imaginary part of the first moments induces an increment in the normalization (which often, in ideal cases, cancels the DV decay $\mathcal{C}$). Conditional displacements and rotations presented in the Section~\ref{sec:operations} can be used as examples:  (a) a conditional displacement of the vacuum (Eq.~\ref{eq:cond_dis_update} with $r_0 = (0,0)$) has fringes $k_W =  2 \Omega d_{\Delta}$ and the envelope Gaussian is centred at $\bar{d}$, with $\mathcal{N} = 2/\pi$; or (b) by a conditional rotation from a coherent state with $r_0 = (x_0, p_0)$ has $k_W =  - 2\sin\!\left(\frac{\Delta\theta}{2}\right) r_0$ smeared by a normalized Gaussian ($\mathcal{N} = 2/\pi$) of minimal uncertainty centred at $\cos\!\left(\frac{\Delta\theta}{2}\right) r_0$. This allows the interpretation and picturization of Fig.~\ref{fig:conditional_dis}.

Secondly, when $\sigma^{\text{off}} \in \mathds{C}$, the imaginary part of the inverse of the covariance matrix ($\sigma_{I}^{-1}$) induces non-linear fringes that are quadratic in the phase space variables, via the term $(\xi-\bar r_R)^{\mathsf T} \sigma_{I}^{-1} (\xi-\bar r_R)$.
It further includes a shift of the Gaussian envelope (of the vector $\sigma_{I}^{-1}  \bar r_I$) and a change in the normalization ($\exp( \bar r_R^{\mathsf T} \sigma_{I}^{-1}  \bar r_I )$). 
For example, in the case of the conditional rotation of a squeezed state (Sec.~\ref{sec:operations}, with $\bar{\theta} = 0$), the resulting covariance matrix is the one of a squeezed state tilted in the complex plane of an angle $\psi$ (Eq.~\ref{eq:cov_rotation}) as shown in Fig.~\ref{fig:conditional_rot}. This implies that $\sigma^{-1}_R = \cos \psi \text{diag} \left(1/\zeta , \zeta  \right) $ and $\sigma^{-1}_I =  \sin \psi \text{diag} \left( - 1/\zeta , \zeta \right) $. The difference in the sign of the $\sigma^{-1}_I$ diagonal elements implies that the quadratic term of the fringes are hyperbolic~\cite{nicacio_phase_2010}, as $\Phi_W $ contains a term $\sin \psi  \left( - \tilde{x}^2 /\zeta + \tilde{p}^2 \zeta \right) $.

\section{Measurements \& \\ Expectation Values \label{sec:measure}}

Having solved a large class of deterministic dynamics and operations, one may be interested in the behaviour of GBCS under quantum measurements. As in the case of dynamics and operations, the phase-space description of the system allows for easy computations and for the compact expression of expectation values, measurement probabilities, and post-measurement quantum states in terms of the phase-space parameters $\sigma_{jk}$, $\bar{r}_{jk}$, and $\varrho_{jk}^q$. 

Let us start by giving an overview of computing expectation values of GBCSs, by considering the general observables of the joint system $  \hat{\mathcal{O}}^m\otimes\hat{\mathcal{O}}^q$. The CV observables can be canonical operators $\hat{r}^k$ (where $k$ labels the $k$-th element of the $\hat{r}$ vector), products of these (e.g., the product $\hat{r}^{k_1} \hat{r}^{k_2}$ or higher order terms), or observables in the ladder operator basis (such as the phonon number $a^\dag a$, which can be always computed via the transformation of Eq.~\ref{eq:change_basis}). Any DV observable can be written in the coupling basis as $\hat{\mathcal{O}}_q = \sum_{j, k  = 0}^{N-1} O^{q}_{jk} \ket{j} \bra{k}$. 

In order to evaluate the joint system expectation value one can first perform the trace on the modes degrees of freedom. Generically, the product of $n$ canonical observables $\mathcal{O}^m = \hat{r}^{k_1} \hat{r}^{k_2}... \hat{r}^{k_n} $ can be computed from the repetitive use of the generating function
\begin{equation}
\label{eq:expectation_characterstic}
    \braket{\hat{\mathcal{O}}^m_{jk}} = (-i )^n \left. \frac{\partial^{n} \chi_{jk}(\tilde{\xi})}{\partial \tilde{\xi}^{k_1} \partial \tilde{\xi}^{k_2} ... \partial \tilde{\xi}^{k_n}} \right|_{r=0} \;,
\end{equation}
via the partial derivative of the corresponding phase space variable. Eq.~(\ref{eq:expectation_characterstic}) can always be evaluated in terms of $\sigma_{jk}$, $\bar{r}_{jk}$, and $s_{jk}$. Specifically, for general CV observables, one finds $\braket{\hat{r}}_{jk} := \text{Tr} (\hat{\varrho}_{jk} \hat{r}) = \varrho_{jk}^q \bar{r}_{jk}$ and similarly $\braket{\hat{r} \hat{r}^{\rm T}}_{jk} := \text{Tr} (\hat{\varrho}_{jk}\hat{r} \hat{r}^{\rm T}) = \varrho_{jk}^q  \left( \sigma_{jk} - \bar{r}_{jk} \bar{r}_{jk}^{\rm T} \right)$, representing indeed the first and second moments of each element of the density matrix, respectively. Finally, one can compute the joint observable by tracing over the DV Hilbert space, such that $\braket{\hat{\mathcal{O}}^m\otimes\hat{\mathcal{O}}^q} = \sum_{jk} \braket{\hat{\mathcal{O}}^m_{jk}} \mathcal{O}_{jk}^q$. Two specific cases can be derived from this general treatment. The expectation value of any observable of the qubit subsystem can be computed via the QRDM, as Eq.~(\ref{eq:expectation_characterstic}) simplifies to $\chi_{jk} (0) = \varrho_{jk}^q$, implying $\braket{\mathds{1} \otimes \hat{\mathcal{O}}_q } = \sum_{jk} O_{jk} \varrho^{q}_{jk}  $. Any observables of only the CV subsystem can be computed as the statistical mixture (weighted according to the QRDM diagonal terms, i.e. probabilities) of the expectation values computed with the Gaussian state of each branch: for example, the vector of expectation values reads $\braket{\hat{r} \otimes \mathds{1}} = \sum_{i} \varrho_{ii}^q r_{ii} $, while the matrix of correlations is $\braket{\hat{r}\hat{r}^{\rm T} \otimes \mathds{1}} = \frac{1}{2}\sum_{i} \varrho_{ii}^q \left( \sigma_{ii} - r_{ii} r_{ii}^{\rm T} \right)$.

Let us now move on to describing the granular outcomes of measurements and their associated conditional states, starting from single-subsystem measurements and concluding with joint measurements.

An ideal projective measurement of an observable $\mathcal{O}^q$ has $N^2$ outcomes (which may be degenerate) given by the eigenvalues ${\rm e}^i$. From the associated eigenvectors $\ket{{\rm e}^i}_q$ it is possible to derive the positive operator valued measure (POVM), i.e., in this case, the set of projectors $\{M^i =\ket{{\rm e}^i}_q \bra{{\rm e}^i}_q= M^i_{jk} \ket{j}\bra{k} \}$, where $\ket{j}$ and $\ket{k}$ refer to the coupling basis, and such that $\sum_i M^i = \mathds{1}$. The probability of the $i$-th  outcome can be computed as the expectation value using the QRDM as previously described for observables of the qubit's subsystem, i.e., as $P(i) = \sum_{jk} M^i_{jk} \varrho_{jk}^q$. The post-measurement state after the outcome $i$ has the characteristic function 
\begin{equation}
    \chi^{\text{post}}_i (\xi) = \frac{1}{P(i)} \sum_{j, k  = 0}^{N-1}  M^{i}_{jk} \chi_{jk} (\xi) \;,
\end{equation}
 which may well show fringes of GBCS and negativity in the Wigner function (see Sec.~\ref{sec:physical_examples}). 

The class of Gaussian measurements of the CV subsystem is described by the ``general-dyne'' measurements (with the POVM of Eq.~\ref{eq:generadyne}). These measurements are described by sets of projectors $\Pi_{r_o}$ labelled by the continuous labels of possible outcomes $r_o$, which represent the first moments of the post-measurement Gaussian state $\hat{\varrho}_o$ of covariance $\sigma_o$, i.e., $\Pi_{r_o} = \hat{D}_{-r_o} \hat{\varrho}_g \hat{D}_{r_o}/(2 \pi)^n$, where $\varrho_{g}$ is the Gaussian state with null first moments and covariance matrix $\sigma_{o}$. As in the case of expectation values, it is convenient to first trace over the CV subsystem, which gives the unnormalized post-measurement QRDM associated with the outcome $r_o$, i.e.,  $ \varrho^{q,\; \text{post}} (r_o) = \text{Tr}_m [\Pi_{r_o} \hat{\varrho}_{jk} ]$. Similarly to Eq.~(\ref{eq:prob_generadyne_gaussian}), its elements can be easily computed as the overlap between a GBCS and the measurement Gaussian state,\footnote{This follows from the orthogonality of Weyl operators under trace, $\text{Tr}_m [\hat{D}_{r} \hat{D}_{-s}] = (2 \pi)^{n} \delta^n (r - s)$ and from the Fourier--Weyl representation of the involved states.} such that
\begin{equation}
\label{eq:qubit_PMST}
    \varrho_{jk}^{q,\; \text{post}} (r_o) = \varrho^q_{jk}  \frac{{\rm e}^{ - (\bar{r}_{jk} - r_o)^{\rm T} ( \sigma_{jk} + \sigma_o )^{-1}  (\bar{r}_{jk} - r_o) }}{ \pi^n \sqrt{ \text{det} [ \sigma_{jk} + \sigma_o ] }} \;, \nonumber
\end{equation}
while the CV state has collapsed into the Gaussian state that has been measured. The probability distribution of outcomes $r_o$ can be found by tracing out the unnormalized post-measurement QRDM 
\begin{equation}
\label{eq:prob_generadyne}
    P(r_o) = \sum_{i} \varrho^q_{ii}  \frac{{\rm e}^{ - (r_{ii} - r_o)^{\rm T} ( \sigma_{ii} + \sigma_o )^{-1}  (r_{ii} - r_o) }}{\pi^n  \sqrt{ \text{det} [ \sigma_{ii} + \sigma_o ] }} \;,
\end{equation}
which is the statistical mixture of the general-dyne measurement of $N$ Gaussian states, weighted according to the QRDM diagonal elements (see Fig.~\ref{fig:example_2}).

This treatment encompasses ideal and noisy homodyne and heterodyne detection schemes, which we will describe for a single mode only but are trivially generalisable to $n$ modes. Specifically, homodyne detections measure the observable $\hat{x}_{\phi} = \cos (\phi) \hat{x} + \sin(\phi) \hat{p}$, by mixing the system state with a probing coherent state of high intensity at the same frequency ($\ket{\alpha} = |\alpha| {\rm e}^{i \phi}$ with $\alpha \gg 1$). For finite detection efficiency $0 \leq \eta \leq 1$, homodyne detection is described by the general-dyne POVM with covariance matrix 
\begin{equation}
\label{eq:homodyne}
    \sigma_o = \lim_{z \to 0} R^{\rm T} (\phi) \begin{pmatrix}
        z^{2} & 0  \\
        0 & 1/z^2 
    \end{pmatrix} R (\phi) + (\tan \theta)^2 \mathds{1} \;,
\end{equation}
where we define $\theta = \cos^{-1}(\sqrt{\eta})$. The heterodyne detection scheme is similar to the homodyne one but the reference signal of the probe has a different frequency from the measured mode. In this case, $\sigma_o = (1 + 2( \tan(\theta))^2) \mathds{1}$. From Eq.~(\ref{eq:prob_generadyne}), it is possible to calculate the probability distribution of these two specific measurements, which are  apt to detect statistical mixtures of Gaussian states (see Sec.~\ref{sec:physical_examples}). The ideal case can be recovered by setting $\eta = 1$ (i.e., $\theta =0$).

Finally, the measurement of the joint system described by a POVM $\Pi_{r_m} \otimes M^i$ has $N^2$ probability distributions of $r_m$, labelled by the discrete outcome $i$, which can be expressed as
\begin{equation}
    P^{i}(r_o) = \sum_{j, k  = 0}^{N-1} M^{i}_{jk} \frac{\varrho^q_{jk} {\rm e}^{ - (\bar{r}_{jk} - r_o)^{\rm T} ( \sigma_{jk} + \sigma_o )^{-1}  (\bar{r}_{jk} - r_o) }}{\pi^n  \sqrt{ \text{det} [ \sigma_{jk} + \sigma_o ] }}\; . \nonumber
\end{equation}
Such probability distributions are the fringe pattern often related to Schr\"odinger cat states, which in the CV-DV systems arise under joint measurement (see Fig.~\ref{fig:example_1}).

\section{Perturbation Theory \label{sec:perturbation}}

While the operator-valued interactions are very common in practice, often the Hamiltonians and noises of hybrid systems have additional uncoupled non-diagonal DV terms, which one would like to incorporate into the presented formalism. In the following (and Appendix~\ref{app:perturbation}), we develop a perturbation theory that exploits the analytical solution of GBCS to account for Hamiltonian and Lindbladian terms involving DV operators that do not commute with $\hat{J}_z$ (of the form $\hat{V} = \mathds{1}_m \otimes \sum_{jk} V_{jk} \ket{j} \bra{k}$). The perturbative analysis is indeed required in these cases, as the algebra of operators does not close under commutation (for instance in the BCH expansion).

For the following, it may be useful to vectorize the DV Hilbert space (to dimension $N^2$), according to the known practices~\cite{fundamentals_gyamfi_2020}: for instance, the state of Eq.~(\ref{rhojk}) is $\dket{\varrho} = \sum_{j, k  = 0}^{N-1} \hat{\varrho}_{jk} \ket{j} \otimes \ket{k}$, with the specific case of a qubit $\dket{\varrho} = (\hat{\varrho}_{11}, \hat{\varrho}_{01}, \hat{\varrho}_{10}, \hat{\varrho}_{00})^{\rm T}$. 
The solutions of quantum dynamics presented in Sec.~\ref{sec:dynamics} rely on solving $N^2$ uncoupled homogeneous PDEs for the characteristic functions: $\left( \partial_\tau - \mathcal{L} \right) \dket{\chi} = 0$, where $\dket{\chi}$ is the vectorial form of branched characteristic functions and $\mathcal{L}$ is a diagonal superoperator (a matrix $N^2 \times N^2$). Each element of $\mathcal{L}$ is a complex differential operator in the phase space, quadratic in the variables $\xi$ and their partial derivatives. 

If some additional non-diagonal DV operators are present in the quantum dynamics, they result in additional inhomogeneous terms that couple the PDEs (as a time-dependent source term~\cite{Antonakakis_close_2013}): the phase-space evolution reads $\left( \partial_\tau - \mathcal{L} \right) \dket{\chi} = \mathds{V}\dket{\chi}$, 
where  $\mathds{V}$ is a Liouvillian DV superoperator independent from~$\xi$. As shown in Appendix~\ref{app:perturbation}, the solutions presented in Sec.~\ref{sec:dynamics} can be formulated in terms of the diagonal matrix of spatial Green's functions  $\mathcal{G}(\tau,  \tilde{\xi}, \tilde{\xi}')$, for instance, for a qubit, $\mathcal{G}= \text{diag} (\mathcal{G}_{00}, \mathcal{G}_{01},\mathcal{G}_{10},\mathcal{G}_{11})$, such that one finds the solution of the homogeneous equations presented in Sec.~\ref{sec:dynamics} as $\dket{\chi^{h} (\tau, \tilde{\xi})} = \int_{\mathds{R}^{2n}} \text{d} \tilde{\xi}' \mathcal{G}(\tau,  \tilde{\xi}, \tilde{\xi}') \dket{\chi_0(\tilde{\xi}')}$, with the vectorized initial conditions $\dket{\chi_0}$. Then, the solution of the coupled PDEs can be formulated as a recursive integral~\cite{Gripenberg1990}
\begin{equation}
\label{eq:volterra}
    \dket{\chi(\tau, \tilde{\xi})} = \dket{\chi^{h} (\tau, \tilde{\xi})}  + \int_0^\tau\hspace*{-2mm} \text{d} t \hspace*{-1mm}\int_{\mathds{R}^{2n}}\hspace*{-4mm} \text{d} \tilde{\xi}' \mathcal{G}(\tau - t ,  \tilde{\xi}, \tilde{\xi}') \mathds{V} \dket{\chi(t,\tilde{\xi}')},  \nonumber
\end{equation}
where the additional term is a particular solution of the inhomogeneous PDE. 
Define the kernel $\mathds{k}(\tau, \tilde{\xi}, \tilde{\xi}') \equiv \mathcal{G}(\tau,\tilde{\xi},\tilde{\xi}')\,\mathds{V}\ $ and the phase-space convolution as 
\begin{equation}
    (\mathds{k}*\mathds{k})(\tau, \tilde{\xi}, \tilde{\xi}') = \int_0^\tau \text{d} t \int_{\mathds{R}^{2n}} \text{d} \tilde{\xi}'' \mathds{k}(\tau- t, \tilde{\xi}, \tilde{\xi}'')\mathds{k}(t, \tilde{\xi}'', \tilde{\xi}') \nonumber
\end{equation}
and its $k$-th iteration as $\mathds{k}^{* k}.$
Under these definitions, the integrand reads $\dket{\chi(\tau,\tilde \xi)}=\dket{\chi^{h}(\tau,\tilde \xi)}+(\mathds{k}*\dket{\chi})(\tau,\tilde \xi)$, where $*$ also labels convolution between a superoperator and a vector. These types of integrals can be solved via an iterative series~\cite{Gripenberg1990}
\begin{equation}
    \label{eq:series}
    \dket{\chi(\tau,\tilde{\xi})} = \dket{\chi^{h} (\tau,\tilde{\xi})} + \sum_{k=1}^{\infty} \big[\mathds{k}^{*k}*\dket{\chi^{h}}\big](\tau,\tilde{\xi})\ ,
\end{equation}
which is the phase-space representation of the Dyson series for super-operators and the nested convolutions respect time-ordering. In the power of $\mathds{V}$, the zeroth order is the homogeneous solution, the first order is explicitly given by
\begin{equation}
    \label{eq:order1}
    \dket{\chi^{(1)}(\tau,\tilde{\xi})} = \int_0^\tau {\rm d}t \int_{\mathds{R}^{2n}} {\rm d}\tilde{\xi}'\ \mathcal{G}(\tau-t,\tilde{\xi},\tilde{\xi}')\,\mathds{V}\,\dket{\chi^{h}(t,\tilde{\xi}')} \;.  \nonumber
\end{equation}

\subsection{DV Decay Noise: $T_1$}

Decay in the DV subsystem has to be considered to give a realistic analysis of quantum systems in which the decay time ($T_1$) is similar to the dephasing one  ($T_2$), for instance, in some trap ion and superconducting qubit architectures with large coherence time~\cite{Wang2021,bland20252dtransmonslifetimescoherence}. The additional Lindbladian term which, in the most general case of an $N$-level system, describes decay is  
\begin{equation}
\label{eq:Nlevel_lindblad}
    D_{T_1} [\varrho] = \sum_{j=0}^{N-1} \hat{L}_j \varrho\, \hat{L}_j^\dagger - \tfrac{1}{2} \{ \hat{L}_j^\dagger \hat{L}_j, \varrho \}, 
\end{equation}
with jump operators $\hat{L}_{j} = \sqrt{\Gamma_{j}} \ket{j-1}\bra{j}$, for $j \in [0,N-1]$. The second term of  Eq.~(\ref{eq:Nlevel_lindblad}) can be reabsorbed in $\gamma_{jk}$ (increasing the de-phasing rates of the off-diagonal terms and including decay of the diagonal one). The first term is the source term, which couples the evolution of $\hat{\varrho}_{j-1,j-1}$ to $\hat{\varrho}_{j,j}$ via the $j \to j-1$ transition. Observe that, in the case of decay, the truncation of the series occurs exactly as soon as all the excitations in the DV sector have decayed to the lowest excitation state. This is manifested via the fact that $\mathds{V}$ is lower diagonal and $\mathds{k}^{*N} = 0$. 

To illustrate how the perturbation theory can be applied to this context, let us consider the single-qubit state of Eq.~\ref{eq:density_one_qubit} under qubit decay, i.e.  $\hat{L}_{1} = \sqrt{\Gamma_{1}} \hat{\sigma}_-$.  In the case of the qubit,  the evolving state is described by the vector of characteristic functions
\begin{equation}
    \dket{\chi^{T_1} (\tau, \tilde{\xi})} = \begin{pmatrix}
        {\rm e}^{-\Gamma_{1}\tau  } \; \chi^{\text{on}}_+ (\tau, \tilde{\xi})    \\
{\rm e}^{-\Gamma_{1}\tau  /2 }\; \chi^{\text{off}}  (\tau, \tilde{\xi})  \\
{\rm e}^{-\Gamma_{1}\tau /2 } \;\chi^{* \text{off}} (\tau, \tilde{\xi})    \\
\chi^{\text{on}}_- (\tau, \tilde{\xi})   + \chi^{(1)}  (\tau, \tilde{\xi})
    \end{pmatrix}  ,
\end{equation} 
where the perturbation arises only in the DV ground state  (the only state where the excited state can decay) and it has the additional characteristic function
\begin{equation}
\label{eq:decay_qubit}
    \chi^{(1) } (\tau, \tilde{\xi})  =  p_+ \Gamma_1 \int_0^{\tau}  {\rm d} t {\rm e}^{-\frac{1}{4} \tilde{\xi}^{\rm T} \sigma^{\Gamma_1} ( \tau, t) \tilde{\xi}  +  i  \tilde{\xi}^{\rm T} \bar{r}^{\Gamma_1} ( \tau, t) - \Gamma_1 t } ,
\end{equation}
with the ``two times'' covariance matrix and first moments given by
\begin{align}
\label{eq:two_times_parameters}
    \sigma^{\Gamma_1} ( \tau, t) &= S_-(\tau - t) S_+(t) \sigma_0 S^{\rm T}_+(t) S^{\rm T}_-(\tau - t) \\
    \bar{r}^{\Gamma_1} ( \tau, t) &= S_-(\tau - t) \left[ S_+(t) r_0 - \left(  S_+(t) - \mathds{1} \right) r_+ \right] \nonumber \\
    &\hspace{2cm} - \left(  S_-(\tau - t) - \mathds{1} \right) r_-   \nonumber
\end{align}
where, to illustrate the point, we considered the unitary case in the CV ($S_{\pm}$ are given in Eq.~\ref{eq:symplectic_pm}, and $r_{\pm} = r_m \pm r_q$), while the open dynamics is discussed in Appendix~\ref{app:perturbation}. The change in the qubit populations can be computed from the diagonal elements of the QRDM, which undergoes decays at rate $\Gamma_1$ for the excited states ($\varrho^q_{11} = p_+ {\rm e}^{-\Gamma_1 \tau}$), while the ground state increases as expected
\begin{align}
    \varrho^q_{00} = \chi^{h}_- (0,\tau) + \chi^{(1)}(0,\tau) &  = p_- + p_+ \Gamma_1 \int_0^{\tau} dt {\rm e}^{-\Gamma_1 t} \nonumber\\
    &= p_- + p_+ (1 - {\rm e}^{-\Gamma_1 \tau}) \;, \nonumber
\end{align}

 The presented solution finds a clear interpretation as a quantum jump: in the interval of time $[t, t + dt]$, there is a probability $P(t) - P(t+dt) = p_+  {\rm e}^{-\Gamma_1 (t+dt)} - p_+  {\rm e}^{-\Gamma_1 t} \approx p_+  \Gamma_1 {\rm e}^{-\Gamma_1 t} dt $ that a jump occurs. When the jump does not occur, the branched characteristic functions $\dket{ \chi}$ evolve according to their given dynamics, and are described by their homogeneous solution (with the inclusion of the diagonal part of $T_1$ noise which gives the decays). When the jump occurs, $\chi^{\text{on}}_+ (\tilde{\xi},t)$ is the new initial condition for the dynamics of the $-$ branch (given by Eq.~\ref{eq:two_times_parameters}), and all other branches are annihilated. After integrating over all possible jump times, one finds the total state as, indeed, the one of Eq.~(\ref{eq:decay_qubit}). An example of the Wigner function arising from $\chi^{(1) } (\tau, \tilde{\xi})$ is given in Sec.~\ref{sec:physical_examples}. In a similar way, one can describe the decay of an $N$-level system as $N$ possible jumps, as this noise only induces change in populations on the diagonal term and additional decay in the off-diagonal one.


\section{Physical Examples \\ 
\& Comparative Analysis \label{sec:physical_examples}}

In this section we present two examples that showcase our method's capability of producing predictions for realistic experiments. We shall cover (a) a qubit (or atom) dispersively interacting with a resonator undergoing dynamical squeezing induced by a parametric amplifier; (b)~a levitated nanoparticle undergoing Stern-Gerlach non-Gaussian matter-wave interferometry. 
The former example represents an instance of quadratic operator-valued CV-DV interaction, reproducing, in the steady state, experimentally detected quantities~\cite{eddins_high_2019}. The latter showcases the results of linear CV-DV interactions in open dynamics, providing analytical predictions for sensing with levitated systems in the open non-Gaussian regime. Both these examples will allow us to illustrate the physical roles of the phase-space parameters and how they relate to dynamical properties of realistic quantum systems (via the element of Table~\ref{tab:open}). This allows us to discuss transitional behaviours of these systems, generating Wigner negativities which, to the best of our knowledge, have not yet been detected in experimental settings.


\begin{figure*}[!t]
    \centering
    \includegraphics[width=0.9\linewidth]{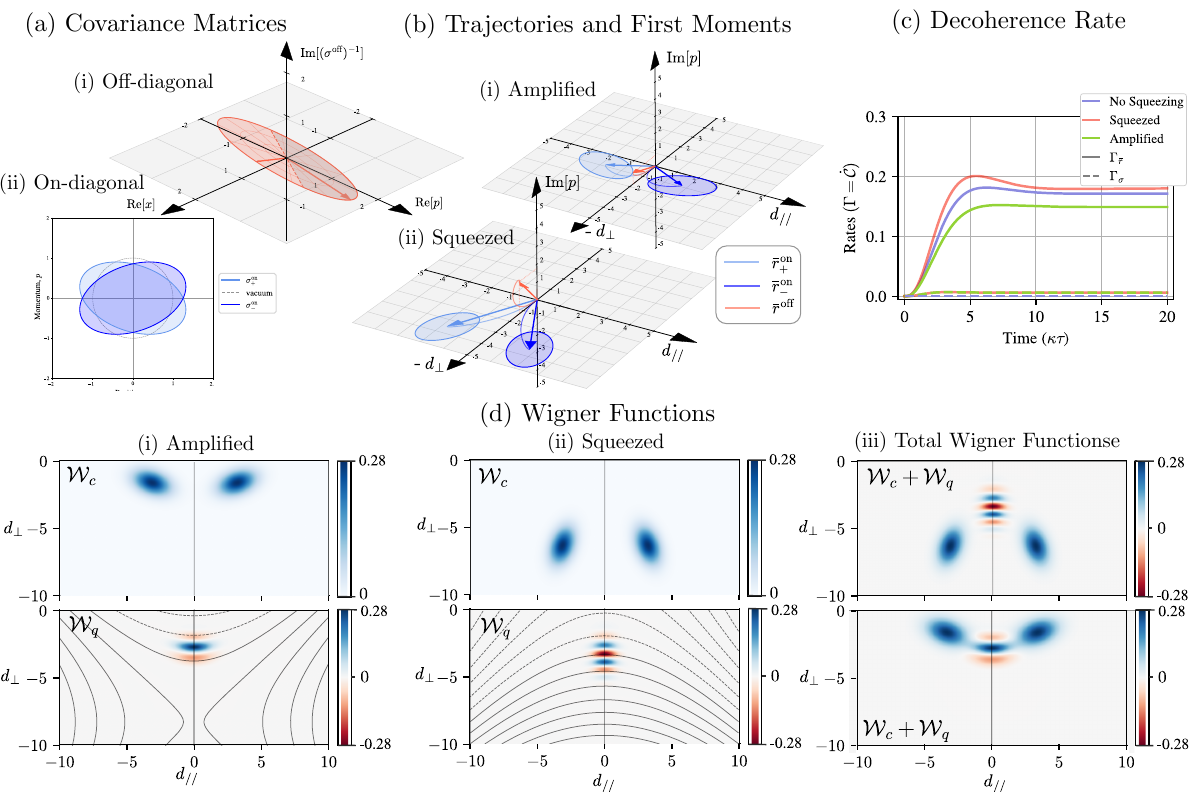}
    \caption{A qubit dispersively coupled to a resonator with a paramteric amplifier ($\kappa = 1$, $\mu = 0.4$, $ \lambda = 0.3$, $\alpha_{\text{in}} = 1$): \\
    (a) the steady state covariance matrices for the (i) off-diagonal (under polar decomposition, where the tilting angles are the complex phases) and (ii) on-diagonal branches; (b) trajectories in reference to the driving input field for (i) the amplified and (ii) squeezed regime;  (c) time evolution of the decoherence rate  $\Gamma_{\sigma}$ and $\Gamma_{\bar{r}}$ for the amplified and squeezed regime alongside the no-squeezing  regime ($\lambda =0$); (d) Wigner negativities of $\sigma_x$ post-measurement states.} 
    \label{fig:example_1}
\end{figure*}

\newpage

\subsection{Transient Behaviours of a Qubit Dispersively Coupled to a Driven Parametric Amplifier \label{subsec:atom}}

The coherent absorption and emission of a cavity photon by and to a qubit is central to quantum computation. Examples of architectures featuring such a behaviour are found in superconducting Josephson junctions~\cite{strong_wallraff_2004,coherent_chiorescu_2004,coupling_majer_2007,blais_quantum_2007}, as well as dopant spins with magnetic~\cite{strong_kubo_2010,strong_chen_2018,realization_liu_2023} and electric~\cite{strong_samkharadze_2018,strong_samkharadze_2018,coherent_mi_2018,coherent_landig_2018} coupling.  Resonators, on the other hand, range from optical cavities to LC superconducting circuits and nanomechanical resonators~\cite{xiang_hybrid_2013}.
In these schemes, the dispersive coupling between the qubit and the cavity creates a distinguishable superposition of the CV state, that can be used for implmenting conditional rotations.  Squeezing the resonator is known to provide enhancement  in many protocols~\cite{Entanglement_Wahyu_2008,Sete2013,Szorkovszky_2014,didier_fast_2015,walter_rapid_2017,kam_fast_2024} and it has been experimentally demonstrated by embedding a qubit in a driven parametric amplifier that dynamically squeezes the state of the resonator~\cite{eddins_high_2019}.


Let us consider a realistic qubit coupled to a resonator. The ladder operators of the cavity of frequency $\omega$ are $a$ and $a^\dag$, $\sigma_j$ with $j\in\{x,y,z,\pm\}$ are the Pauli operators of the qubit, and $\omega_q$ is the frequency of the qubit. Under the rotating-wave approximation, the interaction between qubit and resonator is described by the Jaynes-Cummings model with $g$ being the vacuum Rabi splitting. 
In the off-resonant regime (also known as the dispersive regime), i.e.,~for $\omega \gg \Delta \gg g$, where $\Delta = |\omega - \omega_q|$ is the detuning frequency, the effective Hamiltonian of the system reads
\begin{align}
\label{eq:example_1_hamitonian}
    \frac{\hat{H}_1}{\hbar \omega}  =  &  \left(  \Delta + \frac{\mu}{2} \left( \hat{a}^\dag \hat{a} +\frac{1}{2} \right) \right) \otimes \hat{\sigma}_z  + \frac{i \lambda}{2} \left( \hat{a}^{\dag 2} - \hat{a}^2 \right),
\end{align} 
where we considered the rotating frame at frequency $\omega$. The first term represents a frequency shift conditional on the qubit state of strength given by the dispersive coupling $\mu = 4 g^2 / \Delta$ and the second is the squeezing of strength $\lambda$ (which is related to the gain of the amplifier). 
The inclusion of realistic noise is done through the master equation 
\begin{align}
\label{eq:example_1_master}
    \hspace{-0.1cm}\frac{\partial \hat{\varrho}}{\partial \tau} =&  \frac{i}{\hbar \omega} [ \hat{\varrho}, \hat{H}_1] +  i \sqrt{\kappa}  \left[ \hat{\varrho}, \alpha_{in} \hat{a}^{\dag} + \alpha^*_{in} \hat{a} \right]   \\&  \nonumber + \kappa \mathcal{D}[\hat{a}] (\hat{\varrho})  + \frac{1}{T_2}\mathcal{D}[\hat{\sigma}_z] (\hat{\varrho}) ,
\end{align}
which describes qubit dephasing, cavity losses and a driving term~\cite{eddins_high_2019}. For this example, we shall take the approximation that qubit decay can be omitted. It is possible to recast this dynamics to our formalism via the Hamiltonians $H_q^{(0)} = 2 \mu j $,  $r_j = (0,0)^{\rm T} $, and $H_j= \lambda \sigma_x + j \mu \mathds{1}$ for $j \in \{-1,1\}$; for the noise terms, via $D = \kappa \mathds{1}$, $E = \kappa/2 \Omega_2$ and $\gamma_z = \sigma_x/ T_2 $. The driving term is, in canonical coordinates, $d = |\alpha_{in}| \sqrt{2 \kappa}(\cos \phi_d, \sin \phi_d )$, where $\alpha_{in}=|\alpha_{in}|{\rm e}^{i\phi_d}$. The initial state is a tensor product of the cavity ground state and $\ket{+}  = (\ket{1}+\ket{-1})/\sqrt{2}$.




For the off-diagonal element, the dynamics can be solved via the doubling method of the Riccati equation as outlined in Sec.~\ref{sec:dynamics} in terms of the phase-space parameters which are given in Appendix~\ref{app:examples}. The frequency shift induced by the qubit rotates the initial coherent state of the cavity at two frequencies conditionally on the qubit state. The squeezing and decaying of the cavity create a GBCS with a stable steady-state, both superposed in the first moments as well as the squeezing parameter. This leads, on the diagonal elements, to distinct $\sigma^{\text{on}}_{\pm}$ and $\bar{r}^{\text{on}}_{\pm}$. By varying the driving quadrature, one can tune between squeezed ($\phi_d = \pi/2$) and amplified ($\phi_d = 0$) regimes, creating a superposition of first moments along a phase space direction either orthogonal or parallel to the squeezing direction~\cite{eddins_high_2019}. In Fig.~\ref{fig:example_1},  we present the GBCS in the steady state for these two regimes. 

The dynamical creation of a superposition in the CV induces dephasing in the qubit state via two processes with rates immediately apparent from inspection of the QRDM ODE of Eq.~(\ref{eq:general_unitary_ODE_off}): on the one hand, the difference in squeezing between the branches induces an effective dephasing at the time-dependent rate 
\begin{equation}
\label{eq:rate_sigma}
\Gamma_{\sigma}(\tau) =  \,\text{Im}\!\left[ \frac{1}{2}\text{Tr}\!\left[H_q \sigma^{\text{off}} \right] \right] = \frac{\mu}{2} \text{Tr} \left[ \sigma_\text{I} (\tau) \right]  \; ; \end{equation} 
on the other hand, the superposition in first moments gives a dephasing rate 
\begin{equation}
\label{eq:rate_r}
\Gamma_{\bar{r}} (\tau)= \text{Im}\left[ (\bar{r}^{\text{off}})^{\rm T}  H_q \bar{r}^{\text{off}} \right] = 2 \mu \bar{r}_{\text{R}}^{\rm T}(\tau) \bar{r}_{\text{I}}(\tau).
\end{equation}
Their time evolution is given in Appendix~\ref{app:examples} and plotted in Fig.~\ref{fig:example_1}. The former is unavoidable as it stems from the non-commuting nature of squeezing and conditional rotations, while the latter depends on the driving, and can be minimized by choosing $\phi_d = 0$. The total decoherence rate of the off-diagonal element of the QRDM is then given by $\Gamma_{\text{tot}} (\tau) = \Gamma_{\sigma}(\tau) + \Gamma_{\bar{r}} (\tau) +\frac{1}{T_2}$, from which one can compute the total decay $\mathcal{C}_{\text{tot}} (\tau) = \int_{0}^\tau \text{d}t \Gamma_{\text{tot}} (t)$. As shown in Fig.~\ref{fig:example_1}, it transits to a maximal decoherence rate before reaching a steady state solution. As $\tau \to \infty$, we recover the steady state decoherence rates, measured in experimental setting, of~\cite{eddins_high_2019}. Thus, our formalism naturally captures these transient processes and provides physical intuition of dynamical decoherence in terms of the time-dependent generalised complex moments: the imaginary part of the off-diagonal covariance matrix captures the non-overlapping features of the superposition at the level of the uncertainties, while the coupling between the real and imaginary components of the off-diagonal first moment at the level of the expectation values. 

Furthermore, one can inquire if this experimental set up can be used to detect the  hyperbolic Wigner negativities, which until now have been only theoretically predicted~\cite{nicacio_phase_2010}. The presence of this behaviour can be concluded from Fig.\ref{fig:example_1} (a) (i), as the principal axes of the complex component of the inverse off-diagonal covariance matrix at the steady state have different signs (as in the case of conditional rotations of squeezed states of Sec.~\ref{sec:operations}). As explained in the next example, non-diagonal measurements of the qubits can be used to witness the quantum fringes $\mathcal{W}_q$ of Fig.~\ref{fig:example_1}. This may allow for the first detection of hyperbolic-type of Wigner negativities.



\begin{figure*}[!t]
    \centering
    \includegraphics[width=0.9\linewidth]{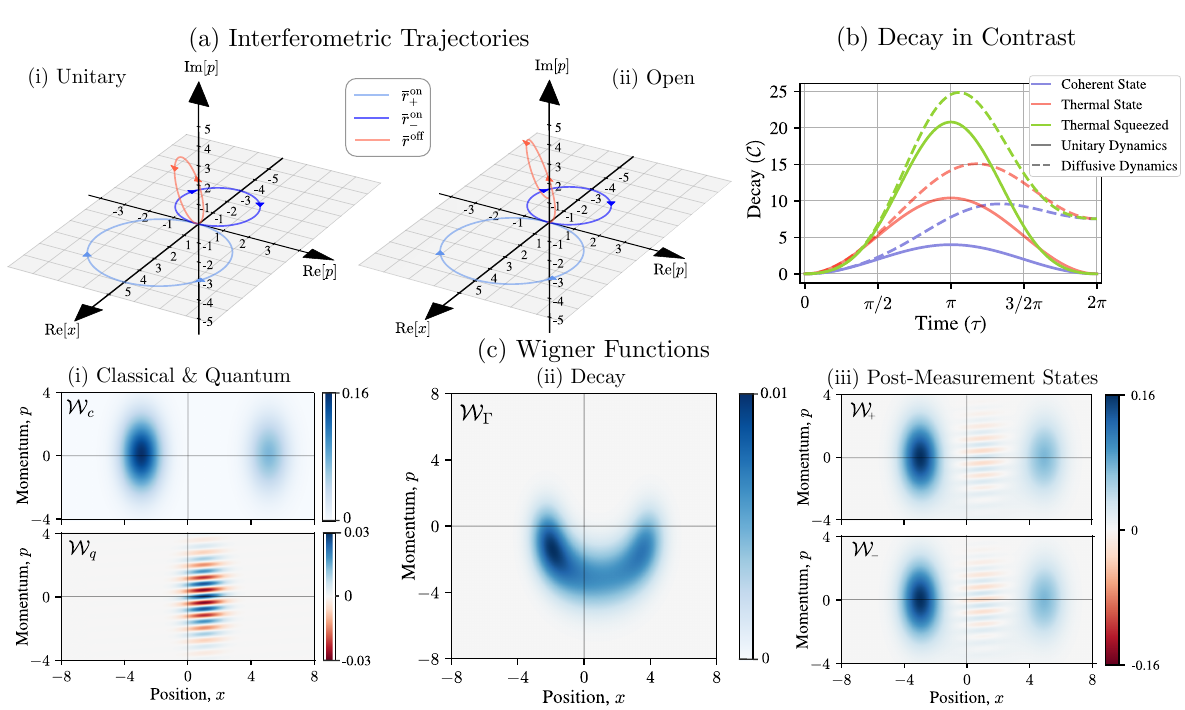}
    \caption{A Stern-Gerlach matter-wave interferometer undergoing diffusive dynamics with initial squeezed-thermal state in the presence of an unknown force ($f_q = 2, \; f_u = 1/2$, $N_p = 0.8$, $s = 1/2$, $\Gamma_x = 0.05$, $\Gamma_1 = 0.15$, $\Gamma_2 = 0$): \\
    (a) phase-space trajectories for an initial coherent state (generalising the plot of Ref.~\cite{scala_matter_2013}) exemplifying the difference between (i) unitary and (ii) open dynamics for $r^{\text{off}}$; (b) dynamical decay in contrast ($\mathcal{C}$) of a $\sigma_x$ measurement with initial coherent state, thermal state, and thermal squeezed state; (c) Wigner functions at the maximum superposition ($\tau = \pi$) of (i) the classical and quantum components, (ii) the decayed contribution, and \\
    (iii) the two post-measurement states of the $\sigma_x$  } 
    \label{fig:example_2}
\end{figure*}

\subsection{Stern-Gerlach Matter-Wave Interferometry: \\Squeezing, Diffusion, and Wigner Negativity \label{subsec:mass}}

The creation of non-classical states of levitated nanoparticles (NPs) is an active area of research, aiming to detect forces and entanglement for both applied and fundamental objectives~\cite{Romero_Isart_quantum_2011,bose_spin_2017,marletto_gravitationally-induced_2017}. In non-Gaussian matter-wave interferometry, large superpositions in the CV may be experimentally achieved via the linear coupling -- known as Stern-Gerlach interaction or spin-dependent force -- between a qubit and the NP, forming and recombining a GBCS. This interaction may be used to generate non-classical states in levitated masses with application to sensing forces and entanglement~\cite{bose_qubit_2006,Romero_Isart_quantum_2011,scala_matter_2013,geraci_sensing_2015,qvarfort_gravimetry_2018,marshman2020mesoscopic,eva_requirements_2023,barker2022entanglement,toros_relative_2021,bose2016matter,bose_spin_2017, marletto_gravitationally-induced_2017}.  The systems that are candidate for this type of couplings are dopant spins (such as Nitrogen-Vacancy centres) in diamagnetic NPs~\cite{scala_matter_2013,yin_large_2013,wan_free_2016,marshman_constructing_2022}, superconducting flux qubits with magnetic NPs~\cite{romero_isart_quantum_2012,nair_massive_2023} and atoms coupled to NPs via optical fields~\cite{toros_creating_2021}. While great advances have been made in the noise analysis of these systems~\cite{scala_matter_2013, pedernales_motional_2020,toros_loss_2020,toros_relative_2021,rijavec_decoherence_2021,tilly_qudits_2021,ma_torque_2021,schut_improving_2022,henkel_internal_2022,henkel_universal_2023,fragolino_decoherence_2023}, a full dynamical description of the experiments, with the inclusion of state preparation and diffusion noise in the CV degrees of freedom (one of the main sources of noise) is, to the best of our knowledge, still lacking in the literature. 

Here, the CV observables $\hat{X}$ and $\hat{P}$ describe the position and momentum of the centre of mass of the levitated NP, respectively. This degree of freedom is cooled to a thermal state (with $\sigma_T = (1+2N_p) = \coth({\hbar} \omega_t/ (2 k_B T_m))$ -- where $N_p$ is the number of phonons, $k_B$ is the Boltzmann constant and $T_m$ the temperature~\cite{serafini_quantum_2017,schlosshauer_quantum_2007}) of a \textit{tight} harmonic trap -- for instance, an optical tweezer of frequency $\omega_t$~\cite{delic_cooling_2020,magri_real_2021}. At $\tau = 0$, the cooling trap is rapidly switched off and the mass undergoes Stern-Gerlach interferometry in a dark trap, with frequency $\omega < \omega_t$, whose ground-state spread is $x_0= \sqrt{\hbar/(2 M \omega)}$.  A qubit (of natural frequency $ \omega_q$) applies an operator-valued force of strength $F_q$ to the mass. The total Hamiltonian is 
\begin{equation}
\label{eq:hamiltonian_1}
    \hat{H}_{2} = \frac{\hbar \omega_q}{2} \hat{\sigma}_z + \frac{\hat{P}^2}{2M} + \frac{1}{2} M \omega^2 \hat{X}^2    -  F_q \hat{X}  \otimes \hat{\sigma}_z - F_u \hat{X},
\end{equation}
where the first term is the free evolution of the qubit, the second and third terms are the kinetic energy and harmonic trapping of the mass, and the last two are the operator-valued and unknown forces ($F_u$), respectively. 

  The centre of mass is initially centred with respect to the dark trap and, given the change of frequency and imperfect cooling, the initial state is a thermal squeezed state of $N_p$ phonons and squeezing parameter $s = \omega/\omega_t$, i.e., its initial first and second moments are $r_0 = (0,\; 0)^{\rm T}$ and $\sigma_0 = (1+2N_p) \;\text{diag}(s,1/s)$. The qubit is initialised in the state $\varrho_q (0) = \ket{+}\bra{+}$.
By moving to dimensionless canonical observables, we can define the unitless forces $f_i = \sqrt{2} x_0 F_i/(\hbar \omega)$ with $i \in u, q$. The Hamiltonian of Eq.~(\ref{eq:hamiltonian_1}) can be written as  Eq.~(\ref{eq:Hamiltonian_force_only})  by setting $H_m = \mathds{1}$, $H_q = 0$, and $r_j = r_m + j r_q$, with $r_m=(f_u,\;0)^{\rm T}$ and $r_q = (f_q, \; 0)^{\rm T}$. During the evolution,  the qubit decays at rate $\Gamma_1$ and dephases at a rate $\Gamma_2$, while the CV undergoes diffusive dynamics in an Ohmic bath of thermal photons at high temperature $T_{p} \gg \hbar \omega/k_B$, such that the rate is at rate $\Gamma_x = 2 \gamma_0 k_B T_p/(\hbar \omega^2)$, where $\gamma_0$ is customarily measured in experimental settings~\cite{schlosshauer_quantum_2007}. Diffusive noise is equivalent to averaging over stochastic random forces experienced by the NP~\cite{weiss_large_2021}. The dynamics is then described by the Lindbladian 
\begin{align}
    \frac{\partial \hat{\varrho}}{\partial \tau} =  \frac{i}{\hbar \omega} [ \hat{\varrho}, \hat{H}_{2}] &+ \Gamma_x  \mathcal{D}[\hat{x}] (\hat{\varrho})  + \Gamma_1 \mathcal{D}[\hat{\sigma}_-] (\hat{\varrho}) + \Gamma_2 \mathcal{D}[\hat{\sigma}_z] (\hat{\varrho}) 
\end{align}
At first, let us solve the homogeneous case (Table~\ref{tab:open}) and consider the qubit decay as a perturbation theory (Sec.~\ref{sec:perturbation}). In this case, the dynamics is a specific case of Eq.~(\ref{eq:linbladian}), with 
\begin{equation}
    B = \begin{pmatrix}
         \Gamma_x &0\\
        0&0
    \end{pmatrix} \implies  D =  \begin{pmatrix}
        0 & 0\\
        0 &  2 \Gamma_x
    \end{pmatrix} .
\end{equation}
and 
\begin{equation}
    \begin{pmatrix}
        \frac{1}{T}_1 & \frac{1}{2 T}_1+\frac{1}{T}_2 \\
        \frac{1}{2 T}_1+\frac{1}{T}_2 & 0
    \end{pmatrix}
\end{equation}
where we included the diagonal component of DV decay. Given that the Hamiltonian has only linear interaction between the centre of mass and the qubit ($H_q = 0$) and that $E =0$ (i.e., the noise has only the symmetric diffusive term $D$), the dynamics can be analytically solved via the single symplectic transformation $S_m(\tau) = {\rm e}^{\tau \Omega H_m}$ (as detailed in Sec.~\ref{sec:open}), which represents a phase space rotation of angle~$\tau$ (since $ H_m = \mathds{1}$). The time evolutions of the phase-space parameters are given in Appendix \ref{app:examples}. The single covariance matrix evolves as a squeezed thermal state according to a Gaussian diffusive process, while the superposed first moments undergo the trajectories of Fig.~\ref{fig:example_2}. Furthermore, as discussed in Appendix \ref{app:examples}, the $T_1$ decay noise can also be treated exactly since the corresponding perturbation theory naturally truncates at first order.


Let us now investigate a $\sigma_x$-measurement, which can be used to detect the unknown force in terms of a qubit phase estimation protocol or generate Wigner negativities in the CV. The measurement outcomes $\pm1$ have associated projectors $\Pi_\pm = \ket{\pm}\bra{\pm}$, with $ \ket{\pm} = (\ket{1}\pm\ket{-1})/\sqrt{2}$ being the eigenstates of $\sigma_x$. The time evolution of the outcomes' probabilities is given by
\begin{align}
\label{eq:probability_SG}
    P_x(\pm, \tau) = \text{Tr}_q\left[\Pi_\pm \hat{\varrho}^q(\tau)\right] = \frac{1}{2} \left[ 1 \pm {e\rm}^{-\mathcal{C} (\tau)} \cos  \phi (\tau)  \right], \nonumber
\end{align}
where $\phi(\tau) = 2 f_q f_u (\tau -\sin \tau ) + \frac{1}{2} \omega_q \tau $ and
\begin{equation}
    \mathcal{C} (\tau) = \int_{0}^{\tau} \text{d}t \left( - 2 \text{Im} [r_q^{\rm T} r^{\text{off}} (t) ] + \frac{1}{2 T_1}  + \frac{1}{T_2}  \right) \;,
\end{equation}
whose analytical form (which can be computed in general as in Table~\ref{tab:open}) is given in Appendix \ref{app:examples} and is plotted in Fig.~\ref{fig:sg_trajectory_decay} for different initial states and dynamics. The time evolution of $\mathcal{C}$ shows the apparent decoherence of the qubit due to the entanglement with the CV degree of freedom, its value increasing at early times as the centre of mass is superposed at two distinct locations. However, the interference contrast shows a revival of coherence, when $\mathcal{C}$ decreases in value, reaching a local minimum at $\tau = 2 \pi$. At this time, as in the case of unitary dynamics of initial thermal states~\cite{scala_matter_2013}, the wave-matter interferometer performs one loop, also for the case of open dynamics with a squeezed thermal state. This can be concluded as $\bar{r}^{\text{on}}_{\pm} (2 \pi) = (0,0)^{\rm T}$, thus representing a full recombination of the interferometer with the mass is recombined and centreed in the trap again (without apparent ``Humpty-Dumpty effect''~\cite{englert_is_1988, japha_quantum_2023, xiang2024phononinducedcontrastmatter, zhou2024gyroscopicstabilitynanoparticlessterngerlach}).  However, due to the inclusion of the diffusion term, the first moments of the off-diagonal terms are $\bar{r}^{\text{off}} (2 \pi) = (0,2 i \pi f_q \Gamma_x)^{\rm T}$. Being not zero and complex, this leads to an exponential decay of the visibility of the qubit measurement, caused by the diffusion process, with $\mathcal{C}(2 \pi) = 6 \pi  f_q^2 \Gamma_x + \pi \Gamma_1 + 2 \pi \Gamma_2 $.  Interestingly, $\mathcal{C}(2 \pi)$ is found to be independent of the squeezing parameter $s$ and the initial number of phonons $N_p$, but linear in the diffusion and dephasing rates $\Gamma_i$ (with $i\in x,z$) and quadratic in the superposition size $f_q$.  It is remarkable how our formalism allows one to pinpoint the origin of such effects affecting the dynamics: similarly to the previous example, for linear couplings, the imaginary part of the off-diagonal first moments gives the time-dependent decoherence rate for the qubit $\Gamma (\tau) = -2 r_q^{\rm T} r_{\text{I}} (\tau)$.

The maximum superposition is achieved at $\tau = \pi$, given by $\delta X = \bar{r}^{\text{on}}_{+}(\pi) -  \bar{r}^{\text{on}}_{-}(\pi) = 4 f_q$ in ground spread units. At this moment, if a $\sigma_x$ measurement occurs, the loss of visibility is $ \mathcal{C} (\pi) = f_q^2 \left(4 (1 +2 N_p)/s + 3 \pi \Gamma_x \right) + \pi  \Gamma_1/2 +  \pi \Gamma_2 $, with $P_x (\pm, \pi) \approx 1/2$ when the ground state spread is resolvable.  Summing over elements of Eq.~(\ref{eq:wigner_jk}), the post-measurement states of the mode have Wigner functions $\mathcal{W}_{\pm} = \mathcal{W}_c +  \mathcal{W}_\Gamma \pm \mathcal{W}_q $, allowing one to detect, via appropriate measurements, the non-classical state of the CV. $\mathcal{W}_c$ and $\mathcal{W}_q$ are the classical and quantum components discussed in Sec.~\ref{sec:wigner} and analytically computed in Appendix~\ref{app:examples} for this dynamics. The interesting quantity that should be noted is the Wigner phase
\begin{equation}
    \phi_W = 4 f_q \left(2 \tilde{p} - \frac{4 \Gamma_x  \left(\tilde{x}-2 f_u\right)}{\pi  \Gamma_x+  (1 +2 N_p)s}- \frac{\pi  \Gamma_x  \tilde{p} }{\pi  \Gamma_x + (1 + 2 N_p)/s }\right) \nonumber,
\end{equation}
illustrating how the fringes' spacing decreases with the superposition size (proportional to $f_q$). One then can envision the application of this analytical solution to identify a coupling $f_q$, which is a function of experimentally accessible quantities, that creates superpositions large enough to resolve the two-Gaussian statistical mixture $\mathcal{W}_c$ from the fringes, but also small enough to keep the fringes resolvable under the considered measurement apparatus (for instance, homodyne detection of scattered light). Finally, $\mathcal{W}_\Gamma$ is the component of the Wigner function due to the DV decay, and it is the Fourier transform of the inhomogeneous characteristic function $\chi^{(1)} (\tau = \pi, \tilde{\xi})$  (given in Appendix~\ref{app:examples} as a series of Bessel functions). In Fig.~\ref{fig:example_2}, we plot the different contributions of the Wigner functions and the total Wigner functions for the post-measurement states.

\subsection{Comparative Analysis with Known Methods}

Having showcased the ubiquitous appearance of GBCS among quantum protocols and experiments, we shall discuss GBCS's parametrisation in the context of existing techniques for the modelling of CV-DV hybrid systems, which are either performed numerically or analytically (see Ref.~\cite{hybrid_Liu_2026} for a complete discussion). 

Numerical solutions rely on truncating the CV Hilbert space in the $\ket{n}$-basis or on other appropriate finite-dimensional bases, making each mode, effectively, a DV variable subsystem of dimensionality $N_t$ (where $N_t$ is the truncation number)~\cite{qutip1,Killoran2019strawberryfields}. The numerical nature of these methods equips them with great flexibility, as they can solve, at least in principle, any hybrid dynamics with any initial state. However, when either the states involved approach high quantum numbers (as is always the case with significant squeezing) or the Hamiltonian approaches instabilities or large differences between couplings, the required $N_t$ to accurately simulate the system rapidly increases. To approximately solve, in the open setting, the dynamics of a composite system of $n$ CV modes and an $N$-level DV subsystem, the number of parameters scales as $\mathcal{O}( N_t^{2 n} \times N^2)$ for full density-matrix treatments, or $\mathcal{O}( N_t^{n} \times N^2)$ for trajectory methods (which, however, require sampling). By restricting to the class of operator-valued Gaussian Hamiltonians, which often are the non-linearities common in experimental practice, the formalism presented in this work shows that it is possible to parametrise the state with $\mathcal{O}( n^2 \times N^2)$ variables. Hence, the parametrisation of GBCS shows an exponential improvement in parameter efficiency in the number of modes. 

Furthermore, Hilbert-space truncations are generally prone to issues and complications~\cite{stornati2024variationalquantumsimulationusing,ashhab26}. These still stand even if displaced Fock bases are adopted (following the classical trajectories, as in~\cite{grosso_controlled_2025}), since the latter can approximately take care of high excitation numbers due to displacements, but not to squeezing and entanglement. Our methods avoid these issues as our solutions are either exact or semi-analytical. By the latter, we refer to cases where the ODEs are solved via numerical solutions -- which, even though the original ODEs are non-linear, can always be mapped to a linear system. This allows for a much more accurate quantification of errors (stability of numerical solution of systems of ODEs is extensively studied in the literature), than the outcome of Hilbert space truncated dynamics. Hence, our method, exclusively tailored on operator-valued quadratic dynamics, enjoys the combined advantages of an efficient parametrisation and of controlled and reliable results.

The types of interactions and states presented in this work have been extensively studied in analytical treatments present in the literature, but restricting to specific examples. General statements have been achieved for linear-coupling whose resulting unitary dynamics can be decomposed into a Gaussian unitary and a conditional displacement (equivalent to Table~\ref{tab:force})~\cite{molmer1999,sorensen2000}. This is due to the fact that in the unitary case of linear couplings, expansions like the Magnus one~\cite{blanes2009}, or Lie algebraic approaches, naturally truncate~\cite{qvarfort_time-evolution_2020}. Yet, these analyses fail to account for Gaussian noise in the general setting (Table~\ref{tab:open}).  In the case of quadratic couplings, to the best of our knowledge, no general treatment prior to the present one was ever developed, and the instances known in literature are limited to specific Hamiltonians and initial states\footnote{Let us also mention that, after the appearance of the first version of this study, analogous equations were derived in the context of generalised master equations (without, however, drawing connections with hybrid systems)~\cite{albarelli2026efficientevaluationfundamentalsensitivity} and quantum field theory (treating composition rules of arbitrary Gaussian unitaries)~\cite{sun2026representationtheoryinhomogeneousgaussian}.}. 
Our state parametrisation encompasses all these specific dynamics, alongside the action of a vast range of experimentally accessible measurements and operations.

\section{Conclusions\label{sec:conclusion}}

Quantum experiments and computational tasks are often achieved via a succession of dynamics, measurements, and operations, which creates and manipulates a superposition of CV Gaussian states entangled with a DV subsystem. Our work presents a formalism to analytically treat these processes and discusses some of the many instances where these states arise in the quantum world. The presented state parametrisation of GBCS is a complex generalisation of first moments and covariance matrices. These quantities are operationally meaningful, in the sense that the state is defined via expectation values and uncertainties often measured in experimental practice. Their real and imaginary parts are directly related to coherence and phases of the DV state and to the defining properties of Wigner functions of superposed Gaussian states. This allows visualisation of the CV parameters, as in Gaussian quantum mechanics, by extending the phase space to a complex space. We have illustrated the applicability to modelling experimental settings through two case studies: (a) dynamical effects in dispersive qubit read-out experiments via a parametric amplified resonator, and (b) the evolution of a levitated nanoparticle in a Stern-Gerlach interferometer with diffusive environment for phase estimation and the certification of Wigner negativity.

Given its flexibility and clear operational grounding, our formalism lends itself to bespoke experimental predictions, spanning a vast range of systems relevant to quantum technologies and fundamental physics, which calls for further investigations. Our results open a systematic avenue to analytically or semi-analytically treat processes which were before computed numerically. For instance, the open-driven dynamics may allow for optimization procedures to minimize unwanted noise via pulse design (for instance, superconducting or ion-based quantum systems). The quadratic scaling of the parameters in the number of resonators, while having the same scaling for the number of qubits, could lead to large simulation of analogue implementations of multi-qubit gates. As shown in the first example, the dynamical analysis may shine light on transitional behaviour of quantum systems which have been, until now, described at steady state. On the fundamental applicability, quantum field theories with superposed Gaussian states could be described via our methodology via a truncation to a high number of modes (as it has been routinely done for Gaussian QFTs). 

A dedicated Python software for the solutions of all the dynamics presented (all tables) and measurements, together with the examples presented, is shared on \href{https://github.com/LorenzoBracciniUCL/GBCS}{GitHub}. Furthermore, the analytical results of Tables~\ref{tab:force} are also provided as Mathematica Notebooks. 

To summarise, our work represents a generalisation of Gaussian techniques to non-Gaussian processes, i.e. superpositions of Gaussian processes, by pointing out that cats have non-linear complex moments.


\acknowledgments

We thank Harvey Weinberger, Marco Genoni, Francesco Albarelli, Eva Kilian-Rademacher, and Sofia Qvarfort for several insightful conversations. L.B. would like to acknowledge Engineering and Physical Sciences Research Council (EPSRC) grants (EP/R513143/1 and EP/W524335/1). The research of
The work of S.B. is funded in part by the Gordon and Betty Moore Foundation through
Grant GBMF12328, DOI 10.37807/GBMF12328, and by the Alfred P. Sloan Foundation under Grant
No. G-2023-21130. A.S. acknowledges financial support from the Leverhulme Trust through
Research Project Grant RPG-2024-287.
 
\bibliography{apssamp}

\appendix

\onecolumngrid

\section{Map from Hilbert to Phase Space \label{app:map_phase_space}}

By repetitive use of Eqs.~(\ref{eq:phase_space_corresp1},\ref{eq:phase_space_corresp2}), one finds 
\begin{equation}
\label{eq:map_chara_first}
   i \hat{r}^m \hat{\varrho}  \longleftrightarrow  \left(  \partial_{\tilde{\xi}^m} + \frac{i}{2} \Omega^{mn} \tilde{\xi}^n \right) \chi (\tilde{\xi}) \hspace{1cm}  i  \hat{\varrho} \hat{r}^m \longleftrightarrow  \left(  \partial_{\tilde{\xi}^m} - \frac{i}{2} \Omega^{mn} \tilde{\xi}^n \right) \chi (\tilde{\xi})
\end{equation}
which will be useful for the Hamiltonians linear terms. Instead, for the quadratic ones, one can note that  
\begin{align*}
     \partial_{\tilde{\xi}^n} \partial_{\tilde{\xi}^m} \chi (\tilde{\xi}) &\longleftrightarrow  - \frac{1}{4} \left(\hat{r}^n ( \hat{r}^m  \hat{\varrho} + \hat{\varrho} \hat{r}^m  ) +  (\hat{r}^m  \hat{\varrho} + \hat{\varrho} \hat{r}^m ) \hat{r}^n  \right)  
\end{align*}
\begin{align*}
    \tilde{\xi}^n \partial_{\tilde{\xi}^m} \chi (\tilde{\xi}) &\longleftrightarrow  \frac{i}{2} \Omega^{on} \left(\hat{r}^o ( \hat{r}^m  \hat{\varrho} + \hat{\varrho} \hat{r}^m  ) -  (\hat{r}^m  \hat{\varrho} + \hat{\varrho} \hat{r}^m ) \hat{r}^o  \right)  
\end{align*}
\begin{align*}
    \tilde{\xi}^n \tilde{\xi}^m \chi (\tilde{\xi}) &\longleftrightarrow  \Omega^{on} \Omega^{pm} \left(\hat{r}^o ( \hat{r}^p  \hat{\varrho} - \hat{\varrho} \hat{r}^p  ) -  (\hat{r}^p  \hat{\varrho} - \hat{\varrho} \hat{r}^p ) \hat{r}^o  \right) 
\end{align*}
such that inverting 
\begin{align*}
     (a) & \;\; \hat{r}^n  \hat{r}^m  \hat{\varrho} + \hat{\varrho} \hat{r}^m  \hat{r}^n + \hat{r}^n  \hat{\varrho}  \hat{r}^m  +  \hat{r}^m  \hat{\varrho} \hat{r}^n \longleftrightarrow   - 4 \partial_{\tilde{\xi}^n} \partial_{\tilde{\xi}^m} \chi (\tilde{\xi})  \\
      (b) & \;\; \hat{r}^n  \hat{r}^m  \hat{\varrho} - \hat{\varrho} \hat{r}^m  \hat{r}^n + \hat{r}^n  \hat{\varrho}  \hat{r}^m  -  \hat{r}^m  \hat{\varrho} \hat{r}^n   \longleftrightarrow   - 2 i \Omega^{n o} \tilde{\xi}^o  \partial_{\tilde{\xi}^m} \chi (\tilde{\xi}) \\
       (c) &\;\; \hat{r}^n  \hat{r}^m  \hat{\varrho} + \hat{\varrho} \hat{r}^m  \hat{r}^n -  \hat{r}^n  \hat{\varrho}  \hat{r}^m  -  \hat{r}^m  \hat{\varrho} \hat{r}^n    \longleftrightarrow \Omega^{n o} \Omega^{m p} \tilde{\xi}^o \tilde{\xi}^p \chi (\tilde{\xi}) 
\end{align*}
Multiplying each expression by a symmetric matrix $A$, i.e. contracting with  $A^{mn}$:
\begin{align}
\label{eq:sim_phase_master}
     A \cdot &(a) \implies \;\; A^{mn} \left( \hat{r}^n  \hat{r}^m  \hat{\varrho} + \hat{\varrho} \hat{r}^m  \hat{r}^n + 2 \hat{r}^n  \hat{\varrho}  \hat{r}^m \right) \longleftrightarrow   - 4 A^{mn} \partial_{\tilde{\xi}^n} \partial_{\tilde{\xi}^m} \chi (\tilde{\xi}) \nonumber \\ 
      A \cdot &(b) \implies \;\;  A^{mn} \left( \hat{r}^n  \hat{r}^m  \hat{\varrho} - \hat{\varrho} \hat{r}^m  \hat{r}^n  \right) \longleftrightarrow   - 2 i  A^{mn} \Omega^{n o} \tilde{\xi}^o  \partial_{\tilde{\xi}^m} \chi (\tilde{\xi})  \nonumber\\
       A \cdot &(c) \implies \;\;  A^{mn} \left( \hat{r}^n  \hat{r}^m  \hat{\varrho} + \hat{\varrho} \hat{r}^m  \hat{r}^n - 2 \hat{r}^n  \hat{\varrho}  \hat{r}^m   \right)  \longleftrightarrow A^{mn} \Omega^{n o} \Omega^{m p} \tilde{\xi}^o \tilde{\xi}^p \chi (\tilde{\xi}) 
\end{align}
We arrive at the equality needed to describe operator-valued Hamiltonian terms:
\begin{equation}
\label{eq:map_chara_second_1}
    A \cdot (a + 2 b + c) \implies A^{mn} \hat{r}^n  \hat{r}^m  \hat{\varrho} \leftrightarrow   A^{mn} \left[ -\partial_{\tilde{\xi}^n}\partial_{\tilde{\xi}^m}   -  i   \Omega^{n o} \tilde{\xi}^o  \partial_{\tilde{\xi}^m}  + \frac{1}{4} \Omega^{n o} \Omega^{m p} \tilde{\xi}^o \tilde{\xi}^p  \right] \chi (\tilde{\xi}) 
\end{equation}
\begin{equation}
\label{eq:map_chara_second_2}
    A \cdot (a - 2 b + c) \implies A^{mn}  \hat{\varrho} \hat{r}^m  \hat{r}^n   \leftrightarrow   A^{mn} \left[ -\partial_{\tilde{\xi}^n}\partial_{\tilde{\xi}^m}   +  i   \Omega^{n o} \tilde{\xi}^o  \partial_{\tilde{\xi}^m}  + \frac{1}{4} \Omega^{n o} \Omega^{m p} \tilde{\xi}^o \tilde{\xi}^p  \right] \chi (\tilde{\xi}) 
\end{equation}
For diffusive dynamics, it will be useful to multiply by a constant skew-symmetric matrix $B$ with elements $B^{mn}$ 
\begin{align}
\label{eq:antsim_phase_master}
      B \cdot (b) \implies \;\;  B^{mn} \left( \hat{r}^n  \hat{r}^m  \hat{\varrho} - \hat{\varrho} \hat{r}^m  \hat{r}^n  + 2 \hat{r}^n  \hat{\varrho}  \hat{r}^m \right) \longleftrightarrow   - 2 i  B^{mn} \Omega^{n o} \tilde{\xi}^o  \partial_{\tilde{\xi}^m} \chi (\tilde{\xi}) 
\end{align}

\section{Derivation of ODEs for Unitary Evolution \label{app:unitary_general}}

The unitary dynamics in frequency-unit time ($\tau$) is given by the von Neumann equation which, recalling the definitions of the state in Eq.~(\ref{rhojk}) and the Hamiltonian of Eq.~(\ref{eq:Hamiltonian_max}), can be written in the qubit basis as
\begin{equation}
\label{eq:vonNeumann_master}
    \frac{\partial \hat{\varrho}}{\partial \tau} =  \frac{i}{\hbar \omega} \left[ \hat{\varrho} , \hat{H} \right] \Leftrightarrow  \frac{\partial \hat{\varrho}_{jk}}{\partial \tau} = \frac{i}{2} \left(  H_k^{mn} \hat{\varrho}_{jk} \hat{r}^m \hat{r}^n  - H_j^{mn} \hat{r}^m \hat{r}^n \hat{\varrho}_{jk}  \right) + i r_j^m \hat{r}^m \hat{\varrho}_{jk} - i r_k^m  \hat{\varrho}_{jk}  \hat{r}^m - \frac{i}{2} (H_j^0-H_k^0)\hat{\varrho}_{jk} \;.
\end{equation}
As claimed in the main text, we seek a solution of the form 
\begin{equation*}
    \hat{\varrho}(\tau) = \frac{1}{(2\pi)^n}\sum_{j, k  = 0}^{N-1}  \int_{\mathds{R}^{2n}} \text{d} \xi \; \chi_{jk} (\xi ,\tau)  D_{r} \otimes \ket{j} \bra{k}
\end{equation*}
where $\chi_{jk} (\xi ,\tau) = \text{Tr} [ D_{- r}  \hat{\varrho}_{jk} (\tau)] $ are four characteristic functions. Define $\tilde{\xi} = \Omega r$. From Eqs.~(\ref{eq:map_chara_second_1}) and~(\ref{eq:map_chara_second_2}), one finds that 
\begin{equation*}
    \frac{i}{2} H^{mn}_j \hat{r}^n  \hat{r}^m  \hat{\varrho} \longleftrightarrow  \frac{1}{2} H^{mn}_j \left( - i \partial_{\tilde{\xi}^n} \partial_{\tilde{\xi}^m} + \Omega^{n o} \tilde{\xi}^o  \partial_{\tilde{\xi}^m} + \frac{i}{4} \Omega^{n o} \Omega^{m p} \tilde{\xi}^o \tilde{\xi}^p \right)  \chi (\tilde{\xi})
\end{equation*}
\begin{equation*}
    \frac{i}{2} H^{mn}_k \hat{\varrho}  \hat{r}^n  \hat{r}^m  \longleftrightarrow \frac{1}{2}  H^{mn}_k \left( - i \partial_{\tilde{\xi}^n} \partial_{\tilde{\xi}^m} -  \Omega^{n o} \tilde{\xi}^o  \partial_{\tilde{\xi}^m} + \frac{i}{4} \Omega^{n o} \Omega^{m p} \tilde{\xi}^o \tilde{\xi}^p \right)  \chi (\tilde{\xi})
\end{equation*}
and, from Eq.~(\ref{eq:map_chara_first}), 
\begin{equation*}
   i r_j^m \hat{r}^m \hat{\varrho}  \longleftrightarrow  r_j^m \left(  \partial_{\tilde{\xi}^m} + \frac{i}{2} \Omega^{mn} \tilde{\xi}^n \right) \chi (\tilde{\xi}) \hspace{2cm}  i  r^m_k \hat{\varrho} \hat{r}^m \longleftrightarrow  r^m_k \left(  \partial_{\tilde{\xi}^m} - \frac{i}{2} \Omega^{mn} \tilde{\xi}^n \right) \chi (\tilde{\xi})
\end{equation*}
Thus, from Eq.~(\ref{eq:vonNeumann_master}), the time evolution of the characteristic functions is described by the  PDE
 \begin{align}
 \label{eq:PDE_chara_general}
     \dot{\chi}_{jk} &= \Big[ \frac{i}{2} (H_j - H_k)^{mn} \partial_{\tilde{\xi}^n} \partial_{\tilde{\xi}^m}  - \left(  \frac{1}{2}  ( H_j + H_k )^{mn} \Omega^{n o} \tilde{\xi}^o  - (r_j - r_k)^m \right)  \partial_{\tilde{\xi}^m} \nonumber \\
     \hspace{2cm } &\hspace{3cm}- \frac{i}{8} (H_j - H_k)^{mn} \Omega^{n o} \Omega^{m p} \tilde{\xi}^o \tilde{\xi}^p  + \frac{i}{2} (r_j + r_k)^m  \Omega^{mn} \tilde{\xi}^n - \frac{i}{2} (H_j^0-H_k^0) \Big] \chi_{jk} . 
 \end{align}
Let us consider the ansatz
\begin{equation}
    \chi_{jk} (\tilde{\xi}, \tau) =  \exp \left( -\frac{1}{4} \tilde{\xi}^{\rm T} \sigma_{jk} (\tau) \tilde{\xi}  +  i  \tilde{\xi}^{\rm T} \bar{r}_{jk} (\tau) + s_{jk} (\tau)   \right) 
\end{equation}
such that its derivatives are
\begin{equation}
    \partial_{\tilde{\xi}^n} \chi_{jk} (\tilde{\xi}, \tau) = - \left( \frac{1}{2} \sigma_{jk}^{nm} \tilde{\xi}^{m} -  i r^n_{jk}  \right) \chi_{jk} (\tilde{\xi}, \tau)
\end{equation}
\begin{align*}
    \partial_{\tilde{\xi}^n} \partial_{\tilde{\xi}^m} \chi_{jk} (\tilde{\xi}, \tau) &= - \frac{1}{2} \delta^{mn} \sigma_{jk}^{mn} \chi_{jk} (\tilde{\xi}, \tau) + \left( \frac{1}{2} \sigma_{jk}^{pm} \tilde{\xi}^{p} -  i r^m_{jk}  \right) \left( \frac{1}{2} \sigma_{jk}^{on} \tilde{\xi}^{o} -  i r^n_{jk}  \right) \chi_{jk} (\tilde{\xi}, \tau) \\ 
    &= \left[\frac{1}{4} \sigma_{jk}^{pm}  \sigma_{jk}^{on} \tilde{\xi}^{p} \tilde{\xi}^{o} - \frac{i}{2} (\sigma_{jk}^{pm}  r^n_{jk} +  \sigma_{jk}^{pn}  r^m_{jk} ) \tilde{\xi}^{p}  - \left(\bar{r}_{jk}^m \bar{r}_{jk}^n + \frac{1}{2}  \delta^{mn} \sigma_{jk}^{mn} \right) \right]\chi_{jk}  
\end{align*}
\begin{equation*}
    \partial_{\tau} \chi_{jk} = \left( -\frac{1}{4} \tilde{\xi}^{\rm T} \dot{\sigma}_{jk} (\tau) \tilde{\xi}  +  i \tilde{\xi}^{\rm T} \dot{\bar{r}}_{jk} (\tau) + \dot{s}_{jk}(\tau) \right) \chi_{jk} (r, \tau)
\end{equation*}
and, substituting, one gets
 \begin{align*}
     \dot{\chi}_{jk} &= \Big[ \frac{i}{2} (H_j - H_k)^{mn} \left[ \frac{1}{4} \sigma_{jk}^{pm}  \sigma_{jk}^{on} \tilde{\xi}^{p} \tilde{\xi}^{o} - \frac{i}{2} (\sigma_{jk}^{pm}  r^n_{jk} +  \sigma_{jk}^{pn}  r^m_{jk} ) \tilde{\xi}^{p}  - \left(\bar{r}_{jk}^m \bar{r}_{jk}^n + \frac{1}{2}  \delta^{mn} \sigma_{jk}^{mn} \right) \right] 
     \\
     &\hspace{1cm}
    + \left(  \frac{1}{2}  ( H_j + H_k )^{mn} \Omega^{n o} \tilde{\xi}^o  - (r_j - r_k)^m \right)  \left( \frac{1}{2} \sigma_{jk}^{pm} \tilde{\xi}^{p} -  i r^m_{jk}  \right)  
     \\
     &\hspace{2cm }
     - \frac{i}{8} (H_j - H_k)^{mn} \Omega^{n o} \Omega^{m p} \tilde{\xi}^o \tilde{\xi}^p  + \frac{i}{2} (r_j + r_k)^m  \Omega^{mn} \tilde{\xi}^n - \frac{i}{2} (H_j^0-H_k^0) \Big] \chi_{jk} 
     \\
\end{align*}
\begin{align*}
     \dot{\chi}_{jk} &= \Big[ +  \frac{1}{4} 
     \left[ ( H_j + H_k )^{mn} \Omega^{no}  \sigma_{jk}^{pm}  +    \frac{i}{2} (H_j - H_k)^{mn} \left( \sigma_{jk}^{pm}  \sigma_{jk}^{on} +  \Omega^{n o} \Omega^{pm}  \right)   \right] \tilde{\xi}^{p} \tilde{\xi}^{o}  
     \\
     &\hspace{1cm} 
     +   i \tilde{\xi}^p  \left[   \frac{1}{2} \Omega^{p  n} ( H_j + H_k )^{nm}  \bar{r}_{jk}^m    - \frac{1}{2}\Omega^{p m}  (r_j + r_k)^m  - i  \left( \frac{1}{2} (H_j - H_k)^{mn}  \sigma_{jk}^{pm}  r^n_{jk}  -   \frac{1}{2} \sigma_{jk}^{pm} (r_j - r_k)^m \right) \right]  
     \\
     &\hspace{2cm} 
     -  i \left(  \frac{1}{2} (H_j - H_k)^{mn} \left(\bar{r}_{jk}^m \bar{r}_{jk}^n + \frac{1}{2}  \delta^{mn} \sigma_{jk}^{mn} \right) - (r_j - r_k)^m  r^m_{jk}  + \frac{1}{2} (H_j^0-H_k^0)  \right) \Big] \chi_{jk} \\
    &= \left( -\frac{1}{4} \tilde{\xi}^{\rm T} \dot{\sigma}_{jk} (\tau) \tilde{\xi}  +  i \tilde{\xi}^{\rm T} \dot{\bar{r}}_{jk} (\tau) + \dot{s}_{jk}(\tau) \right) \chi_{jk} 
\end{align*}
where $\dot{s}_{jk} = - \dot{\mathcal{C}}_{jk} (\tau) + i \dot{\phi}_{jk} (\tau)$ is understood for the reduced density matrix. The last equality must hold for all $\tilde{\xi}$. Thus, one finds the set of ODEs of Table \ref{tab:unitary} by equating the coefficients of each power in $\tilde{\xi}^p$ of the last equality. 

\section{Integration of Diagonal Terms and Linear Operator-Valued Interactions: Unitary Dynamics \label{app:only_force}}

In order to integrate the operator-valued force, we need to recall some properties: a general symplectic transformation is of the form $S_H(\tau) = \exp( \Omega H \tau)$, such that  $S_H^{-1} (\tau) = \exp (- \Omega H \tau) $ and $S_H^{\rm T} (\tau) = \exp (- H \Omega \tau) $. The symplectic form $\Omega$ respects $\Omega^{-1} = \Omega^{\rm T} = - \Omega$. Furthermore, from the definition of symplectic group (as the set of real matrices $S \in Sp_{2, \mathds{R}}$ such that $S \Omega S^{\rm T} = \Omega$), that $S \Omega  = \Omega (S^{\rm T})^{-1}$, $ \Omega S^{\rm T}  = S^{-1}  \Omega$, and $(S^{-1})^{\rm T}  \Omega = \Omega S $, from which it is possible to derive one of the central identities of the following proofs:
\begin{equation}
\label{eq:identity_usefull}
    \left( S^{\rm T}_H(\tau) - \mathds{1} \right)\Omega =  S^{\rm T}_H(\tau) \left( \mathds{1} - (S^{\rm T}_H)^{-1}(\tau)\right)\Omega = - S^{\rm T}_H(\tau) \Omega (S_H(\tau) - \mathds{1})
\end{equation}
and so its transpose reads $\Omega^{\rm T} \left( S_H(\tau) - \mathds{1} \right) =- (S_H^{\rm T}(\tau) - \mathds{1})\Omega^{\rm T} S_H(\tau) $. Let us also compute the integral
\begin{equation}
    \int_0^\tau \text{d}t S_H(t)  =\int_0^\tau \text{d}t {\rm e}^{ \Omega H t}  =  ( \Omega H)^{-1} \left(S_H(\tau) - \mathds{1} \right) = - H^{-1} \Omega  \left(S_H(\tau) - \mathds{1} \right),
\end{equation}
which implies  
\begin{equation}
    \int_0^\tau S_H^{-1}(t) dt =   \left(S_H^{-1}(\tau) - \mathds{1} \right) H^{-1} \Omega \hspace{1cm} \int_0^\tau S_H^{\rm T}(t) dt =     \left(S_H^{\rm T}(\tau) - \mathds{1} \right) \Omega  H^{-1}
\end{equation}

 Regarding the on-diagonal terms, Eq.~(\ref{eq:general_unitary_ODE_on}) is readily integrable:  letting $H \to H_\pm = H_m \pm H_q$, one has
\begin{align*}
    \bar{r}^{\text{on}}_{\pm 1, \pm 1} &= S_{\pm} (\tau)  \bar{r}^{\text{on}} -  S_{\pm} (\tau)  \int_{0}^{\tau} \text{d} t S_{\pm}^{-1} (t) \Omega (r_m \pm r_q) = S_{\pm} (\tau)  \bar{r}^{\text{on}} - S_{\pm} (\tau)  \left( S_{\pm}^{-1} (\tau)  - \mathds{1} \right) H^{-1}_{\pm} \Omega \Omega (r_m \pm r_q) \\
     &= S_{\pm} (\tau)  \bar{r}^{\text{on}} -   \left( S_{\pm} (\tau)  - \mathds{1} \right) H^{-1}_{\pm} (r_m \pm r_q), 
\end{align*}
which is equivalent to Eq.~(\ref{eq:time_diagonal_general}). This result holds also for the on-diagonal case of only operator-valued force with $H_\pm \to H_m$. 

Let us focus on the case of only operator-valued forces, where the ODEs are integrable also for the off-diagonal case.  As stated in the main text, all the covariance matrices evolve under a single symplectic transformation $ S_m (\tau) = \exp(\Omega H_m \tau)$ such that
\begin{equation}
\label{eq:cov_t_force}
   \sigma (\tau) := \sigma_{jk}(\tau)  = S_m (\tau) \sigma_0 S_m^{\rm T} (\tau).
\end{equation}
Then we need to solve 
\begin{align}
    \dot{\bar{r}}_{jk} =     \Omega  H_m  \bar{r}_{jk}  -   \frac{1}{2}\Omega (r_j + r_k) + \frac{i}{2}  \sigma (\tau) (r_j - r_k) ,
\end{align}
obtaining
\begin{align}
\label{eq:app_first_moment_ODE}
    \bar{r}_{jk} (\tau) &= S_m (\tau) r_0  -   S_m(\tau) \frac12 \int_0^{\tau} \text{d} t  S_m^{-1}(t) \left( \Omega (r_j + r_k) - i  \sigma (t) (r_j - r_k)  \right) \nonumber  \\ 
     &= S_m (\tau) r_0 - \frac{1}{2} \left[ (\tilde{r}_j(\tau) - \tilde{r}_j) + (\tilde{r}_k(\tau) - \tilde{r}_k) \right] - \frac{i}{2}  S_m(\tau) \sigma_0 \int_0^{\tau} \text{d} t  S_m^{\rm T}(t)  (r_j - r_k) \nonumber  \\
     &= S_m (\tau) r_0 - \frac{1}{2} \left[ (\tilde{r}_j(\tau) - \tilde{r}_j) + (\tilde{r}_k(\tau) - \tilde{r}_k) \right] + \frac{i}{2}  S_m(\tau) \sigma_0 \left(S_m^{\rm T}(\tau) - \mathds{1} \right)  \Omega  H^{-1} (r_j - r_k) \nonumber \\
     &= S_m (\tau) r_0 - \frac{1}{2} \left[ (\tilde{r}_j(\tau) - \tilde{r}_j) + (\tilde{r}_k(\tau) - \tilde{r}_k )\right] - \frac{i}{2}  S_m(\tau) \sigma_0 S_m^{\rm T}(\tau)   \Omega \left( S_m(\tau)-  \mathds{1}   \right)  H^{-1}_m (r_j - r_k)  \nonumber\\
     &= r_0(\tau) - \frac{1}{2}  \left( \tilde{r}_j(\tau) - \tilde{r}_j + \tilde{r}_k(\tau) - \tilde{r}_k  \right) -  \frac{i}{2} \sigma (\tau)   \Omega  \left[ (\tilde{r}_j (\tau) - \tilde{r}_j)  - (\tilde{r}_k (\tau) - \tilde{r}_k ) \right] 
\end{align}
where the second equality follows from the time evolution of the covariance matrix Eq.~(\ref{eq:cov_t_force}) and the fourth from Eq.~(\ref{eq:identity_usefull}). From this, it is possible to derive Eq.~(\ref{eq:r_on_force_sol}). We are left to integrate  
\begin{align*}
    \dot{s}_{jk} &= - \dot{\mathcal{C}}_{jk}  + i \dot{\phi}_{jk} =    i  (r_j - r_k)^{\rm T}   \bar{r}_{jk}  - \frac{i}{2} (H_j^0-H_k^0) 
\end{align*}
By substituting Eq.~(\ref{eq:app_first_moment_ODE}) and splitting the real and imaginary parts, one finds
\begin{align}
\label{eq:ode_C_appendix}
    \dot{\mathcal{C}}_{jk} &=  -  \frac{1}{2}  (r_j - r_k)^{\rm T}   \sigma (\tau)   \Omega  \left( (\tilde{r}_j (\tau) - \tilde{r}_j)  - (\tilde{r}_k (\tau) - \tilde{r}_k ) \right) =   \frac{1}{2}  (r_j - r_k)^{\rm T}  S_m(\tau) \sigma_0 \left(S_m^{\rm T}(\tau) - \mathds{1} \right)  \Omega  H^{-1} (r_j - r_k)  \\
    \label{eq:ode_phi_appendix}
    \dot{\phi}_{jk} &=  (r_j  - r_k )^{\rm T}\left[  r_0(\tau) - \frac{1}{2} \left( \tilde{r}_j(\tau) - \tilde{r}_j + \tilde{r}_k(\tau) - \tilde{r}_k \right)  \right] - \frac{1}{2} (H_j^0-H_k^0) 
\end{align}

In order to integrate the former, one may note that, for a general matrix $A$,  $(r_j - r_k)^{\rm T} A (r_j - r_k)  = \frac{1}{2} (r_j - r_k)^{\rm T} (A+A^{\rm T}) (r_j - r_k)$. Then, define the matrices $M_1(t) = (H^{-1} )^{\rm T}  \Omega^{\rm T} \left(S_m(t) - \mathds{1} \right)  $ and $M_2(t) = \sigma_0 \left(S_m^{\rm T}(t) - \mathds{1} \right)  \Omega H^{-1}$, such that $\dot{M}_1(t)= S_m(t)$ and $ \dot{M}_2(t) =  \sigma_0 S_m^{\rm T} (t)$. Note that $(\dot{M}_1(t) M_2(t))^{\rm T} = M_1(t) \dot{M}_2(t)$, such that 
\begin{align}
    \mathcal{C}_{jk} &=  \frac{1}{2} (r_j - r_k)^{\rm T} \left[ \int_0^{\tau} \text{d} t \dot{M}_1(t) M_2(t)  \right] (r_j - r_k)^{\rm T} =  \frac{1}{4} (r_j - r_k)^{\rm T} \left[ \int_0^{\tau} \text{d} t  \dot{M}_1(t) M_2(t) + (\dot{M}_1(t) M_2(t))^{\rm T}  \right] (r_j - r_k)^{\rm T} \nonumber \\
    &=   \frac{1}{4} (r_j - r_k)^{\rm T} \left[ \int_0^{\tau} \text{d} t  \dot{M}_1(t) M_2(t) + M_1(t) \dot{M}_2(t) \right] (r_j - r_k)^{\rm T}  =   \frac{1}{4} (r_j - r_k)^{\rm T} ( M_1(\tau) M_2(\tau) - M_1(0) M_2(0))  (r_j - r_k)^{\rm T}\nonumber \\
    &=  \frac{1}{4} (r_j - r_k)^{\rm T} H^{-1}  \Omega^{\rm T} \left(S_m(\tau) - \mathds{1} \right)  \sigma_0 \left(S_m^{\rm T}(\tau) - \mathds{1} \right)  \Omega H^{-1} (r_j - r_k) \label{eq:integral_app_contr} \\
    &=  \frac{1}{4} (r_j - r_k)^{\rm T} H^{-1}  \left(S^{\rm T}_m(\tau) - \mathds{1} \right)  \Omega^{\rm T} S_m(\tau) \sigma_0 S^{\rm T}_m(\tau)\Omega  \left(S_m^{\rm T}(\tau) - \mathds{1} \right)  H^{-1} (r_j - r_k)\nonumber \\
    &=  \frac{1}{4} \left[ (\tilde{r}_j (\tau) - \tilde{r}_j)  - (\tilde{r}_k (\tau) - \tilde{r}_k ) \right]^{\rm T} \Omega^{\rm T} \sigma (\tau)   \Omega  \left[ (\tilde{r}_j (\tau) - \tilde{r}_j)  - (\tilde{r}_k (\tau) - \tilde{r}_k ) \right] 
\end{align}
where we used the fact that $M_1(0) M_2(0) = 0$ and Eq.~(\ref{eq:identity_usefull}). Finally, we integrate the phases
\begin{align}
\label{eq:phase_appendix_unitary}
    \phi_{jk} &=  (r_j - r_k)^{\rm T} \left[ \int_0^{\tau} \text{d} t S_m (t) \right] \left( \tilde{r}_0  - \frac{1}{2} (\tilde{r}_j + \tilde{r}_k) \right) + \frac{\tau}{2} \left( (\tilde{r}_j - \tilde{r}_k)^{\rm T} H_m (\tilde{r}_j + \tilde{r}_k) -  (H_j^0-H_k^0) \right) \nonumber \\ 
    &=  - (r_j - r_k)^{\rm T} H^{-1} \Omega \left( S_m (\tau)  - \mathds{1} \right) \left( \tilde{r}_0  - \frac{1}{2} (\tilde{r}_j + \tilde{r}_k) \right) + \frac{\tau}{2} \left( (\tilde{r}_j - \tilde{r}_k)^{\rm T} H_m (\tilde{r}_j + \tilde{r}_k) -  (H_j^0-H_k^0) \right) \\ 
    & =  - (\tilde{r}_j - \tilde{r}_k)^{\rm T} \Omega \left[ ( \bar{r}_0  (\tau) - \bar{r}_0 ) - \frac{1}{2} \left( \tilde{r}_j (\tau) - \tilde{r}_j + \tilde{r}_k (\tau) -  \tilde{r}_k  \right) \right] + \frac{\tau}{2} \left[ (\tilde{r}_j - \tilde{r}_k)^{\rm T} H_m (\tilde{r}_j + \tilde{r}_k) -  (H_j^0-H_k^0) \right] \nonumber
\end{align}
This concludes the derivation in Table \ref{tab:force}.


\section{Maps of Additional Terms for Open Dynamics \label{app:noise}}

First of all, let us find out how the ODEs change with the addition of the noise terms of Eq.~(\ref{eq:linbladian}) in the $\hat{r}$ basis. One finds that
\begin{align*}
    \mathcal{L}_B (\hat{r}) &:=  B^{mn} \left(  \hat{r}^m \hat{\varrho} \hat{r}^n  - \frac{1}{2} \{ \hat{\varrho}, \hat{r}^m  \hat{r}^n \} \right) =     \left(\frac{1}{2}  (\Omega^{\rm T} D \Omega)^{mn} -  i  {\rm e}^{mn} \right) \left(  \hat{r}^m \hat{\varrho} \hat{r}^n  - \frac{1}{2} \{ \hat{\varrho}, \hat{r}^m  \hat{r}^n \} \right) 
\end{align*}
where we used the fact that $B$ is Hermitian, and we divided the symmetric real part ($\Omega^{\rm T} D \Omega$, where $D$ is symmetric) and antisymmetric complex part ($i  E$, where $E$ is \textit{skew}-symmetric). From Eq.~(\ref{eq:sim_phase_master}) and Eq.~(\ref{eq:antsim_phase_master}), one finds that  
\begin{equation}
    \mathcal{L}_{\hat{r}}  (\hat{\varrho}) \longleftrightarrow - 
    \left[ \frac{1}{4} (\Omega^{\rm T} D \Omega)^{mn}  \Omega^{n o} \Omega^{m p} \tilde{r}^o \tilde{r}^p   -   {\rm E}^{mn} \Omega^{n o} \tilde{r}^o  \partial_{\tilde{r}^m}  \right] \chi (\tilde{r}) =  - 
    \left[ \frac{1}{4} D^{op} \tilde{r}^o \tilde{r}^p  -  {\rm E}^{mn} \Omega^{n o} \tilde{r}^o  \partial_{\tilde{r}^m}  \right] \chi (\tilde{r}) 
\end{equation}
which represents the additional noise terms of the PDE of the time evolution of the branched characteristic functions (Eq.~\ref{eq:PDE_chara_general}). By taking the ansatz for $\chi_{jk}$ (Eq.~\ref{eq:chara_jk}), one finds 
\begin{align}
    \mathcal{L}_{\hat{r}} (\hat{\varrho}) \longleftrightarrow &-  \left[ \frac{1}{4}  D^{mn} \tilde{r}^m \tilde{r}^n  + {\rm e}^{mn} \Omega^{no}  \left( \frac{1}{2} \sigma_{jk}^{mp} \tilde{r}^{p} -  i r^m_{jk}  \right) \tilde{r}^o  \right] \chi_{jk} (\tilde{r}) \nonumber \\
    &=  \left[ - \frac{1}{4} \left( D^{po} + 2 \sigma_{jk}^{pm} {\rm e}^{mn}  \Omega^{no}     \right) \tilde{r}^p \tilde{r}^o  +  i   {\rm e}^{mn} \Omega^{no}  r^m_{jk}   \tilde{r}^o  \right] \chi_{jk} (\tilde{r}) \nonumber 
\end{align}
By symmetrizing the first term and noticing that $ (\sigma E  \Omega )^{\rm T} = -  \Omega {\rm e}^{\rm T} \sigma $, the corrections to the ODEs for $\sigma_{jk}$ and $\bar{r}_{jk}$ are
\begin{equation}
\dot{\Sigma}_{jk} = \left( \dot{\sigma}_{jk} \right) + \Omega E  \sigma_{jk}  -  \sigma_{jk} E^{\rm T} \Omega + D \hspace{2cm} \dot{\bar{r}}_{jk} = \left( \dot{\bar{r}}_{jk} \right) + \Omega E \bar{r}_{jk} + \Omega d
\end{equation}
where the inclusion of the driving term ($d$) is easily concluded by noticing that it is just an additional force. We note that noise and driving of the modes do not affect the ODE for $s_{jk}$ as there are no constant terms in the PDE of $\chi_{jk}$. 
The addition of these terms, with the inclusion of the DV dephasing, leads to the ODEs of Table \ref{tab:open}.

\section{Integration of Linear Operator-Valued Interactions: Open Dynamics \label{app:only_force_open}}

In the case of operator-valued forces and open dynamics, it is possible to write the general solution of the ODEs (Table~\ref{tab:open}) in integral form (Table~\ref{tab:open}), and by restricting to only diffusive noise ($E=0$), one finds a more compact form (Table \ref{tab:open_app}). In the case of only linear coupling, $H_j = H_m \forall j$, such that it is possible to define a transformation $S_A(\tau) = \exp(  \Omega A \tau)$ where $A =H_m + E$ is no longer symmetric and $S_A$ is not symplectic. We note that $S_A^{-1} (\tau) = \exp ( - \Omega A \tau) $ and $S_A^{\rm T} (\tau) = \exp (- A^{\rm T} \Omega \tau) $, and 
\begin{equation}
    \int_0^\tau \text{d}t S_A(t)  =\int_0^\tau \text{d}t {\rm e}^{ \Omega A \tau}  =  ( \Omega A)^{-1} \left(S_H(\tau) - \mathds{1} \right) = - A^{-1} \Omega  \left(S_H(\tau) - \mathds{1} \right)
\end{equation}
which implies 
\begin{equation}
    \int_0^\tau S_A^{-1}(t) dt =    \left(S_A^{-1}(\tau) - \mathds{1} \right) A^{-1} \Omega \hspace{1cm} \int_0^\tau S_A^{\rm T}(t) dt =     \left(S_A^{\rm T}(\tau) - \mathds{1} \right) \Omega  (A^{\rm T})^{-1}
\end{equation}

\begin{table*}
    \centering
    \renewcommand{\arraystretch}{1.7}
    \begin{tabular}{ |p{4.0cm}||P{7cm}|P{7cm}|}
       \hline
        Phase-Space parameters & \multicolumn{2}{c|}{Operator-Valued Linear Interactions (Symmetric Case)} \\ 
        \hline
        Covariance Matrices   & \multicolumn{2}{c|}{\ensuremath{
   \sigma (\tau) := \sigma_{jk}(\tau) = 
    S_m(\tau) \sigma_0 S_m^{\rm T}(\tau) + \int_{0}^{\tau} \text{d}t S_m(\tau - t) D S^{\rm T}_m(\tau - t) }}\\
    \hline
       Vectors or First Moments & \multicolumn{2}{c|}{
       \ensuremath{\bar{r}_{jk} (\tau) = r_0(\tau)  + d(\tau) - d - \frac{1}{2}  \left( \tilde{r}_j(\tau) - \tilde{r}_j + \tilde{r}_k(\tau) - \tilde{r}_k  \right) - \frac{i}{2}  \sigma (\tau)  \Omega  \left[ (\tilde{r}_j (\tau) - \tilde{r}_j)  - (\tilde{r}_k (\tau) - \tilde{r}_k ) \right] }} \\
           & \multicolumn{2}{c|}{
       \ensuremath{ \hspace{2cm}  - \frac{i}{2}  \int_0^{\tau} \text{d}t  D (\tau-t) \Omega  \left[ (\tilde{r}_j(\tau - t) - \tilde{r}_j(\tau) ) -  (\tilde{r}_k(\tau - t) - \tilde{r}_k(\tau) ) \right]  }} \\
       \hline
     QRDM Contrasts  & \multicolumn{2}{c|}{ \hspace{-0.1cm} \ensuremath{  \mathcal{C}_{jk} (\tau)  =    \frac{1}{4} \left[ (\tilde{r}_j (\tau) - \tilde{r}_j)  - (\tilde{r}_k (\tau) - \tilde{r}_k ) \right]^{\rm T} \Omega^{\rm T} \sigma (\tau)   \Omega  \left[ (\tilde{r}_j (\tau) - \tilde{r}_j)  - (\tilde{r}_k (\tau) - \tilde{r}_k ) \right] + \tau \left[\frac{1}{2} (\Gamma_z^{jj} + \Gamma_z^{kk} ) - \Gamma_z^{jk} \right] }} \\
 & \multicolumn{2}{c|}{\ensuremath{ \hspace{1cm} + \frac{1}{4}     \int_0^\tau \text{d} t   \Big\{ \left[ (\tilde{r}_j (\tau - t) - \tilde{r}_j(\tau))  - (\tilde{r}_k (\tau - t) - \tilde{r}_k(\tau))  \right]^{\rm T}  \Omega^{\rm T}  D(\tau - t) \Omega  }} \\
 & \multicolumn{2}{c|}{\ensuremath{ \hspace{2.5cm} 
 \left[ (\tilde{r}_j (\tau - t)  + \tilde{r}_j(\tau) - 2 r_j)  - (\tilde{r}_k (\tau - t) + \tilde{r}_k(\tau) - 2 r_k)  \right]  \Big\} }} \\
\hline 
    \end{tabular}
    \caption{Open evolution of a GBCS phase-space quantity under an operator-valued linear Hamiltonian (Eq. \ref{eq:Hamiltonian_force_only}) and Gaussian diffusive noise ($D$ and $E=0$) of an initial Gaussian state with initial first and second moments $r_0$ and $\sigma_0$, where $\tilde{r}_j = H^{-1}_m r_j$, $S_m (\tau) = \exp (  \Omega H_m \tau)$, and $\tilde{r}_a (\tau) = S_m (\tau) \tilde{r}_a$, with $a \in 0,j,k$. The phases are the same as in Table~\ref{tab:force}.}
    \label{tab:open_app}
\end{table*}

Under such definitions, the ODEs for $\sigma_{jk}$ (Table \ref{tab:open}) and their solutions read,
\begin{align}
    \dot{\sigma}_{jk} &=   \sigma_{jk}  A \Omega  -   \Omega  A^{\rm T}   \sigma_{jk}  + D   \implies \sigma_{jk} = \sigma(\tau) = S_A(\tau) \sigma_0 S_A^{\rm T}(\tau) + \int_{0}^{\tau} \text{d}t S_A(\tau - t) D S^{\rm T}_A(\tau - t)
\end{align}
which is the usual solution of Lyapunov equations, to which all second moments evolve. Thus 
\begin{align}
     \dot{\bar{r}}_{jk} &=    \Omega A  \bar{r}_{jk}  + \frac{1}{2} \Omega (2 d - r_j - r_k)  +  \frac{i}{2}    \sigma(\tau) (r_j - r_k)   
\end{align}
which can be integrated 
\begin{align*}
    \bar{r}_{jk} (\tau) &= S_A (\tau) r_0  + \frac{1}{2}   S_A(\tau)  \int_0^{\tau} \text{d} t  S_A^{-1}(t) \left( \Omega (2d - r_j - r_k) + i  \sigma (t) (r_j - r_k)  \right)  \\ 
    &= S_A (\tau) r_0  + \frac{1}{2}   S_A(\tau)  \left(S_A^{-1}(\tau) - \mathds{1} \right) A^{-1} \Omega \Omega (2d - r_j - r_k) \\
    &\hspace{1cm} +  \frac{i}{2} S_A(\tau)  \sigma_0 \int_0^{\tau} \text{d} t  S_A^{T}(t)  (r_j - r_k)   + \frac{i}{2} S_A(\tau) \int_0^{\tau} \text{d}t \int_{0}^{t}  \text{d} t'  S_A^{-1}(t) S_A(t-t')D S^{\rm T}_A(t - t')   (r_j - r_k)    \\ 
    &= S_A (\tau) r_0   + \frac{1}{2}     \left(S_A(\tau) - \mathds{1} \right) A^{-1}  (2d - r_j - r_k) + 
    \frac{i}{2} S_A (\tau) \sigma_0 \left( S_A^{\rm T} (\tau) - \mathds{1}  \right) \Omega (A^{\rm T})^{-1}  (r_j - r_k)   \\
    &\hspace{1cm} + \frac{i}{2} \int_0^{\tau} \text{d}t'  S_A(\tau- t') D \int_{t'}^{\tau}  \text{d} t S^{\rm T}_A(t- t')(r_j - r_k)  \\
      &=S_A (\tau) r_0  + \frac{1}{2}     \left(S_A(\tau) - \mathds{1} \right) A^{-1}  ( 2 d - r_j - r_k) + 
    \frac{i}{2} S_A (\tau) \sigma_0 \left( S_A^{\rm T} (\tau) - \mathds{1}  \right) \Omega (A^{\rm T})^{-1}  (r_j - r_k)   \\
    &\hspace{1cm} + \frac{i}{2}  \int_0^{\tau} \text{d}t'  S_A(\tau- t') D \left( S^{\rm T}_A(\tau - t') - \mathds{1} \right) \Omega (A^{\rm T})^{-1} (r_j - r_k) 
\end{align*}
The final parameters for the QRDM can be integrated
\begin{align*}
    \dot{\phi}_{jk} &= - (r_j  - r_k )^{\rm T} \left[ S_A (\tau) r_0  +     \left(S_A(\tau) - \mathds{1} \right) A^{-1} \left[ d - \frac{1}{2}  (r_j + r_k) \right] \right] + \frac{1}{2} (H_j^0-H_k^0) 
\end{align*}
which implies that
\begin{align}
    \phi_{jk} (\tau) =  - (r_j - r_k)^{\rm T} A^{-1} \Omega \left( S_A (\tau)  - \mathds{1} \right) \left[ r_0  + A^{-1} \left( d  - \frac{1}{2} (r_j + r_k) \right) 
    \right] + \tau\left( (r_j - r_k)^{\rm T} A^{-1} \left(d - \frac{1}{2} (r_j + r_k) \right) -  \frac{1}{2} (j-k) H_q^0 \right) \nonumber
\end{align}
from similarity of Eq. \ref{eq:ode_C_appendix} (i.e. the replacement $H_m \to A$ and addition of $d$). For the contrasts one needs to integrate
\begin{align*}
    \dot{\mathcal{C}}_{jk} &=   \frac{1}{2}  (r_j - r_k)^{\rm T}  S_A(\tau) \sigma_0 \left(S_A^{\rm T}(\tau) - \mathds{1} \right)  \Omega  (A^{\rm T})^{-1} (r_j - r_k)  \\
    &\hspace{1cm} +  \frac{1}{2}  (r_j - r_k)^{\rm T} \int_0^{\tau} \text{d}t'  S_A(\tau- t') D \left( S^{\rm T}_A(\tau - t') - \mathds{1} \right) \Omega (A^{\rm T})^{-1} (r_j - r_k)
\end{align*}
such that
\begin{align*}
     \mathcal{C}_{jk} (\tau)  &=  \frac{1}{4} (r_j - r_k)^{\rm T} A^{-1}  \Omega^{\rm T} \left( S_A(\tau) - \mathds{1} \right) \sigma_0 \left(S^{\rm T}_A(\tau) - \mathds{1} \right) \Omega  (A^{\rm T})^{-1} (r_j - r_k)   \nonumber \\
     &\hspace{1cm} + \frac{1}{2}  (r_j - r_k)^{\rm T} \int_0^\tau \text{d} t \int_0^{t} \text{d}t'  S_A(t- t') D \left( S^{\rm T}_A(t - t') - \mathds{1} \right) \Omega (A^{\rm T})^{-1} (r_j - r_k) 
\end{align*}
where the first line comes from Eq. \ref{eq:integral_app_contr}. For the second term one can approach the integration similarly to Appendix \ref{app:only_force}, by noticing that $(r_j - r_k)^{\rm T} M (r_j - r_k)  = \frac{1}{2} (r_j - r_k)^{\rm T} (M+M^{\rm T}) (r_j - r_k)$, for any matrix $M$.  Then, define the matrices $N_1(t, t') = A^{-1}  \Omega^{\rm T} \left(S_A(t - t') - \mathds{1} \right)  $ and $N_2(t, t') =  D \left(S_A^{\rm T}(t - t') - \mathds{1} \right)  \Omega (A^{\rm T})^{-1} $, such that 
$\frac{d N_1}{d t} =   S_A(t - t')$ and $\frac{d N_2}{d t}=  D S_m^{\rm T} (t - t')$. Note that 
\begin{equation*}
    \left(\frac{d N_1}{d t}  N_2 \right)^{\rm T}  =  - \left( S_A(t - t')D \left(S_A^{\rm T}(t - t') - \mathds{1} \right)  \Omega^{\rm T} (A^{\rm T})^{-1} \right)^{\rm T}   =  N_1 \frac{d N_2}{d t}
\end{equation*}
and so the second line reads 
\begin{align*}
    &=  \frac{1}{2} (r_j - r_k)^{\rm T} \int_0^\tau \text{d} t' \int_{t'}^{\tau} \text{d}t  S_A(t- t') D \left( S^{\rm T}_A(t - t') - \mathds{1} \right) \Omega (A^{\rm T})^{-1} (r_j - r_k) \\
     &=   \frac{1}{2} (r_j - r_k)^{\rm T} \int_0^\tau \text{d} t' \int_{t'}^{\tau} \text{d}t   \left[ \frac{d N_1}{d t}  N_2\right]  (r_j - r_k) =  \frac{1}{4} (r_j - r_k)^{\rm T} \int_0^\tau \text{d} t' \int_{t'}^{\tau} \text{d}t  \left[ \frac{d N_1}{d t}  N_2 +  \left(\frac{d N_1}{d t}  N_2 \right)^{\rm T} \right] (r_j - r_k)    \\ 
      &=  \frac{1}{4} (r_j - r_k)^{\rm T} \left[ \int_0^\tau \text{d} t'    N_1(\tau, t') N_2(\tau, t') - N_1(t',t') N_2(t',t') \right] (r_j - r_k)  \\
      &=  \frac{1}{4} (r_j - r_k)^{\rm T}  A^{-1}  \Omega^{\rm T} \left[  \int_0^\tau \text{d} t'    \left(S_A(\tau - t') - \mathds{1} \right) D \left(S_A^{\rm T}(\tau - t') - \mathds{1} \right)   \right]  \Omega (A^{\rm T})^{-1} (r_j - r_k) 
\end{align*}
where we integrated by parts. The entire expression reads
\begin{align*}
     \mathcal{C}_{jk} (\tau)  &=  \frac{1}{4} (r_j - r_k)^{\rm T}  A^{-1} \Omega^{\rm T} \left( S_A(\tau) - \mathds{1} \right) \sigma_0 \left(S^{\rm T}_A(\tau) - \mathds{1} \right) \Omega  (A^{\rm T})^{-1}  (r_j - r_k)   \\
     &\hspace{1cm} + \frac{1}{4} (r_j - r_k)^{\rm T}  A^{-1}   \Omega^{\rm T} \left[  \int_0^\tau \text{d} t'    \left(S_A(\tau - t') - \mathds{1} \right) D \left(S_A^{\rm T}(\tau - t') - \mathds{1} \right)   \right]  \Omega (A^{\rm T})^{-1}  (r_j - r_k) 
\end{align*}

In the case of $E=0$,  $A^{\rm T} = A = H_m$  and $S_A$ is symplectic. In this case, it is possible to use the identity of Eq.~(\ref{eq:identity_usefull}). By defining $\tilde{r}_j = H^{-1}_m r_j$ and $\tilde{d} = H^{-1}_m d$, and their symplectic evolution $\tilde{r}_j (\tau)= S_A(\tau)\tilde{r}_j$ and $\tilde{d} (\tau)= S_A(\tau)\tilde{d}$, the following further simplification is found
\begin{align*}
\bar{r}_{jk} (\tau) &= r_0(\tau) + d(\tau) - d - \frac{1}{2}  \left( \tilde{r}_j(\tau) - \tilde{r}_j + \tilde{r}_k(\tau) - \tilde{r}_k  \right) - 
    \frac{i}{2} S_A (\tau) \sigma_0 S_A^{\rm T} (\tau) \Omega \left( S_A (\tau) - \mathds{1}  \right) (\tilde{r}_j - \tilde{r}_k)   \\
    &\hspace{1cm}  - \frac{i}{2}  \int_0^{\tau} \text{d}t'  S_A(\tau- t') D S^{\rm T}_A(\tau - t') \Omega \left( S_A(\tau - t') - \mathds{1} \right) (\tilde{r}_j - \tilde{r}_k) \\
     &= r_0(\tau)  + d(\tau) - d - \frac{1}{2}  \left( \tilde{r}_j(\tau) - \tilde{r}_j + \tilde{r}_k(\tau) - \tilde{r}_k  \right) -
    \frac{i}{2} S_A (\tau) \sigma_0 S_A^{\rm T} (\tau) \Omega  \left[ (\tilde{r}_j(\tau) - \tilde{r}_j) - (\tilde{r}_k(\tau) - \tilde{r}_k)  \right]    \\
    &\hspace{1cm}  - \frac{i}{2}  \int_0^{\tau} \text{d}t'  S_A(\tau- t') D S^{\rm T}_A(\tau - t') \Omega \left( S_A(\tau - t') - S_A(\tau) + S_A(\tau) - \mathds{1} \right)  (\tilde{r}_j - \tilde{r}_k) 
 \end{align*}
\begin{align*}
\bar{r}_{jk} (\tau)    &= r_0(\tau)  + d(\tau) - d - \frac{1}{2}  \left( \tilde{r}_j(\tau) - \tilde{r}_j + \tilde{r}_k(\tau) - \tilde{r}_k  \right) \\
    &\hspace{1cm}- 
    \frac{i}{2} \left( S_A (\tau) \sigma_0 S_A^{\rm T} (\tau) + \int_0^{\tau} \text{d}t'  S_A(\tau- t') D S^{\rm T}_A(\tau - t') \right)\Omega  \left[ (\tilde{r}_j(\tau) - \tilde{r}_j) - (\tilde{r}_k(\tau) - \tilde{r}_k)  \right]    \\
    &\hspace{1cm}  - \frac{i}{2}  \int_0^{\tau} \text{d}t'  S_A(\tau- t') D S^{\rm T}_A(\tau - t') \Omega \left( S_A(\tau - t') - S_A(\tau) \right)  (\tilde{r}_j - \tilde{r}_k) \\
      &=  r_0(\tau)  + d(\tau) - d - \frac{1}{2}  \left( \tilde{r}_j(\tau) - \tilde{r}_j + \tilde{r}_k(\tau) - \tilde{r}_k  \right) - \frac{i}{2}  \sigma (\tau)  \Omega  \left[ (\tilde{r}_j (\tau) - \tilde{r}_j)  - (\tilde{r}_k (\tau) - \tilde{r}_k ) \right] \\
      & \hspace{2cm}  - \frac{i}{2}  \int_0^{\tau} \text{d}t  D (\tau-t) \Omega  \left[ (\tilde{r}_j(\tau - t) - \tilde{r}_j(\tau) ) -  (\tilde{r}_k(\tau - t) - \tilde{r}_k(\tau) ) \right] 
\end{align*}
where $D(\tau) = S_A(\tau)D S_A^{\rm T}(\tau)$. Similarly, 
\begin{align*}
C_{jk} (\tau) &=   \frac{1}{4} (\tilde{r}_j - \tilde{r}_k)^{\rm T}   \left(S^{\rm T}_A(\tau) - \mathds{1} \right)  \Omega^{\rm T} S_A(\tau) \sigma_0 S^{\rm T}_A(\tau)\Omega  \left(S_A(\tau) - \mathds{1} \right)   (\tilde{r}_j - \tilde{r}_k)   \nonumber \\ 
& \hspace{1cm}+ \frac{1}{4} (\tilde{r}_j - \tilde{r}_k)^{\rm T}  \left[  \int_0^\tau \text{d} t'    \left(S^{\rm T}_A(\tau - t') - \mathds{1} \right) \Omega^{\rm T} S_A(\tau - t')   D S_A^{\rm T}(\tau - t') \Omega \left(S_A(\tau - t') - \mathds{1} \right)   \right]   (\tilde{r}_j - \tilde{r}_k) \\
&=   \frac{1}{4} (\tilde{r}_j - \tilde{r}_k)^{\rm T}   \left(S^{\rm T}_A(\tau) - \mathds{1} \right)  \Omega^{\rm T} S_A(\tau) \sigma_0 S^{\rm T}_A(\tau)\Omega  \left(S_A(\tau) - \mathds{1} \right)   (\tilde{r}_j - \tilde{r}_k)   \nonumber \\ 
& \hspace{0.4cm} + \frac{1}{4} (\tilde{r}_j - \tilde{r}_k)^{\rm T}   \left[  \int_0^\tau \text{d} t'    \left(S^{\rm T}_A(\tau - t') - S^{\rm T}_A(\tau) +  S^{\rm T}_A(\tau) - \mathds{1} \right) \Omega^{\rm T}  D(\tau - t') \Omega \left(S_A(\tau - t')  - S_A(\tau) +  S_A(\tau) - \mathds{1} \right)   \right]   (\tilde{r}_j - \tilde{r}_k) \\
&=   \frac{1}{4} (\tilde{r}_j - \tilde{r}_k)^{\rm T}   \left(S^{\rm T}_A(\tau) - \mathds{1} \right)  \Omega^{\rm T} \left[  S_A(\tau) \sigma_0 S^{\rm T}_A(\tau) + \int_0^\tau \text{d} t D(\tau - t)  \right] \Omega  \left(S_A(\tau) - \mathds{1} \right)   (\tilde{r}_j - \tilde{r}_k)   \nonumber \\ 
& \hspace{0.4cm} + \frac{1}{4} (\tilde{r}_j - \tilde{r}_k)^{\rm T}   \left[  \int_0^\tau \text{d} t'    \left(S^{\rm T}_A(\tau - t') - S^{\rm T}_A(\tau)  \right) \Omega^{\rm T}  D(\tau - t') \Omega \left(S_A(\tau - t')  - S_A(\tau)  \right)   \right]   (\tilde{r}_j - \tilde{r}_k) \\
& \hspace{0.4cm} + \frac{1}{2} (\tilde{r}_j - \tilde{r}_k)^{\rm T}   \left[  \int_0^\tau \text{d} t'    \left(S^{\rm T}_A(\tau - t') - S^{\rm T}_A(\tau)  \right) \Omega^{\rm T}  D(\tau - t') \Omega \left(S_A(\tau)  - \mathds{1} \right)   \right]   (\tilde{r}_j - \tilde{r}_k) \\
&=   \frac{1}{4} \left[ (\tilde{r}_j (\tau) - \tilde{r}_j)  - (\tilde{r}_k (\tau) - \tilde{r}_k ) \right]^{\rm T} \Omega^{\rm T} \sigma (\tau)   \Omega  \left[ (\tilde{r}_j (\tau) - \tilde{r}_j)  - (\tilde{r}_k (\tau) - \tilde{r}_k ) \right]  \nonumber \\ 
& \hspace{2cm} + \frac{1}{4}     \int_0^\tau \text{d} t   \Big\{ \left[ (\tilde{r}_j (\tau - t) - \tilde{r}_j(\tau))  - (\tilde{r}_k (\tau - t) - \tilde{r}_k(\tau))  \right]^{\rm T}  \Omega^{\rm T}  D(\tau - t) \Omega \\&\hspace{4cm}\left[ (\tilde{r}_j (\tau - t)  + \tilde{r}_j(\tau) - 2 r_j)  - (\tilde{r}_k (\tau - t) + \tilde{r}_k(\tau) - 2 r_k)  \right]  \Big\}
\end{align*}
The phase is the same as in Table \ref{tab:unitary}, thus, concluding all the derivations of Table \ref{tab:open_app}.

\section{Doubling Solutions of Riccati Equations \label{app:doubling}}

Consider the ansatz for the covariance matrices $\sigma_{jk} = \mathds{Y}_{jk} \mathds{X}_{jk}^{-1}$, where $\mathds{X}_{jk}, Y_{jk}$ are two time-dependent $2n$-dimensional matrices which are solutions to some coupled linear ODEs: 
$\dot{X}_{jk} = A_{jk} \mathds{X}_{jk} + B_{jk} \mathds{Y}_{jk} $ and   $\dot{Y}_{jk} = C_{jk} \mathds{X}_{jk} + D_{jk} \mathds{Y}_{jk} $. The initial condition $\sigma (0) = \sigma_0$ can be found as $\mathds{Y}_{jk} = \sigma_0$ and $\mathds{X}_{jk} = \mathds{1}$. Thus, computing  
\begin{align}
    \dot{\sigma}_{jk} &= \dot{\mathds{Y}}_{jk} \mathds{X}_{jk}^{-1}- \mathds{Y}_{jk} \mathds{X}_{jk}^{-1} \dot{\mathds{X}}_{jk} \mathds{X}_{jk}^{-1} = \left(  C_{jk} \mathds{X}_{jk} + D_{jk} \mathds{Y}_{jk} \right) \mathds{X}_{jk}^{-1} - \mathds{Y}_{jk} \mathds{X}_{jk}^{-1} \left(  A_{jk} \mathds{X}_{jk} + B_{jk} \mathds{Y}_{jk} \right) \mathds{X}_{jk}^{-1} \\
    & = C_{jk} + D_{jk} (\mathds{Y}_{jk} \mathds{X}_{jk}^{-1}) - (\mathds{Y}_{jk} \mathds{X}_{jk}^{-1})A_{jk}  - (\mathds{Y}_{jk} \mathds{X}_{jk}^{-1} ) B_{jk} (\mathds{Y}_{jk} \mathds{X}_{jk}^{-1}) \\
    & = D_{jk} {\sigma}_{jk} -  {\sigma}_{jk} A_{jk} - {\sigma}_{jk}^{\rm T} B_{jk} {\sigma}_{jk} + C_{jk} 
\end{align}
Comparing to the ODE of the Table~\ref{tab:open},  one can infer that 
\begin{align}
    \frac{\text{d}}{\text{d} \tau}\begin{pmatrix}
        \mathds{X}_{jk} \\
        \mathds{Y}_{jk}
    \end{pmatrix} &= \begin{pmatrix}
         A_{jk} & B_{jk}\\
        C_{jk} & D_{jk}
    \end{pmatrix}
    \begin{pmatrix}
        \mathds{X}_{jk} \\
        \mathds{Y}_{jk}
    \end{pmatrix} =  \frac{1}{2} \begin{pmatrix}
          ( H_j + H_k   + 2 {\rm E}^{\rm T} ) \Omega  &  i ( H_j - H_k) \\
          i \Omega ( H_j - H_k)  \Omega^{\rm T } + D & \ \Omega  ( H_j + H_k   + 2 E )
    \end{pmatrix} \begin{pmatrix}
        \mathds{X}_{jk} \\
        \mathds{Y}_{jk}
    \end{pmatrix} = \mathds{A}_{jk}
    \begin{pmatrix}
        \mathds{X}_{jk} \\
        \mathds{Y}_{jk}
    \end{pmatrix} \nonumber
\end{align}
where we also included general Gaussian noise. One can find the solution of these linear ODEs in terms of the exponential of the $4n*4n$ matrix $S_{\mathds{A}_{jk}}(\tau) = \exp(\tau \mathds{A}_{jk})$ and  $\begin{pmatrix}
    \mathds{X}_{jk} (\tau)\\
    \mathds{Y}_{jk}(\tau)
\end{pmatrix}  = S_{\mathds{A}_{jk}}(\tau) \begin{pmatrix}
    \mathds{1} \\
    \sigma_0
\end{pmatrix} $. 
One can now turn to solve the first-moment ODE
\begin{equation}
    \dot{\bar{r}}_{jk} =\frac{1}{2}\left[ \left(  \Omega ( H_j + H_k +2 E )  - i   \sigma_{jk} (H_j - H_k) \right)  \bar{r}_{jk} 
     + \Omega (2d - r_j - r_k) + i  \sigma_{jk} ( r_j - r_k) \right] 
\end{equation}
via the propagator 
\begin{equation}
\label{eq:propagator_r_off}
    \mathds{P}_{jk}(\tau) = \mathcal{T} \left[ \exp \left(\frac{1}{2} \int_{0}^{\tau} \text{d} t \;  \Omega ( H_j + H_k +2 E)  - i   \sigma_{jk}(t) (H_j - H_k) \right)  \right] 
\end{equation}
where $ \mathcal{T}$ is the time ordering, such that
\begin{equation}
    \bar{r}_{jk} = \mathds{P}_{jk}(\tau) r_0 + \frac{1}{2} \mathds{P}_{jk}(\tau) \int_{0}^{\tau} \text{d} t \mathds{P}^{-1}_{jk}(t)  \left[ i  \sigma_{jk} (t) (r_j - r_k) + \Omega (2d - r_j - r_k) \right]
\end{equation}
from which one can compute the time evolution of the QRDM exponent. We should note that, often, the exponent of $\mathds{P}_{jk}(\tau) $ commutes at different times, which makes the time ordering trivial. The QRDM elements $s_{jk}$ can then be computed by direct integration of the final trivial ODE.

\section{Protocols and Operations \label{app:operation}}

In this section, we derive the maps of GBCS under Gaussian operations on the CV and common conditional operations on the qubit states. 

\subsection{Gaussian Operations and Deterministic CP Maps}

A Gaussian completely positive map $\Phi_g$ is uniquely determined by two matrices $X$ and $Y$, such that $Y+ i \Omega \geq i X \Omega X^{\rm T}$, to preserve the Heisenberg uncertainty principle. The action on a GBCS can be computed via the dual map $\Phi_g^*$, defined as $\text{Tr}_m \left[\Phi_{g} [\hat{A}]  \hat{B}\right] = \text{Tr}_m \left[\hat{A}  \Phi_{g}^* [\hat{B}]\right]$, on the displacement operators basis, such that one can use the well-known identity 
 \begin{equation}
     \Phi_g^* \left(\hat{D}_{\Omega r} \right) = {\rm e}^{-\frac{1}{4} r^{\rm T}  Y  r }\hat{D}_{\Omega X^{\rm T} r} \implies \Phi_g^* \left(\hat{D}_{- r} \right) = {\rm e}^{-\frac{1}{4} r^{\rm T} \Omega  Y \Omega^{\rm T}  r }\hat{D}_{- \Omega X^{\rm T} \Omega^{\rm T} r}
 \end{equation}
From definition, the branched characteristic functions of a GBCS it follows the update on the characteristic function under the general Gaussian map 
\begin{align}
    \chi[\Phi_g(\hat{\varrho}_{jk})] (r) &= \text{Tr}[\Phi_g[\hat{\varrho}_{jk}] \hat{D}_{-r}] = \text{Tr}[\hat{\varrho}_{jk} \Phi^*_g[\hat{D}_{- r}]] = {\rm e}^{-\frac{1}{4} r^{\rm T} \Omega  Y \Omega^{\rm T}  r } \text{Tr}[\hat{\varrho}_{jk} \hat{D}_{- \Omega X^{\rm T} \Omega^{\rm T} r} ] =  {\rm e}^{-\frac{1}{4} r^{\rm T} \Omega  Y \Omega^{\rm T}  r } \chi_{jk} (\Omega X^{\rm T} \Omega^{\rm T} r) \nonumber \\
    &=   
  \exp \left( -\frac{1}{4}  r^{\rm T}  \Omega^{\rm T}  \left( X^{\rm T} \sigma_{jk}  X + Y \right) \Omega r  +  i  r^{\rm T} \Omega^{\rm T} X \bar{r}_{jk} + s_{jk}   \right) \;,
\end{align} 
which is a GBCS (Eq.~\ref{eq:chara_jk}) under the update rules for the phase-space parameters given in Eq.~(\ref{eq:Gaussian_CP_update}).

Specific examples present in laboratories are given in the main text, and here we report other examples~\cite{serafini_quantum_2017}. Specifically, common coherent operations are the phase shift and beam splitter, defined according to the transformations
\begin{equation}
    R_{\varphi} = \begin{pmatrix} \cos\varphi & \sin\varphi \\ -\sin\varphi & \cos\varphi \end{pmatrix} \hspace{1cm}
    R_{\theta} = \begin{pmatrix} \cos\theta & 0 & \sin\theta & 0 \\ 0 & \cos\theta & 0 & \sin\theta \\ -\sin\theta & 0 & \cos\theta & 0 \\ 0 & -\sin\theta & 0 & \cos\theta \end{pmatrix}
\end{equation}
and the single- and two-mode squeezing transformations are given by 
\begin{equation}
    S_{1m} = \begin{pmatrix} s & 0 \\ 0 & 1/s \end{pmatrix} \hspace{1cm}
    S_{2m} = \begin{pmatrix} \cosh r & 0 & \sinh r & 0 \\ 0 & \cosh r & 0 & -\sinh r \\ \sinh r & 0 & \cosh r & 0 \\ 0 & -\sinh r & 0 & \cosh r \end{pmatrix}
\end{equation}

\subsection{Conditional Rotations}


For an initial squeezed state, one finds the solution of the complex Riccati equation using the method outlined in Sec.~\ref{subsec:riccati}, with the two matrices (at $\tau = 1$)
\begin{equation}
    X = \left(
\begin{array}{cc}
 \cos (\bar{\theta}) \left(\cos \theta_\Delta +\frac{i}{s} \sin \theta_\Delta \right) & \sin (\bar{\theta}) (\cos \theta_\Delta +i s \sin \theta_\Delta ) \\
 -\frac{1}{s} \sin (\bar{\theta}) (s \cos \theta_\Delta +i \sin \theta_\Delta ) & \cos (\bar{\theta}) (\cos \theta_\Delta +i s \sin \theta_\Delta ) \\
\end{array}
\right)\nonumber
\end{equation}
\begin{equation}
   Y =  \left(
\begin{array}{cc}
 \frac{1}{s}\cos (\bar{\theta}) (\cos \theta_\Delta +i s \sin \theta_\Delta ) & \sin (\bar{\theta}) (s \cos \theta_\Delta +i \sin \theta_\Delta ) \\
 -\frac{1}{s}\sin (\bar{\theta}) (\cos \theta_\Delta +i s \sin \theta_\Delta ) & \cos (\bar{\theta}) (s \cos \theta_\Delta +i \sin \theta_\Delta ) \\
\end{array}
\right)\nonumber
\end{equation}
such that one finds 
\begin{equation}
    \sigma^{\text{off}} = \left(
\begin{array}{cc}
 \frac{\cos ^2(\bar{\theta}) (\cos \theta_\Delta +i s \sin \theta_\Delta )}{s \cos \theta_\Delta +i \sin \theta_\Delta }+\frac{\sin ^2(\bar{\theta}) (s \cos \theta_\Delta +i \sin \theta_\Delta )}{\cos \theta_\Delta +i s \sin \theta_\Delta } & \frac{\left(s^2-1\right) \sin (\bar{\theta}) \cos (\bar{\theta})}{(s \cos \theta_\Delta +i \sin \theta_\Delta ) (\cos \theta_\Delta +i s \sin \theta_\Delta )} \\
 \frac{\left(s^2-1\right) \sin (\bar{\theta}) \cos (\bar{\theta})}{(s \cos \theta_\Delta +i \sin \theta_\Delta ) (\cos \theta_\Delta +i s \sin \theta_\Delta )} & \frac{\cos ^2(\bar{\theta}) (s \cos \theta_\Delta +i \sin \theta_\Delta )}{\cos \theta_\Delta +i s \sin \theta_\Delta }+\frac{\sin ^2(\bar{\theta}) (\cos \theta_\Delta +i s \sin \theta_\Delta )}{s \cos \theta_\Delta +i \sin \theta_\Delta } \\
\end{array}
\right)
\end{equation}
which can be rewritten in the form of Eq.~(\ref{eq:cov_rotation}), i.e. $\sigma^{\text{off}} = R_{\bar{\theta}} D (\theta_\Delta) R_{\bar{\theta}}^{\rm T}$, where $D (\theta_\Delta) = \text{diag} ({\rm e}^{i \psi (\theta_\Delta)} |\zeta (\theta_\Delta)|, {\rm e}^{-i \psi(\theta_\Delta)} /|\zeta(\theta_\Delta)|) $ is diagonal in its polar decomposition. Hence, by moving in the rotating frame with respect to the angle $\bar{\theta}$, i.e. $R_{\bar{\theta}} \bar{r}^{\text{off}} \to \bar{r}^{\text{off}}$, and changing the variable from $\tau$ to $\theta_\Delta$, one finds that the ODE for the first moment is 
\begin{equation}
   \frac{ \text{d} \bar{r}^{\text{off}}}{\text{d} \theta_{\Delta}} =   - i  D (\theta_\Delta) \bar{r}^{\text{off}} \implies \bar{r}^{\text{off}} = \exp \left( - i \int_{0}^{\theta_{\Delta}} \text{d} t D (t) \right) r_0 = \begin{pmatrix}
\frac{s  x_0}{s \cos \theta_\Delta +i \sin \theta_\Delta }  \\
 \frac{p_0}{\cos \theta_\Delta +i s \sin \theta_\Delta )}  
   \end{pmatrix}
\end{equation}
Given that the trace of the covariance and the norm of the first moments are both invariant under rotations, the QRDM ODE in the rotating frame reads
\begin{align}
    \frac{ \text{d} \bar{r}^{\text{(0) off}}}{\text{d} \theta_{\Delta}}   &= - i \left[ \frac{ x_0^2}{\left(  \cos \theta_\Delta +i / s \sin \theta_\Delta  \right)^2 } +\frac{p_0^2}{\left( \cos \theta_\Delta +i s \sin \theta_\Delta \right)^2} + \frac{1}{2} \left(\frac{s \cos \theta_\Delta +i \sin \theta_\Delta }{\cos \theta_\Delta +i s \sin \theta_\Delta }+\frac{\cos \theta_\Delta +i s \sin \theta_\Delta }{s \cos \theta_\Delta +i \sin \theta_\Delta }\right) \right] \nonumber
\end{align}
The off-diagonal exponent reads
\begin{align}  
s^{\text{off}} = -i \sin  \theta_\Delta \left(\frac{p_0^2}{\cos  \theta_\Delta+i s \sin  \theta_\Delta}+\frac{s x_0^2}{s \cos  \theta_\Delta+i \sin  \theta_\Delta}\right)-\frac{1}{2} \log \left(\frac{1}{s}(s \cos  \theta_\Delta+i \sin  \theta_\Delta) (\cos  \theta_\Delta+i s \sin  \theta_\Delta)\right)+\log(q_0)\nonumber
\end{align}
where $q_0$ is the initial state of the QRDM that, after the rotation, reads 
\begin{equation}
    \varrho_q^{\text{off}} = q \frac{\sqrt{s} \exp \left( -i \sin  \theta_\Delta \left(\frac{p_0^2}{\cos  \theta_\Delta+i s \sin  \theta_\Delta}+\frac{s x_0^2}{s \cos  \theta_\Delta+i \sin  \theta_\Delta}\right) \right) }{ \sqrt{s \cos  \theta_\Delta+i \sin  \theta_\Delta} \sqrt{\cos  \theta_\Delta+i s \sin  \theta_\Delta }} 
\end{equation}

\subsection{Conditional Displacements}

The case of conditional displacements, with a set of vectors $d_j$, one can directly compute 
\begin{align}
    \Phi^d[\hat{\varrho}] &= \frac{1}{(2 \pi)^n} \sum_{j, k  = 0}^{N-1}  \int_{\mathds{R}^{2n}} \text{d} r \; \chi_{jk} (r)  \hat{D}_{d_j} \hat{D}_{r} \hat{D}_{d_k}^\dag \otimes \ket{j} \bra{k}  \\
    &= \frac{1}{(2 \pi)^n} \sum_{j, k  = 0}^{N-1} \int_{\mathds{R}^{2n}} \text{d} r \; \chi_{jk} (r)  {\rm e}^{ \frac{i}{2} r^{\rm T} \Omega (d_j + d_k) + \frac{i}{2}  d_j^{\rm T}  \Omega d_k }\hat{D}_{r +d_j - d_k}  \otimes \ket{j} \bra{k} \nonumber
\end{align}
where we repeatedly used $ \hat{D}_{r_1} \hat{D}_{r_2} = {\rm e}^{- \frac{i}{2} r_1^{\rm T} \Omega r_2} \hat{D}_{r_1+r_2} =  {\rm e}^{\frac{i}{2} r_2^{\rm T} \Omega r_1} \hat{D}_{r_1+r_2}$. By changing the integration variable to $r' = r+ r_j - r_k$, one finds that the updated characteristic function after the operation is
\begin{equation}
    \chi_{jk}^{\Phi} (r') = \chi_{jk} (r' - d_j + d_k) \exp \left( { - \frac{i}{2} \bar{r'}^{\rm T} \Omega^{\rm T} (d_j+d_k) - \frac{i}{2}  d_j^{\rm T}  \Omega d_k  } \right)
\end{equation}
Substituting in Gaussian $\chi_{jk}$, it reads
\begin{align}
    \chi_{jk}^{\Phi} (r') =  \exp \Big[ -\frac{1}{4} &r'^{\rm T} \Omega^{\rm T}\sigma_{jk} \Omega r'  +  i  r'^{\rm T} \Omega^{\rm T} \left( \bar{r}_{jk} - \frac{1}{2} (d_j +d_k) - \frac{i}{2} \sigma_{jk} \Omega  (d_j - d_k) \right)  \\
    &+ s_{jk} - \frac{1}{4} \left( d_j - d_k\right)^{\rm T} \Omega^{\rm T} \sigma_{jk}  \Omega \left( d_j - d_k\right) +  i (d_j - d_k)^{\rm T} \Omega \bar{r}_{jk} - \frac{i}{2}  d_j^{\rm T}  \Omega d_k  \Big)
\end{align}
from which one can read the update rules given in Eq.~(\ref{eq:cond_dis_update}).

\section{Perturbation Theory \label{app:perturbation}}

Greek symbols label the components of vectorial representation of some quantity and the Einstein summations are performed up to $N^2$. 
Examples of DV non-orthogonal superoperators that frequently arise are non-orthogonal Hamiltonian term (such as $\hat{\sigma}_x$) and $T_1$ Lindbladian term. As we shall see, this induces a coupling between the homogenous PDEs such as the evolution of the system can be described by PDEs $\left( \partial_\tau - \mathcal{L}_{\alpha} \right) \chi_{\alpha} = V_{\alpha \beta} \chi_{\beta}$. The solution can be formulated in terms of Green's functions, which solve $\left( \partial_\tau - \mathcal{L}_{\alpha} \right) \left[\mathcal{G}_{\alpha \alpha} (\tau,  \tilde{\xi}, \tilde{\xi}') \right]  =0$ under the boundary conditions $\mathcal{G}_{\alpha \alpha} (0,  \tilde{\xi}, \tilde{\xi}') = \delta( \tilde{\xi}- \tilde{\xi}')$. Then, one can write the solution of the coupled PDEs as the sum of the homogeneous solution solving for some initial condition and a particular solution of the inhomogeneous, such that 
\begin{equation}
    \chi_\alpha (\tilde{\xi}, \tau) = \chi^h_\alpha (\tilde{\xi}, \tau) +\chi^p_\alpha (\tilde{\xi}, \tau) =\chi^h_\alpha (\tilde{\xi}, \tau) +  \int_0^{\tau} {\rm d} t \int_{\mathds{R}^{2n}} {\rm d} \tilde{\xi}'  \mathcal{G}_{\alpha \alpha} (\tau - t,  \tilde{\xi}, \tilde{\xi}') V^{\alpha \beta} \chi_{\beta} (\tilde{\xi}', t)
\end{equation}
Given that the homogeneous solution solves $\left( \partial_\tau - \mathcal{L}_{\alpha} \right) \chi^h_{\alpha} = 0$, one finds
\begin{align}
    \left( \partial_\tau - \mathcal{L}_{\alpha} \right) \chi_{\alpha} (\tilde{\xi}, \tau) &= \left( \partial_\tau - \mathcal{L}_{\alpha} \right)  \chi^i_\alpha (\tilde{\xi}, \tau) \nonumber\\  
    &=  \int_{\mathds{R}^{2n}} {\rm d} \tilde{\xi}' \delta(\tilde{\xi} - \tilde{\xi}') V^{\alpha \beta} \chi_{\beta} (\tilde{\xi}', \tau)  + \int_0^{\tau} {\rm d} t \int_{\mathds{R}^{2n}} {\rm d} \tilde{\xi}'  V^{\alpha \beta} \chi_{\beta} (\tilde{\xi}', t)  \left( \partial_\tau - \mathcal{L}_{\alpha} \right) \left[\mathcal{G}_{\alpha \alpha} (\tau - t,  \tilde{\xi}, \tilde{\xi}') \right] = V^{\alpha \beta} \chi_{\beta} (\tilde{\xi}, \tau) \nonumber
\end{align}
thus, recovering the PDE. Let us now take a Gaussian ansatz for the Green's function
\begin{equation}
    \mathcal{G}_\alpha(\tilde{\xi}, \tilde{\xi}', \tau) =   \exp \left( -\frac{1}{4} \tilde{\xi}^{\rm T} \Sigma_\alpha (\tilde{\xi}',  \tau) \tilde{\xi}  +  i  \tilde{\xi}^{\rm T} R_\alpha (\tilde{\xi}', \tau) +S_\alpha (\tilde{\xi}', \tau)\right) 
\end{equation}
to solve for $\left( \partial_\tau - \mathcal{L}_{\alpha} \right) \left[\mathcal{G}_{\alpha \alpha} (\tau,  \tilde{\xi}, \tilde{\xi}') \right]  =0$. From the same argument used in this work to solve any dynamics, in order to be a valid ansatz, $\Sigma_\alpha$, $R_\alpha$ and $S_\alpha$ have to solve the vectorial form of the ODEs derived in the other sections. The boundary condition can be formulated as the limit of a Gaussian of covariance matrix $\tilde{\Sigma}$
\begin{align}
    \mathcal{G}_{\alpha}(\tilde{\xi}, \tilde{\xi}', 0) &= \delta(\tilde{\xi}-\tilde{\xi}') = \lim_{\tilde{\Sigma} \to 0}  \frac{2^n}{\pi^n \sqrt{\text{det} \tilde{\Sigma} }} \exp \left( - \left(\tilde{\xi}- \tilde{\xi}'\right)^{\rm T} \tilde{\Sigma}^{-1} \left(\tilde{\xi}- \tilde{\xi}'\right) \right) 
\end{align}
which implies that one has to compute the dynamics with initial conditions, and 
\begin{equation}
    \Sigma_\alpha (\tilde{\xi}',0) = 4 \tilde{\Sigma}^{-1} \hspace{1cm} R_\alpha (\tilde{\xi}',0) = - i 2  \tilde{\Sigma}^{-1} \tilde{\xi}' \hspace{1cm} R^{0}_{\alpha}(\tilde{\xi}',0)  =  \ln \left[ \frac{2^n}{\pi^n \sqrt{\text{det} \tilde{\Sigma} }} \exp(- \tilde{\xi}'^{ \; \rm T } \tilde{\Sigma}^{-1} \tilde{\xi}') \right]
\end{equation}

The Volterra integral can be solved as von Neumann series with  the $N$-th order given by $0<t_N<\cdots<t_1<\tau$:
\begin{equation}
    \hspace{-1cm}
    \dket{\chi^{(N)}}(\tau,\tilde{\xi}) = \!\!\int\limits_{0<t_N<\cdots<t_1<\tau} \!\!\!\! {\rm d}t_1\cdots{\rm d}t_N \int_{(\mathds{R}^{2n})^N} \!\!\!\! {\rm d}\tilde{\xi}_1\cdots{\rm d}\tilde{\xi}_N\ \mathcal{G}(\tau-t_1,\tilde{\xi},\tilde{\xi}_1)\,\mathds{V}\, \mathcal{G}(t_1-t_2,\tilde{\xi}_1,\tilde{\xi}_2)\,\mathds{V}\cdots \mathcal{G}(t_{N-1}-t_N,\tilde{\xi}_{N-1},\tilde{\xi}_N)\,\mathds{V}\,\dket{\chi^{h}(t_N,\tilde{\xi}_N)}\ ,
\end{equation}

\subsection{DV Decay ($T_1$) \label{app:qubit_decay}}

Let us first consider the case of a single qubit. The jump operator $\hat{L}_1 = \sqrt{\Gamma_1}\, \hat{\sigma}_-$ generates the Lindbladian whose vectorized phase-space representation reads
\begin{equation}
    \Gamma_1 \Big( \sigma_-\varrho\,\sigma_+ - \tfrac{1}{2}\{\sigma_+\sigma_-,\varrho\} \Big) \;\longleftrightarrow\; -\Gamma_1 \frac{2+j+k}{4}\,\chi_{jk}(\tilde{r}) \;+\; \Gamma_1\,\delta_{j,-1}\delta_{k,-1}\,\chi_{+1,+1}(\tilde{r})\,, \qquad j,k \in \{\pm 1\}\,.
\end{equation}
The diagonal (first) term is absorbed into the homogeneous generator $\mathcal{L}$, providing the damping rates $\{\Gamma_1, \Gamma_1/2, \Gamma_1/2, 0\}$ of the four branches, while the source (second) term defines the perturbation which, in the ordered basis $\dket{\chi} = (\chi_{++}, \chi_{+-}, \chi_{-+}, \chi_{--})^{\rm T}$, reads
\begin{equation}
    \mathds{V} = \Gamma_1 \begin{pmatrix} 0&0&0&0 \\ 0&0&0&0 \\ 0&0&0&0 \\ 1&0&0&0 \end{pmatrix}\,.
\end{equation}
It is possible to note that $\mathds{V}\,\mathcal{G}\,\mathds{V} = 0$, hence $\mathds{k}^{*k} = 0$ for all $k \geq 2$, such that the Liouville--Neumann series truncates at first order, reflecting the fact that a single DV excitation supports at most one quantum jump. 

The diagonal terms, which we shall just label with $\pm$,  evolve under the Gaussian differential operators $\mathcal{L}^{\pm} = -\frac{1}{4} D^{op}\, \tilde{r}^o \tilde{r}^p + A^{op}_{\pm}\, \tilde{r}^o \partial_{\tilde{r}^p} + i\, r^m_{\pm} \Omega^{mn} \tilde{r}^n$, with $A_\pm = \Omega(E + H_\pm)$ and $r_{\pm} = r_m \pm r_q$. Inserting the Gaussian ansatz for $\mathcal{G}^{\pm}$ yields the Lyapunov-type ODEs for its covariance matrix and first moments: $\dot{\Sigma}_\pm = A_\pm \Sigma_\pm + \Sigma_\pm A^{\rm T}_\pm + D$ and $\dot{R}_\pm = A_\pm R_\pm - \Omega r_\pm$. Thus, by imposing the initial conditions enforcing $\mathcal{G}^{\pm}(0, \tilde{r}, \tilde{r}') = \delta(\tilde{r} - \tilde{r}')$ previously described, one can write the propagator in the closed form
\begin{align}
\label{eq:green_qubit}
    \mathcal{G}^{\pm}(\tau, \tilde{r}, \tilde{r}') &=  \exp \left( -\frac{1}{4} \tilde{r}^{\rm T}  \left[\int_{0}^{\tau} \text{d}t S_{A_\pm}(\tau - t) D S^{\rm T}_{A_\pm}(\tau - t) \right]  \tilde{r}  - i  \tilde{r}^{\rm T} \left(S_{A_\pm}(\tau) - \mathds{1} \right)  r_{\pm}   \right) \nonumber \\
&\hspace{0.5cm}\cdot \lim_{\tilde{\Sigma} \to 0}  \frac{2^n}{\pi^n \sqrt{\text{det} \tilde{\Sigma} }}  \exp \left( - \tilde{r}^{\rm T} S_{A_\pm}(\tau) \tilde{\Sigma}^{-1} S_{A_\pm}^{\rm T}(\tau) \tilde{r}  + 2  \tilde{r}^{\rm T} S_{A_\pm} (\tau)   \tilde{\Sigma}^{-1} \tilde{r}' - \tilde{r}^{' \; \rm T } \tilde{\Sigma}^{-1} \tilde{r}' \right) \nonumber \\  
&= \exp\!\left( -\tfrac{1}{4}\, \tilde{r}^{\rm T} D_\pm(\tau)\, \tilde{r} \;-\; i\, \tilde{r}^{\rm T} r^f_\pm(\tau) \right) \delta\!\left( S^{\rm T}_{A_\pm}(\tau)\, \tilde{r} - \tilde{r}' \right)\,,
\end{align}
with $S_{A_\pm}(\tau) = {\rm e}^{A_\pm \tau}$, $D_\pm(\tau) = \int_0^\tau {\rm d}t\, S_{A_\pm}(\tau - t)\, D\, S^{\rm T}_{A_\pm}(\tau - t)$ and $r^f_\pm(\tau) = \left( S_{A_\pm}(\tau) - \mathds{1} \right) r_\pm$. The homogeneous term can be re-derived by acting with Eq.~(\ref{eq:green_qubit}) on a Gaussian initial condition $\chi_0(\tilde{r}) = \exp( -\frac{1}{4} \tilde{r}^{\rm T} \sigma_0 \tilde{r} + i \tilde{r}^{\rm T} r_0 )$: the delta function trivializes the phase-space integral and one recovers the closed-form Gaussian solutions of Sec.~\ref{sec:dynamics}, with moments $\sigma_\pm(\tau) = S_{A_\pm} (\tau)\sigma_0 S^{\rm T}_{A_\pm} (\tau)+ D_\pm(\tau)$ and $\bar{r}_\pm(\tau) = S_{A_\pm}  (\tau) r_0 - r^f_\pm(\tau)$. Including its diagonal damping and initial population $p_+$, one finds  $\chi_{+}^{h \text{on}}(t, \tilde{r}) = p_+\, {\rm e}^{-\Gamma_1 t}\, \exp( -\frac{1}{4} \tilde{r}^{\rm T} \sigma_+(t)\, \tilde{r} + i\, \tilde{r}^{\rm T} \bar{r}_+(t) )$. By the structure of $\mathds{V}$, the only non-vanishing component of Eq.~(\ref{eq:order1}) is the ground-state one,
\begin{align}
    \chi^{(1)}(\tau, \tilde{r}) &= \Gamma_1 \int_0^\tau {\rm d}t \int_{\mathds{R}^{2n}} {\rm d}\tilde{r}'\; \mathcal{G}^{-}(\tau - t, \tilde{r}, \tilde{r}')\; \chi^{h}_{++}(t, \tilde{r}') 
    = p_+\, \Gamma_1 \int_0^\tau {\rm d}t\; {\rm e}^{ -\frac{1}{4} \tilde{r}^{\rm T} \sigma^{\Gamma}(\tau, t)\, \tilde{r} \,+\, i\, \tilde{r}^{\rm T} \bar{r}^{\Gamma}(\tau, t) \,-\, \Gamma_1 t }\,,
\end{align}
where the $\tilde{r}'$-integral is again resolved by the delta function of Eq.~(\ref{eq:green_qubit}), giving the moments of the open dynamics
\begin{align}
\label{eq:decay_moments_open}
    \sigma^{\Gamma}(\tau, t) &= S_{A_-}(\tau - t) \left[ S_{A_+}(t)\, \sigma_0\, S^{\rm T}_{A_+}(t) + D_+(t) \right] S^{\rm T}_{A_-}(\tau - t) + D_-(\tau - t)\,, \nonumber \\
    \bar{r}^{\Gamma}(\tau, t) &= S_{A_-}(\tau - t) \left[ S_{A_+}(t)\, r_0 - \left( S_{A_+}(t) - \mathds{1} \right) r_+ \right] - \left( S_{A_-}(\tau - t) - \mathds{1} \right) r_-\,.
\end{align}
Also in the open dynamics case, the state evolves in the excited branch ($+$) up to the jump time $t$, when it is transferred to the ground branch by $\mathds{V}$, and evolves in the ground branch ($-$) for the remaining time $\tau - t$; the diffusion accumulated in each branch is transported accordingly. Setting $D = 0$ and $E = 0$ recovers Eq.~(\ref{eq:decay_moments_open}). 

For an $N$-level system, the decay Lindbladian terms map to the phase space as 
\begin{equation}
    \text{Eq.(\ref{eq:Nlevel_lindblad})} \rightarrow  -\frac{\Gamma_j + \Gamma_k}{2}\, \chi_{jk}(\tilde{r}) \;+\; \delta_{jk}\, \Gamma_{j+1}\, \chi_{j+1,\,j+1}(\tilde{r})\,, \implies  V_{(j,k),(j',k')} = \Gamma_{j+1}\, \delta_{jk}\, \delta_{j'k'}\, \delta_{j',\, j+1}\,,
\end{equation}
with $j,k\in[1,...,N-1]$. The former terms can be reabsorbed in the diagonal solutions, leading to exponential decay of all the entries with the exception of the ground state ($j=k=0$),  while the latter only concerns population transfer between diagonal elements, for which the propagators can be expressed as delta functions, imposing the symplectic transformation update at each jump. On the analogy of the case of a single qubit, for an $N$-dimensional DV, there are $N$ possible jumps (from the most excited state to the ground state) which truncates the series. Each phase space jump can be accounted as phase-space parameters as a recursive Eq.~(\ref{eq:two_times_parameters}).

\section{Phase-Space Parameters in Physical Examples \label{app:examples}}

\subsection{Example 1}

On the diagonal terms, the dynamics can be computed via the two non-symplectic transformations
\begin{equation}
    S^{\text{on}}_\pm = {\rm e}^{\tau \Omega (H_{\pm} + E} ) = {\rm e}^{-\kappa\tau/2}
\begin{pmatrix}
\cosh(\omega_{\mu, \lambda} \tau)+\dfrac{ \lambda}{\omega_{\mu, \lambda}}\sinh(\omega_{\mu, \lambda} \tau) & \pm\dfrac{\mu}{\omega_{\mu, \lambda}}\sinh(\omega_{\mu, \lambda} \tau) \\
\mp\dfrac{\mu}{\omega_{\mu, \lambda}}\sinh(\omega_{\mu, \lambda} \tau) & \cosh(\omega_{\mu, \lambda} \tau)-\dfrac{ \lambda}{\omega_{\mu, \lambda}}\sinh(\omega_{\mu, \lambda} \tau)
\end{pmatrix}
\nonumber
\end{equation}
with $\omega_{\mu, \lambda} = \sqrt{\mu^2- \lambda^2}$. From the solution to the Lyapunov equation  (Table~\ref{tab:open} b), one can compute the time evolution of the covariance matrices and the first moments. At the steady state, one finds that 
  \begin{align}
       \sigma_{\pm}^\text{on} (\tau \to \infty) &=\frac{1}{4  \mu ^2+\kappa ^2-4 \lambda ^2}
\begin{pmatrix}
 4   \mu ^2+\kappa ^2+2 \kappa  \lambda  & \mp 4  \lambda  \mu  \\
 \mp 4  \lambda  \mu  & 4   \mu ^2+\kappa ^2-2 \kappa  \lambda  \\
\end{pmatrix} \\
\bar{r}_{\pm}^\text{on} (\tau \to \infty) &= \frac{2 \alpha  \sqrt{2 \kappa }}{4   \mu ^2+\kappa ^2-4 \lambda ^2}
\begin{pmatrix}
 \pm 2  \mu  \cos (\phi )-(\kappa +2 \lambda ) \sin (\phi ) \\
 (\kappa -2 \lambda ) \cos (\phi )\pm2  \mu  \sin (\phi ) \\
\end{pmatrix}
   \end{align}

Define the two branch frequencies $
    \omega_{\pm} = 
    \frac{1}{4} \left[ (\kappa \pm 2 \lambda)^{2} + 4 i \kappa \mu - 4 \mu^{2} \right] $. Solving the
Riccati equation by the doubling method of Sec.~\ref{subsec:riccati}, with the vacuum
initial condition $\sigma_{0} = \mathds{1}$, gives the diagonal off-diagonal covariance
\begin{equation}
    \sigma^{\text{off}} (\tau) = \text{diag} \left( 
        \frac{(\kappa + 2 \lambda + 2 i \mu) \sinh \left( \tau \sqrt{\omega_{-}} \right)
        + 2 \sqrt{\omega_{-}} \cosh \left( \tau \sqrt{\omega_{-}} \right)}
        {(\kappa - 2 \lambda + 2 i \mu) \sinh \left( \tau \sqrt{\omega_{-}} \right)
        + 2 \sqrt{\omega_{-}} \cosh \left( \tau \sqrt{\omega_{-}} \right)} ,
       \frac{(\kappa - 2 \lambda + 2 i \mu) \sinh \left( \tau \sqrt{\omega_{+}} \right)
        + 2 \sqrt{\omega_{+}} \cosh \left( \tau \sqrt{\omega_{+}} \right)}
        {(\kappa + 2 \lambda + 2 i \mu) \sinh \left( \tau \sqrt{\omega_{+}} \right)
        + 2 \sqrt{\omega_{+}} \cosh \left( \tau \sqrt{\omega_{+}} \right)}
    \right) \nonumber
\end{equation}
The first moments follow from the propagator $\mathds{P}_{jk}$ of Eq.~(\ref{eq:propagator_r_off})
via $\bar{r}^{\text{off}} = \mathds{P}_{jk} (\tau) \int_{0}^{\tau} \text{d} t \,
\mathds{P}^{-1}_{jk} (t) \, \Omega d$, with $d = \sqrt{2 \kappa} \, \alpha (\cos \phi , \sin \phi)^{\rm T}$
and $\bar{r}_0 = (0,0)^{\rm T}$, 
\begin{equation}
    \bar{r}^{\text{off}} (\tau) = 2 \sqrt{2} \, \alpha \sqrt{\kappa}
    \begin{pmatrix}
        \sin \phi \;
        \dfrac{\dfrac{\kappa - 2 \lambda + 2 i \mu}{2 \sqrt{\omega_{-}}}
        \left[ \cosh \left( \tau \sqrt{\omega_{-}} \right) - 1 \right]
        + \sinh \left( \tau \sqrt{\omega_{-}} \right)}
        {(\kappa - 2 \lambda + 2 i \mu) \sinh \left( \tau \sqrt{\omega_{-}} \right)
        + 2 \sqrt{\omega_{-}} \cosh \left( \tau \sqrt{\omega_{-}} \right)} \\[24pt]
        - \cos \phi \;
        \dfrac{\dfrac{\kappa + 2 \lambda + 2 i \mu}{2 \sqrt{\omega_{+}}}
        \left[ \cosh \left( \tau \sqrt{\omega_{+}} \right) - 1 \right]
        + \sinh \left( \tau \sqrt{\omega_{+}} \right)}
        {(\kappa + 2 \lambda + 2 i \mu) \sinh \left( \tau \sqrt{\omega_{+}} \right)
        + 2 \sqrt{\omega_{+}} \cosh \left( \tau \sqrt{\omega_{+}} \right)}
    \end{pmatrix} ,
\end{equation}
The ODE of the QRDM is
\begin{align}
   \dot{s}^{\text{off}} =& - 8 i \alpha^{2} \kappa \mu \Big[ 
    \cos^{2} \phi_d \;
    \frac{\left( \dfrac{\kappa + 2 \lambda + 2 i \mu}{2 \sqrt{\omega_{+}}}
    \left[ \cosh \left( \tau \sqrt{\omega_{+}} \right) - 1 \right]
    + \sinh \left( \tau \sqrt{\omega_{+}} \right) \right)^{2}}
    {\left( (\kappa + 2 \lambda + 2 i \mu) \sinh \left( \tau \sqrt{\omega_{+}} \right)
    + 2 \sqrt{\omega_{+}} \cosh \left( \tau \sqrt{\omega_{+}} \right) \right)^{2}} \\
    &\hspace{2cm}
    + \sin^{2} \phi_d \;
    \frac{\left( \dfrac{\kappa - 2 \lambda + 2 i \mu}{2 \sqrt{\omega_{-}}}
    \left[ \cosh \left( \tau \sqrt{\omega_{-}} \right) - 1 \right]
    + \sinh \left( \tau \sqrt{\omega_{-}} \right) \right)^{2}}
    {\left( (\kappa - 2 \lambda + 2 i \mu) \sinh \left( \tau \sqrt{\omega_{-}} \right)
    + 2 \sqrt{\omega_{-}} \cosh \left( \tau \sqrt{\omega_{-}} \right) \right)^{2}} \Big] \nonumber \\
 &- \frac{i \mu}{2} \left[ 
    \frac{(\kappa - 2 \lambda + 2 i \mu) \sinh \left( \tau \sqrt{\omega_{+}} \right)
    + 2 \sqrt{\omega_{+}} \cosh \left( \tau \sqrt{\omega_{+}} \right)}
    {(\kappa + 2 \lambda + 2 i \mu) \sinh \left( \tau \sqrt{\omega_{+}} \right)
    + 2 \sqrt{\omega_{+}} \cosh \left( \tau \sqrt{\omega_{+}} \right)} 
    + \;  \frac{(\kappa + 2 \lambda + 2 i \mu) \sinh \left( \tau \sqrt{\omega_{-}} \right)
    + 2 \sqrt{\omega_{-}} \cosh \left( \tau \sqrt{\omega_{-}} \right)}
    {(\kappa - 2 \lambda + 2 i \mu) \sinh \left( \tau \sqrt{\omega_{-}} \right)
    + 2 \sqrt{\omega_{-}} \cosh \left( \tau \sqrt{\omega_{-}} \right)} \right] \nonumber.
\end{align}
where the real part of the first two lines is $\dot{\mathcal{C}}_{\bar{r}}$ and  the last line is $\dot{\mathcal{C}}_{\sigma}$.
For conciseness, here, we have set $H_{q}^{0} = 0$, as a pure frequency shift that does not affect the dephasing rate.
In the limit $\tau \to \infty$ one has
$\dot{\mathcal{C}}_{\bar{r}}  \to \text{Re} \left( \sqrt{\omega_{+}} + \sqrt{\omega_{-}} \right) / 2 - \kappa / 2$
and
$\dot{\mathcal{C}}_{\bar{r}}  \to 2 \mu^{2} \kappa^{2} \alpha^{2}
\left( \cos^{2} \phi / | \omega_{+} |^{2} + \sin^{2} \phi / | \omega_{-} |^{2} \right)$,
recovering Eqs.~(2) and~(4) of Ref.~\cite{eddins_high_2019}.

 On the diagonal term, the steady state of the CV can be found by solving 
\begin{align}
0 &=   \left( j \mu \Omega  + \lambda \sigma_x  - \kappa \mathds{1} \right)    \sigma_{jj}^s - \sigma_{jj}^s  \left( j \mu \Omega  + \lambda \sigma_x  - \kappa \mathds{1} \right)  + \kappa \mathds{1} \hspace{1cm} 
        0  = \left( j \mu \Omega  + \lambda \sigma_x  - \kappa \mathds{1} \right)    \bar{r}_{jj}^s + 
      \Omega  d 
   \end{align}
and, for the off-diagonal, for $j= 1$ and $k=-1$,
\begin{align}
0 & = \left(  \lambda \sigma_x  - \kappa \mathds{1} \right)    \sigma_{jk}^s - \sigma_{jk}^s  \left(\lambda \sigma_x  - \kappa \mathds{1} \right)  + \kappa \mathds{1} + i  \mu \left( \sigma_{jk}^s \sigma_{jk}^s - \mathds{1} \right)
\nonumber  \\
        0 & = \left[  \frac{j+k}{2} \mu \Omega  + \lambda \sigma_x  - \kappa \mathds{1} - i  \frac{j-k}{2} \mu \sigma_{jk}^s \right]    \bar{r}_{jk}^s + 
      \Omega  d  
   \end{align}
which agrees with the time-dependent quantities as $\tau \to \infty$.

\subsection{Example 2}

When decay in the DV is not considered, the time evolution of the phase-space parameters of the Stern-Gerlach interferometer undergoing diffusive dynamics with an unknown external force can be computed from Eq.~(\ref{eq:Lyapunov_sol}) and Table~\ref{tab:open_app}. This is achieved by a single symplectic transformation describing the quadratic evolution of the quantum harmonic oscillator of the mass, which reads
\begin{equation}
    S_{\text{qho}}(\tau) := {\rm e}^{\tau \Omega H} = \begin{pmatrix}
        \cos \tau &  \sin \tau \\ 
        -\sin \tau & \cos \tau
    \end{pmatrix} \;,
\end{equation}
describing a rotation of the phase space. Thus, the phase-space parameters of an initial squeezed thermal state with first and second moments $r_0 = (0,0)^{\rm T}$  and $\sigma_0 = (1+2N_p) \text{diag}(s, 1/s)$, respectively, and a qubit in initial state given by the density matrix  $\varrho^q(0) = \ket{+}\bra{+}$ evolves to a GBCS with the time-dependent phase-space parameters are, a single covariance matrix
\begin{equation} \label{eq:covariance_SG}
    \hspace{-1cm}\sigma(\tau) = (1 + 2 N_p) 
\begin{pmatrix}
 s \cos ^2(\tau) +\sin ^2(\tau) /s &  \left(1/s - s \right) \sin \tau  \cos \tau  \\
 \left(1/s - s \right) \sin \tau  \cos \tau   & s \sin ^2(\tau) +\cos ^2(\tau) /s \\    
\end{pmatrix}
+ \Gamma_x
\begin{pmatrix}
     \tau -\sin \tau  \cos \tau  & \sin ^2(\tau)  \\
 \sin ^2(\tau)  & \tau +\sin \tau  \cos \tau  \\
\end{pmatrix}
\end{equation}
where we recognise the first term as the covariance matrix of a squeezed thermal state undergoing a unitary evolution in a QHO and the second being the additional term adding diffusion (with a linear increment in $\tau$). The three independent vectors of first moments are
\begin{equation}
    \bar{r}^{\text{on}}_{\pm} (\tau)
    = \left(f_u \pm f_q\right)
\begin{pmatrix}
1 - \cos \tau  \\
  - \sin \tau  
\end{pmatrix} \nonumber
\end{equation}
\begin{equation}
 \bar{r}^{\text{off}}(\tau) = f_u \begin{pmatrix}
          1 - \cos \tau  \\
        \sin \tau 
    \end{pmatrix} + i f_q 
\left[ (1+2 N_p)
\begin{pmatrix}
  \sin \tau  \left( \left(1/s - s\right) \cos \tau  - 1/s \right) \\
 2 \sin ^2\left(\frac{\tau }{2}\right) \left( \left(s-1/s\right) \cos \tau + s \right) \\
\end{pmatrix}
+ \Gamma_x
\begin{pmatrix}
  - 4 \sin^4(\frac{\tau}{2}) \\
  \tau + \sin \tau (\cos \tau - 2)\\

\end{pmatrix}
\right] \nonumber
\end{equation}
and the QRDM parameters are
\begin{equation}
     \mathcal{C} (\tau) = f_q^2 \left[2 (1 + 2 N_p)  \sin ^2\left(\frac{\tau }{2}\right) \left(\left(s-\frac{1}{s}\right) \cos (\tau )+s+\frac{1}{s}\right) +  \Gamma_x \left( 3 \tau + \sin \tau (\cos \tau - 4) \right) \right] + \Gamma_z \tau + \Gamma_1 \tau/2
\end{equation}
and $\phi(\tau) = 2 f_q f_u (\tau -\sin \tau ) + \frac{1}{2} \omega_q \tau $. 

In order to account for the contribution of the decay in the DV, let us first compute the phase-space parameters of the unravelling of the inhomogeneous term of Eq.~(\ref{eq:decay_qubit}) via Eq.~(\ref{eq:decay_moments_open}). We note that the covariance matrix is given by Eq.~(\ref{eq:covariance_SG}) as the SG coupling, being linear, does not affect the second moments, i.e. $S_{A_\pm} = S_{\text{qho}}$. The first moment can be readily computed using $r_\pm = r_m \pm r_q = (f_u \pm f_q,\;0)^{\rm T}$ and $r_0 = (0,0)^{\rm T}$, such that
\begin{align}
    \bar{r}^{\Gamma}(\tau, t) 
    &= -  \left( S_{\text{qho}}(\tau)  - \mathds{1}  \right) r_m -  \left( S_{\text{qho}}(\tau) -  2 S_{\text{qho}}(\tau - t) + \mathds{1}  \right) r_q \\
    &= f_u \begin{pmatrix}
1 - \cos \tau  \\
   - \sin \tau  
\end{pmatrix} + f_q \begin{pmatrix}
-1 - \cos \tau  \\
  - \sin \tau  
\end{pmatrix} + 2 f_q \begin{pmatrix}
 \cos ( \tau  - t) \\
   \sin (\tau-t)  
\end{pmatrix} =  \bar{r}^{\Gamma}(\tau) + 2 f_q \begin{pmatrix}
 \cos ( \tau  - t) \\
   \sin (\tau-t)  
\end{pmatrix}
\end{align}
The last term is the only quantity which is a function of $t$, such that one can write the characteristic function of the decayed part as
\begin{equation}
    \chi^{(1) } (\tau, \tilde{\xi})  = p_+ \Gamma_1  {\rm e}^{-\frac{1}{4} \tilde{\xi}^{\rm T} \sigma (\tau) \tilde{\xi}  +  i  \tilde{\xi}^{\rm T} \bar{r}^{\Gamma} ( \tau ) }  \ I_\Gamma (\tau, \tilde{\xi}) 
\end{equation}
where 
one can compute the time integral via the Bessel functions $J_{n}(z)$ as 
\begin{align}
\label{eq:I_bessel_series}
    \hspace{-0.5cm}I_\Gamma (\tau, \xi ) &=    \int_0^{\tau}  {\rm d} t {\rm e}^{ 2 i f_q \left(  p \cos(\tau - t) + x \sin(\tau - t)  \right)  - \Gamma_1 t } =  \sum_{n=-\infty}^{\infty} i^n \, J_n(2 f_q |\xi| ) \, {\rm e}^{- i n\phi} \,
\frac{{\rm e}^{in\tau} - {\rm e}^{-\Gamma_1\tau}}{i n + \Gamma_1}  \nonumber
\end{align}
where  $ |\xi| =\sqrt{x^2 + p^2}$ and $\phi =\operatorname{atan2}(x,p) $. It follows that, for the initial state $\varrho_q = \ket{+} \bra{+}$ \begin{equation}
    ke t^q(\tau)= \begin{pmatrix}
        \chi^{\text{on}}_+  (\tau, 0)  & \chi^{\text{off}} (\tau, 0) \\
        \chi^{\text{off}} (\tau, 0)^*  & \chi^{\text{on}}_-  (\tau, 0)  + \chi^{(1) } (\tau, 0) 
    \end{pmatrix} = \frac{1}{2} \begin{pmatrix}
        {\rm e}^{-\Gamma_1 \tau} &  {\rm e}^{ - \mathcal{C} (\tau) + i  \phi (\tau) } \\
        {\rm e}^{ - \mathcal{C} (\tau) - i  \phi(\tau)}   & 2 - {\rm e}^{-\Gamma_1 \tau} \\ 
    \end{pmatrix} \;,
\end{equation}
as $I_\Gamma (\tau, 0 )  =  1 - {\rm e}^{-\Gamma_1 \tau}$ (as for $n\neq 0$, $J_n(0) = 0 $). Thus, the outcome probabilities are given by Eq.~(\ref{eq:probability_SG}).

Specifically, at time $\tau = \pi$, they reduce to 
\begin{equation}
    \sigma(\pi) = \begin{pmatrix}
         s (1 + 2  N_p) + \pi  \Gamma_x & 0\\ 
        0 & \frac{1}{s} (1 + 2 N_p) + \pi  \Gamma_x
    \end{pmatrix} \hspace{1cm}
    \bar{r}^{\text{on}}_{\pm} (\pi)
    =
\begin{pmatrix}
2 \left(f_u \pm f_q\right) \\
 0
\end{pmatrix} \hspace{1cm}
\bar{r}^{\text{off}}(\pi) =  \begin{pmatrix}
          2 f_u - 4 i f_q \Gamma_x \\
       i f_q \left(2 (1 + 2 N_p)/s + \pi \Gamma_x \right)
    \end{pmatrix} \nonumber
\end{equation}
\begin{equation}
    \mathcal{C} (\pi) = f_q^2 \left(4 (1 +2 N_p)/s + 3 \pi \Gamma_x \right) +  \pi \Gamma_z  +  \Gamma_1 \pi /2 \hspace{1cm} \phi(\pi) = 2 \pi f_q f_u +  \frac{\pi}{2} \omega_q  \nonumber
\end{equation}
and at  $\tau = 2 \pi$
\begin{equation}
    \sigma(2 \pi)  = \begin{pmatrix}
         s (1 + 2  N_p) + 2 \pi  \Gamma_x & 0\\ 
        0 & \frac{1}{s} (1 + 2 N_p) + 2 \pi  \Gamma_x
    \end{pmatrix}  \hspace{1cm}
    \bar{r}^{\text{on}}_{\pm} (2 \pi)
    =
\begin{pmatrix}
0 \\
 0
\end{pmatrix} \hspace{1cm}
\bar{r}^{\text{off}}(2 \pi) =  \begin{pmatrix}
   0  \\
        2 i \pi f_q  \Gamma_x
    \end{pmatrix} \nonumber
\end{equation}
\begin{equation}
    \mathcal{C} (2 \pi) = 6 \pi  f_q^2 \Gamma_x + 2 \pi \Gamma_z  +  \Gamma_1 \pi \hspace{1cm} \phi(2 \pi) = 4 \pi f_q f_u +  \pi \omega_q   \nonumber
\end{equation}

One can note that, at $\tau = \pi$, 
\begin{align}
     I_\Gamma (\pi, \xi ) &= \Gamma_1 \sum_{n=-\infty}^{\infty} i^n \, J_n(z) \, {\rm e}^{i n\phi} \,
\frac{(-1)^n - {\rm e}^{-\Gamma_1\pi}}{i n + \Gamma_1}   \\
& =  J_0(z)\big(1-{\rm e}^{-\Gamma\pi}\big) 
+ \sum_{n=1}^{\infty} J_n(z)\,\frac{2 i^n  \big[(-1)^n - {\rm e}^{-\Gamma_1\pi}\big]}{1 + n^2/\Gamma_1^2}
\left[\cos( n\phi ) + n \sin( n\phi ) /\Gamma_1 \right]
\end{align}
where  we used the symmetry of the Bessel functions $J_{-n}(z) = (-1)^{n} J_{n}(z)  $. 

Having solved the dynamics, let us now focus on the Wigner negativity of the system at $\tau = \pi$. The classical contribution ($\mathcal{W}_c$) is the weighted sum of two Gaussians of covariance matrices given by $\sigma^{-1}(\tau) = (1+2N)\text{diag}(s,1/s) + \pi \Gamma_x \mathds{1}$ and centreed at $\bar{r}^{\text{on}}_{\pm}(\pi) =  (2(f_u \pm f_q), 0)^{\rm T}$. They represent the statistical mixture of the Gaussian processes of the diagonal terms of the density matrix.
The quantum contribution leads to the typical fringes of GBCSs given, at time $\tau = \pi$, by
\begin{equation}
   \mathcal{W}_q(r, \pi) =  2 {\rm e}^{-(\mathcal{C}_W (r)+ \mathcal{C} (\pi) )} \cos (\phi_W(r) + \phi (\pi)),
\end{equation}
where the Wigner decay and phase, functions of space variables $\tilde{x}$ and $\tilde{p}$, are
\begin{align}
    \mathcal{C}_W =   (1+2N_p) \frac{8 f_q^2}{s} - \frac{\left(\tilde{x}-2 f_u\right)^2 - (4 \Gamma_x f_q)^2}{\pi  \Gamma_x+  (1 +2 N_p)s} 
     - \frac{\tilde{p}^2 - (\pi \Gamma_x f_q)^2 }{\pi  \Gamma_x + (1 + 2 N_p)/s }\;, \nonumber
\end{align}
and $\phi_W $ is given in Sec.~\ref{sec:physical_examples}.

\end{document}